\documentclass[a4paper,12pt]{article}
\usepackage{times,cite,amsmath,amsfonts,amsthm,fullpage,graphicx}

\newcommand{\bt}{\mathbf{t}}
\renewcommand{\l}{\langle}
\renewcommand{\r}{\rangle}

\theoremstyle{plain}
\newtheorem{Lemma}{Lemma}

\theoremstyle{remark}

\newtheorem{corollary}{Corollary}[section]
\newtheorem{definition}{Definition}[section]

\newtheorem{proposition}{Proposition}[section]

\newtheorem{examps}{Examples}[section]

\newtheorem{lemma}{Lemma}[section]
\newtheorem{remark}{Remark}[section]
\newtheorem{remarks}[remark]{Remarks}

\def\bx{\begin{example}}
\def\ex{\end{example}}
\def\bxs{\begin{examps}. \rm\begin{enumerate}}
\def\exs{\end{enumerate}\end{examps}}
\def\bd{\begin{definition}}
\def\ed{\end{definition}}
\def\bp{\begin{proposition}\rm}
\def\ep{\end{proposition}}
\def\bc{\begin{corollary}}
\def\ec{\end{corollary}}
\def\bl{\begin{lemma}\em}
\def\el{\end{lemma}}
\def\be{\begin{equation}}
\def\ee{\end{equation}}
\def\br{\begin{remark}\rm\small}
\def\er{\end{remark}}
\def\brs{\begin{remarks}.\\ \rm\
\begin{enumerate}}
\def\ers{\end{enumerate}\end{remarks}}
\def\bea{\begin{eqnarray}}
\def\eea{\end{eqnarray}}
\def\bfig{\begin{figure}[!ht]}
\def\efig{\end{figure}}

\def\det{\mathrm {det}}

\def\ln{\mathrm {ln}}

\def\diag{\mathrm {diag}}

\def\&{&{\hskip -20pt}}

\def\AA{{\mathcal A}}

\def\Nb{{\mathbf N}}

\def\Zb{{\mathbf Z}}

\def\time{{\textsc t}}

\def\bt{{\bf t}}
\def\n{|\nu \rangle}
\def\nn{|\nu,n \rangle}
\def\llambda{\langle\lambda |}
\def\lll{\langle \lambda,l |}

\def\lrl{|\lambda,l \rangle}

\def\mlm{\langle\mu,m |}
\def\mr{|\mu \rangle}
\def\mrm{|\mu,m \rangle}

\newcount\YDcount\YDcount=0
\def\YDsize{10pt}

\def\YD#1{%
\ifnum#1=0
 \ifnum\YDcount=0 \ifx\varnothing\undefined\emptyset\else\varnothing\fi
 \else\vskip1.4pt\egroup\YDcount=0\fi
\else
 \ifnum\YDcount=0 \YDcount=1\vcenter\bgroup\vskip1pt
 \else\nointerlineskip\fi
 \vbox{\hrule\hbox{\vrule height\YDsize
 \loop\hskip\YDsize\vrule\ifnum\YDcount<#1\advance\YDcount1\repeat}\hrule
 \kern-0.4pt}\expandafter\YD
\fi}

\begin{document}

\begin{center}
\begin{Large}\fontfamily{cmss}
\fontsize{17pt}{27pt} \selectfont \textbf{Fermionic construction
of tau functions and random processes} \footnote{Work of (J.H.)
supported in part by the Natural Sciences and Engineering Research
Council of Canada (NSERC) and the Fonds FCAR du Qu\'ebec; that of
(A.O.) by the Russian Academy of Science program  ``Fundamental
Methods in Nonlinear Dynamics" and  RFBR grant No 05-01-00498.}
\end{Large}\\
\bigskip
\begin{large} {J. Harnad}$^{\dagger \ddagger}$\footnote{harnad@crm.umontreal.ca}
and {A. Yu. Orlov}$^{\star}$\footnote{orlovs@wave.sio.rssi.ru}
\end{large}
\\
\bigskip
\begin{small}
$^{\dagger}$ {\em Centre de recherches math\'ematiques,
Universit\'e de Montr\'eal\\ C.~P.~6128, succ. centre ville,
Montr\'eal,
Qu\'ebec, Canada H3C 3J7} \\
\smallskip
$^{\ddagger}$ {\em Department of Mathematics and Statistics,
Concordia University\\ 7141 Sherbrooke W., Montr\'eal, Qu\'ebec,
Canada H4B 1R6} \\
\smallskip
$^{\star}$ {\em Nonlinear Wave Processes Laboratory, \\
Oceanology Institute, 36 Nakhimovskii Prospect\\
Moscow 117997, Russia } \\
\end{small}
\end{center}
\bigskip

\begin{abstract}
Tau functions expressed as fermionic expectation values
\cite{DJKM} are shown to provide a natural  and straightforward
description of a number of random processes and statistical models
involving hard core configurations of  identical particles on the
integer lattice, like a discrete version simple exclusion
processes (ASEP), nonintersecting random walkers, lattice Coulomb
gas models and others,  as well as providing a powerful tool for
combinatorial calculations involving paths between pairs of
partitions. We study the decay of the initial step function within
the discrete ASEP (d-ASEP) model as an example.

\end{abstract}

\section{Introduction}

Random processes and integrable systems are often regarded as
opposites extremes in the study of dynamical systems, since the
first, by its very definition involves indeterminacy and
probabilistic considerations, while the second represents
deterministic evolution with the highest degree of coherence in
time and space. However if we consider the simplest example of
motion of a free quantum particle we find that this simplest model
combines integrability with randomness. The Brownian motion and
the motion of quantum particles may be described by similar tools,
say, in terms of correlation functions and integrals over paths.

If we consider quantum integrable systems, beginning with the
earliest studies on spin systems solvable via the Bethe ansatz, it
is clear that, like all quantum systems, these involve
probabilistic considerations, involving averaging and statistical
correlations, and therefore such systems combine integrability
with randomness. In fact, many are solvable using
 the same methods as certain related statistical systems  since
the Hamiltonians may be embedded in the same family of commuting
operators as the transfer matrices governing solvable statistical
models. Later, starting with \cite{GS1}, \cite{GS2}, the success
was achieved in applying methods of Bethe ansatz to various
driven-diffusive and non-equilibrium systems like asymmetric
simple exclusion process (ASEP).

We shall focus on the relations of such systems to classical
integrable systems rather than to quantum ones.

   It was first shown in the work of McCoy, Tracy and Wu on
the Ising model that certain correlation functions in solvable
statistical models satisfy Painleve equations, which naturally
arise in integrable systems under scaling reductions. In the
series of papers by the Kyoto school, a systematic approach to
integrable systems  based on fermionic constructions of tau
functions was developed, in which remarkable links between the
quantum and classical  interpretations appeared.  Classically, a
tau function may be seen as a sort of universal ``potential'' for
families of integrable equations like the KP  and TL hierarchies.
(In the Hamiltonian setting, it may be interpreted as Hamilton's
principle function evaluated along the integral manifolds mapped
out by a complete set of commuting flows.) At the same time, it
has an   interpretation as a certain correlation function for
models of free fermions. Although in the classical integrable
systems context this formulation at first appeared to simply be an
interesting reinterpretation of the previous approaches to
classical integrable systems, based on isospectral flows of linear
operators, inverse spectral theory and the Zakharov-Shabat's
dressing method \cite{ZSh}, it subsequently turned out to be a
powerful tool linking soliton theory with many other fields of
physics and mathematics.

     Tau functions have long been known to play a central role in the
 remarkable links found between random matrix models and integrable
 systems \cite{GMO},  \cite{FIK}, \cite{AvM}. Subsequent applications of
 tau functions were  also found in combinatorics, probability theory, statistical
 physics  and quantum chaos. (See, e.g.,  the series of papers by
 C.Tracy, G. Widom,  A.Okounkov,  A. Borodin, K.Johansson, J. Baik,
 P. Forrester, P. Zinn-Justin, M. Adler,  P. van Moerbeke. In particular see
 \cite{TW},\cite{Ok},\cite{OkR1},\cite{Bor},\cite{BO1},\cite{BO2},\cite{BO3},
 \cite{Johansson},
 \cite{Baik},\cite{F},\cite{FIK},\cite{Forr1},\cite{Forr2},\cite{AMS},\cite{AvM2}).
 Some of these links are quite subtle, while others are still not completely
 understood. They appear however to be quite central
 to the seemingly endless new applications of ideas and methods
 originating in the  modern theory of integrable systems to
these other domains.

In this work, we  present some direct links of the fermionic
construction of tau functions as developed in \cite{DJKM},\cite{JM}
with certain types of random processes and combinatorial problems
that arise in relation to partitions. As in \cite{DJKM},\cite{JM},
we use the language of free fermions, Maya diagrams and Young diagrams.
 This ``quantum-like'' formulation turns out to
be very fruitful in providing natural links between many
stochastic systems and classical integrability, as well as
suggesting new methods for approaching the combinatorial aspects
common to these problems. Suitably interpreted, it
 will be shown to provide a natural and straightforward description of a
 number of random processes and statistical models. This approach is
 based on relating a  natural combinatorially defined basis for the fermionic
 Fock space to states of the random system. To each basis Fock vector
is associated, both a $1D$ configuration of a lattice gas  (via its  ``Maya diagram'')
 and a Young diagram, which may be viewed as a
region of $2D$ hard core particles. This provides a fermionic
setting for well-known models like asymmetric simple exclusion
processes,1D- lattice log gases subject non-intersecting lattice
walkers and models of fluctuating interfaces. Other work along
similar or related lines may be found, e.g. in the work of
Okounkov, Olshansky, Borodin, Reshetihin, and others.

The specific random process or equilibrium statistical model is determined by
specification of the tau function and the interpretation of time variables.
A central tool consists of the well-known \cite{Tinit},\cite{Tak1, Tak2} Schur function
expansions of tau functions,
 \be
 \label{doubleSchur}
\tau(\bf t,{\bar{\bf
t}})=\sum_{\lambda,\mu}\tau_{\lambda,\mu}s_\lambda({\bf
t})s_\mu({\bar{\bt}}),
 \ee
 where the sum is over all pairs of partitions $\lambda, \mu$,
which may be deduced very simply from the fermionic representation.
We show that the tau function $\tau_{\lambda,\mu}$ may be interpreted as a
generating function for the (unnormalized) transition
probabilities between states of certain models of stochastic
systems. After a suitable normalization, the procedure also leads to
 discrete analogs of models of random orthogonal matrices.

In the theory of integrable systems the variables ${\bt}$ and
${\bar{\bt}}$ are known as ``higher times'' describing the simultaneous
evolution under commuting flows, and providing solutions to integrable
``hierarchies'', such as the Toda lattice TL and KP systems.
 In the present work, these variables are either chosen to
 have specific values, thereby defining the statistical weight of
 a given configuration, or just  play the role of formal expansion
parameters which,  through the Schur function series allow the tau
 function to be interpreted as a generating function for transition
 probabilities.

We use the generating tau function for constructing models of
random walks of non-intersecting ("hard core") particles along one
dimensional lattice. We obtain two types of models introduced by
M. Fisher \cite{F}: (a) random turn models and (b) vicious walkers
(whose particular case M.Fisher called lock step models). In
models (a) at each tick of the clock a randomly chosen walker
takes a random step, while in models (b) a certain number of
walkers take random step in each time instant. In both types of
models each site may be occupied by only one walker at the same
time.

The alternative approach to the same models is to consider the
related Young diagrams, which are $2D$ figures which are in
one-to-one correspondence with $1D$ configurations of hard core
particles. This yields models of oscillating domains.

Models (a) will be related to the repeating actions of elements of
the so-called ${\widehat gl}(\infty)$ Lie algebra to Fock vectors
and tau functions. The number of these actions is identified with
the discrete time $\time$ parameter of the stochastic process.
Models (b) will be related to the repeating action of elements of
the related ${\widehat GL}(\infty)$ group. The important role of
group ${\widehat GL}(\infty)$ and its algebra in theory of
integrable systems was discovered in \cite{DJKM}, \cite{JM}.




The structure of the paper is as follows.

Second part of the introductory section is devoted to the general
conception of our approach. Here we introduce a time-dependent
normalization function ${Z}_\nu(\time)$ which plays a central role
in the approach. In section $2$ we consider models of type (a).
After necessary preliminaries we come to subsection 2.3 where we
relate a graph to each Toda lattice tau function and consider a
certain random turn walk along this graph. Here notions of path
and of permutation factor of a path are introduced. In subsection
2.4 we show that each tau function is a generating functions which
counts differences of weights for two types of random motion along
the related graph, which are notable by a value of a permutation
factor of their paths. In subsection 2.5 we consider known models
of Coulomb lattice gas in thermodynamical equilibrium, see
\cite{Forr1} for a review. The Coulomb potential is modified in a
way that the vacuum configuration (the step function presenting
the Fermi surface) has minimal energy. We point out a two fold
relation of these models to classical integrable systems. In
subsection 2.6 we introduce a model of random turn walkers in
external potential which may be viewed as a discrete-time version
of ASEP (d-ASEP) with variable site-depending hopping rates and
compare related normalization function with the partition function
of a model of Coulomb gas in thermodynamical equilibrium. We show
that ${Z}_0(\time)$ may be viewed as a certain correlation
function in a discrete model of a random ensemble of orthogonal
$\time \times \time$ matrices. For constant hopping rate we find
the asymptotic configuration in the large $\time$ limit. We found
that the limiting shape coincides with one obtained by Vershik and
Kerov in \cite{Kerov-Vershik} in the study of a limiting shape of
random partitions with fixed large area which are distributed
according to the so-called Plancherel measure. In our case the
size of asymptotic configuration depends on the hopping rate as it
is given by formula (\ref{R}). In subsection 2.7 we introduce
additional dependence of the hopping rates which now depend on a
mutual position of particles via the modified Coulomb potential
with an arbitrary charge and show how it corrects the size of the
asymptotic configuration. In section 3 we consider models (b). In
subsection 3.1 we show how the so-called determinantal ensembles
\cite{soshnikov} appear as a result of actions of elements of
${\widehat GL}(\infty)$ group on tau functions. In subsection 3.2
 we mark the convenience of usage of multi-component fermions which
allows in a simple way to construct fermionic representations for
models of vicious walkers in an external potential and via Wick's
theorem yields answers for correlation functions in form of
determinants. In subsection 3.3 we consider models of "random
layering" and models of "rocks" (figures which generalize diagrams
of the so-called $3D$ partitions) which we believe were basically
known in certain versions, for instance, as the so-called
polynuclear growth model. Here again we introduce the external
potential which results in a variable site-dependent rate (this is
the hopping rate in $1D$ picture and a gluing rate in $2D$ one).
The normalization function of some of these models coincides with
the tau function of Pfaff lattice \cite{AMS},\cite{KdL-2}.

Except the consideration in subsection 2.6 we restrict ourselves
to the presentation of various models of random processes arising
from integrable systems more than to the analysis of random models
themselves which is mainly an analysis of the large time behavior
of different correlation functions. As it was very successfully
demonstrated in series of papers by A.Okounkov the usage of
fermionic language may be quite helpful for such purposes. This
part of work will be published separately.

\subsection{Action on Fock vectors. Time-dependent normalization
function ${Z}_\nu(\time)$}

We refer the reader to Appendices A.1- A.3 for a compendium useful
facts and conventions regarding fermions, partitions and Schur
functions that are used repeatedly in what follows. For present
purposes, this involves only charged, one-dimensional free
fermions satisfying the standard canonical anti-commutation
relations
 \be
 \label{A-fermions}
 f_i{\bar f_j}+ {\bar f_j}f_i=\delta_{ij},\quad f_i{f_j}+ { f_j}f_i=0,
\quad {\bar f}_i{\bar f_j}+ {\bar f_j}{\bar f_i}=0,  \quad i, j\in {\bf Z}
 \ee
These act on a fermionic Fock space whose ``vacuum state''
$ |0 \rangle$ is, like the Dirac sea, annihilated by all negative component
creation operators $f_i$ and positive component annihilation operators $\bar f_i$
\be
 f_{-i-1}|0\rangle ={\bar f}_{i}|0\rangle  =0,\quad i\ge 0
\ee More generally, we have the ``charge $n$'' vacuum states
$|n\rangle$,
 \be |n\rangle = f_{n-1} \dots f_0 | 0\rangle  \ {\rm for} \
n\ge 0, \quad |n\rangle = \bar f_{n} \dots \bar f_{-1} | 0\rangle
\ {\rm for} \ n< 0, \ee where the integer $n$ denotes the Dirac
``sea level''. We may visualize the level - $n$ Dirac sea as a
diagram consisting of a set of integer sites on a vertical axis,
with all sites below $n$ occupied, while the $n$-th site and all
those above it are empty, as in Fig.1. \bfig
\begin{picture}(50, 50)
\end{picture}
\begin{picture}(150,150)\label{maya-1}
 \put(50,27.5){\circle*{5}}
  \put(50,47.5){\circle*{5}}
   \put(50,67.5){\circle*{5}}
 \put(50,87.5){\circle{5}}
  \put(50,107.5){\circle{5}}
   \put(50,127.5){\circle{5}}
 \put(60,25){n-3}
 \put(60,45){n-2}
 \put(60,65){n-1}
 \put(60,85){n}
 \put(60,105){n+1}
 \put(60,125){n+2}
   \put(15,5){1. Dirac sea of level $n$}
   \end{picture}
\begin{picture}(100,150)\label{sea}
 \put(100,27.5){\circle*{5}}
  \put(100,47.5){\circle*{5}}
   \put(100,67.5){\circle{5}}
 \put(100,87.5){\circle{5}}
 \put(100,107.5){\circle*{5}}
  \put(100,127.5){\circle{5}}
   \put(100,22){.}
   \put(100,20){.}
   \put(100,18){.}
    \put(100,132){.}
   \put(100,134){.}
   \put(100,136){.}
 \put(110,25){n-3}
 \put(110,45){n-2}
 \put(110,65){n-1}
 \put(110,85){n}
 \put(110,105){n+1}
 \put(110,125){n+2}
   \put(60,5){2. One particle one hole}
\end{picture}
\label{fig1}
\caption{Maya diagrams}
\efig

A  complete set of basis vectors for the Fock
space may be associated with {\it Maya diagrams}, which we shall denote by Greek
letters $\lambda,\nu,\dots$, to be more precise, by two letters,
Greek and Latin, the first denotes the configuration of particles,
the second denotes the level of the Dirac sea, which we perturb by
placing a finite number of particles of the Dirac sea to the upper
positions. In this way all basis vectors of Fock space may be
obtained. Since no two particles can occupy the same site, each
Fock basis vector may be encoded by a partition, these are the
ordered sets of weakly decreasing numbers, say,
$\nu=(\nu_1,\nu_2,\dots)$ where $\nu_1 \ge \nu_2,\cdots  \ge 0$,
and by a level, say $n$, related to the coordinates of the
particles, $h_1>h_2>\cdots $, as follows:
 \be\label{nu-h}
\nu_i=h_i+i-n
 \ee
We number (identical) particles in downward direction, the highest
one be the first. Provided we view each Fock vector as being
created by a perturbation of vacuum ('initial') vector, one may
say, that $\nu_i$ measures the deviation of a current
('Lagrangian') coordinate, $h_i$, of the $i$-th particle from its
initial value, $n-i$, in the vacuum state.

The partitions are very useful and conventional notations, that is
why we need them. There are two suitable way to figure partitions:
via Maya diagram, which is just configuration of the particles
$h=(h_1,h_2,\dots)$ placed on the vertical line, and via Young
diagram which is a set of $\nu=(\nu_1,\nu_2,\dots)$ related to
$h=(h_1,h_2,\dots)$ by (\ref{nu-h}). They also yields links with
the topic of growing interfaces, since the diagram of each
partition (Young diagram) may be viewed as a discrete model of a
two-dimensional region. A set of partitions placed one above
another in three-dimensional space may be a discrete model of 3D
region, and so on. A particular case of such $3D$ figure, called
plane partition, was considered in the papers of Okounkov and
Reshetikhin \cite{OkR1}, \cite{OkR2} as a certain statistical
model.

In the present paper we consider a sequence of Fock vectors
 obtained by application of operators $o^{i,i-1}$ to an initial basis
Fock vector, say, $\n$, as follows
 \be\label{fock-sequence}
\nn \to o^{1,0}\nn \to o^{2,1}o^{1,0}\nn \to \cdots  \to
o^{\time,\time-1}\cdots o^{1,0}\nn
 \ee

This sequence may be viewed as a $\time$-step process on the space
of configurations of hard core particles ($1D$ lattice gas) where
each configuration is related to some basis Fock vector and may be
visualized via the so-called Maya diagram, see the picture. The
number $\time$ will be related as discrete time of the process.
Alternatively, (\ref{fock-sequence}) may be viewed as a
$\time$-step process on the Young diagrams (a type of $2D$
regions).

In what follows everywhere, where it will not be not confusing, we
shall omit the label related to the level of configuration (or,
the same, the level of Dirac sea), keeping only Greek letters,
say, $\lambda$, $\mu$ ,$\nu$ for configurations of the hard core
particles, having in mind the pairs
 $(\lambda,l)$, $(\mu,m)$ ,$(\nu,n)$.

The result of the application of \be o(\time):=
o^{\time,\time-1}\cdots o^{1,0} \ee to the initial state  $\nu$
(corresponding to an initial configuration of hard core
particles), is a Fock vector which is a linear combination of
basis Fock vectors (=configurations of hard core particles). One
of the objects of an interest is the relative weight of a given
configuration, say, $\lambda$, \be\label{transition-weight}
 W_{\nu\to\lambda}(\time)=\langle\lambda|  o(\time)\n
 \ee
with respect to the  weight of all possible configurations,
 \be\label{relative-transition-weight}
P_{\nu\to\lambda}(\time)=\frac{W_{\nu\to\lambda}(\time)}{Z_\nu(\time)}
 \ee
where the normalization function in the denominator is
 \be\label{normalization}
Z_\nu(\time)=\sum_\lambda \;\langle\lambda| o(\time)\n
 \ee
the sum ranges over
all possible configurations.

Matrix element
 \be\label{one-step-weight}
o^{j,j-1}_{\lambda,\mu}=\lll o^{j,j-1}\mrm \
 \ee
will be referred as (one-step) transition weight between a
configuration $\mrm$ and a configuration $\lrl$ at moment $j$.

If we want to keep in mind the probabilistic interpretation of the process we ask
transition weights to be positive.

The completeness of the basis Fock vectors implies
$\sum_{\mu,m}\mrm\mlm =1$. We have
 \be\label{intermediate}
\lll o^{\time,\time-1}\cdots o^{1,0}\nn=
\sum_{\{\mu^{(i)},m_i\}} o^{\time,\time-1}_{\lambda,\mu^{(\time-1)}}\cdots
o^{1,0}_{\mu^{(1)},\nu}
 \ee
where one sums over all possible intermediate configurations
$\mu^{(i)},m^{(i)},\ i=1,\dots,\time -1$.

Each monomial non-vanishing term in the sum is related to a certain set of configurations,
 \be\label{one-path}
\nu,n \to \mu^{(1)},m ^{(1)}\ \to \cdots \to \mu^{(\time-1)},m^{(\time-1)} \to\lambda,l \ ,
  \ee
which will be refereed as a {\em path} from the configuration
$\nu,n$ to the configuration $\lambda,l$, having length (or, the
same, duration) $\time$. The value of the monomial is referred as
the weight of the path, which is equal to a product of all
one-step transition weights (\ref{one-step-weight}) along the
path.

Let us notice that it may have sense to evaluate $Z_\nu(\time)$
also for the case where the positivity condition,
\be\label{positivity}
o^{j,j-1}_{\lambda,\mu} \ge 0 ,
 \ee
is violated and therefore we have no a probabilistic
interpretation of (\ref{intermediate}) as a random process on the
space of configurations of the hard core particles.

In certain examples considered at the present paper  transition
weights (\ref{one-step-weight}) take the form
\be
\lll o^{j,j-1}\mrm=e^{-U_{\lambda,l}(j)+U_{\mu,m}(j-1)}
 \ee

The case where the weight of each path depends only on the end
points of the path, i.e. on initial and on final configurations
shall be referred as potential weights. In what follows we shall
consider only potential transition weights.

To compare with, let us assign to each configuration of hard-core
particles, ${\lambda}$, an energy $U_{\lambda}$. What people are
interested in statistical physics is the partition function of the
system
 \be
 \sum_{\lambda} \ e^{-U_{\lambda}}
 \ee
(everywhere we shall put temperature times the Boltzmann constant
to be one).

In case of random process the normalization function
(\ref{normalization}) - which yields the sum of weights of all
pathes of durations $\time$ which start at the configuration $\nu$
- plays the role similar to the role of partition function in
thermodynamics. For potential transition weights the normalization
function takes the form
  \be\label{}
Z_\nu(\time) = \sum_{\lambda} \;
e^{-U_{\lambda}(\time)+U_{\nu}(0)}\
\textsf{N}_{\lambda,\nu}(\textsc{t})
 \ee
where $\textsf{N}_{\lambda,\nu}(\textsc{t})$ counts the number of
pathes of duration $\time$. Given $\time$, this factor may be also
included into the exponent to define a partition function of a
$\time$-dependent statistical model.

\

At the present paper we shall focus to figure out the links of
random processes (\ref{fock-sequence}) to classical integrable
systems, random matrices and growing surfaces. Benefits of these
links will be presented  in our forthcoming papers
\cite{{HOasymp}},\cite{HLO}.

Our main example is  a discrete version of asymmetric simple
exclusion process (d-ASEP), where we present the answer for the
decay of the step function. In short we shall consider
 random walk of $N$ non-intersecting particles and other examples.

\;

The mentioned {\bf  links of random processes with classical
integrable systems} are two-fold. The first one is the usage of
series (\ref{doubleSchur}) as generating functions for transition
weights between different configurations of the hard-core
particles. Then higher times $\bt$ and ${\bar\bt}$ are just formal
parameters. There are different tau functions, related to $A$,
$B$, $D$ and $C$ root systems; the way of construction of these
may involve different type of fermionic (or, bosonic) operators.
At the present paper we deal with $A$ type, which may be realized
in terms of fermions (\ref{A-fermions}). We show that $A$-type tau
function may generate certain classes of transition weights
(\ref{one-step-weight}). We consider two ways of generating
weights, see below. This is one link between random processes and
integrable systems.

The second link is the fact that the normalization functions
$Z_0(\time)$ may be related to a tau function of some other
classical integrable systems. It occurs if
 we consider deformations of transition weights with a set of
deformation parameters ${\tilde t}=({\tilde t}_1,{\tilde
t}_2,\dots)$
 \be
\lll o^{j,j-1}\mrm \to \lll o^{j,j-1}({\tilde t})\mrm =
e^{-U_{\lambda,l}(j,{\tilde t})+U_{\mu,m}(j-1,{\tilde t})}\lll o^{j,j-1}\mrm
 \ee
 specified later by (\ref{ferm-Leur-sol-t})
and caused by transforms  of operators $o^{j,j-1}$ of form
 \be\label{dual-flow}
o^{j,j-1} \to o^{j,j-1}({\tilde t}) =e^{H_0({\tilde
t})}o^{j,j-1}e^{-H_0({\tilde t})}, \quad j=1,\dots,\time
 \ee
(operators $H_0$ are defined in Appendix A.3). These deformation
parameters may be recognized as higher times of different
underlying integrable equations. These are dual integrable
hierarchies in sense of papers \cite{hypsol} and \cite{OS} but not
exactly\footnote{This duality might be related in a general notion
of duality of integrable systems developed in
\cite{AHH},\cite{HdualIso},\cite{HI},\cite{BEH}  }. In many cases,
under flows (\ref{dual-flow}), the normalization function
$Z_0(\time,{\tilde t})$ of
 (\ref{normalization}) is a tau function
of $B$-type, namely, the $B$ type in the realization of Kac-van de
Leur, see \cite{KdL} (also known as Pfaff lattice independently
introduced as a tool for study of random matrices see \cite{AMS},
\cite{L1}). Thus, what we mean is a sort of duality between $A$
and $B$ type of equations.

The last remark is that we may consider a sort of coupling of random processes with
statistical models on the space of configuration,
\be
 Z=\sum_\lambda \ e^{-q^2{\textsc{E}}_\lambda},
\ee
 where $q$ is some constant (charge) as follows.
 We use the energy $q^2{\textsc{E}}_\lambda$ of the model
 to modify each transition weight:
\be
 \llambda o^{j,j-1}\mr \quad \to \quad
 e^{-q^2{\textsc{E}}_\lambda+q^2{\textsc{E}}_\mu}\ \llambda o^{j,j-1}\mr,
\quad j=1,2,\dots,\time \ee Then \be
 Z_\nu(\time)=\sum_\lambda \ \llambda o(\time)\n \quad
\to \quad \sum_\lambda \
e^{-q^2{\textsc{E}}_\lambda+q^2{\textsc{E}}_\nu}\ \llambda
o(\time)\n \ee

 As we shall show in subsection 2.7 this procedure may keep relations with integrable systems
if $q^2{\textsc{E}}_\lambda$ is chosen in a special way.

\section{Action on Fock vectors via ${\widehat{gl}}(\infty)$ algebra}

{\bf Maya diagrams}

Let us consider an infinite set of vertexes labelled by integers.
Each vertex will be sketched as either a white or black disk.
 Following Sato, we shall call it {\it Maya
diagram}, (after a Maya game in which small stones are moved to
occupy free sites according to a certain rule). In what follows we
shall describe certain rules for motion of ``stones''- hard core
particles. The  Maya diagrams for the vacuum state are presented
in Figure 1.2.

We may consider each black ball as a particle and each white ball
as the lack of a particle at this site (empty site).

The Maya diagram is a way of visualization of basis Fock vectors
for fermions on the circle, sites correspond to Fourier components
of the fermions, no two fermions can occupy the same site;
therefore, there is one site for at most one particle (these are
what is called hard core particles). Sites with positive large
enough numbers are supposed to be free, while sites with large
enough negative numbers are supposed to be occupied, that is the
related Maya diagrams possesses this property: all sites to the up
of a certain number are all white, while sites to the down of a
certain number are all black. The Maya diagram where all sites
below $n$-th one are occupied, and all sites above $n-1$ are free,
is called the vacuum Maya diagram of level $n$, see the Figure 1.
It represent the so-called Dirac sea of level $n$.

Each Maya diagram may be obtained from the vacuum diagram of a
certain level, say, $n$, via placing of certain number, say, $k$,
of black balls (particles) to higher empty sites. We say that each
Maya diagram has level $n$.

{\bf Basis Fock vectors, partitions and Maya diagrams}

 By its origin Maya diagrams is a visualization of vectors in the
 fermionic Fock space, see Appendix A.1. Each black ball (a
 particle) located at a site $i$ indicates that the site is occupied.

Let us introduce the following basis vectors in the fermionic Fock
space
 \begin{equation}\label{basisFock-vectors}
|\lambda,n\rangle = (-1)^{\sum_{i=1}^k\beta_i}
 f_{n+\alpha_k}{\bar f}_{n-1-\beta_k}\cdots f_{n+\alpha_1}
 {\bar f}_{n-1-\beta_1}|n\rangle,\quad
 \alpha_1>\cdots>\alpha_k\ge 0,\quad \beta_1>\cdots>\beta_k\ge 0
 \end{equation}
 where $\lambda=(\alpha_1,\dots,\alpha_k|\beta_1,\dots,\beta_k)$ is
the Frobenius notation for the partition $\lambda$, and $n$ is the
level of Dirac sea, see Introduction and Appendix A.1 for more
details.

This vector is one-to-one related to the Maya diagram of level
$n$, where particle are situated as follows. A particle number $i$
(black ball) counted from the top has a coordinate $n+\alpha_i$,
$i=1,\dots,k$. Thus, the upmost particle has the coordinate
$\alpha_1$. Underneath of the sea level all sites are occupied
except free sites (holes) which we count from the bottom. Hole
number $i$ has a coordinate $n-1-\beta_i$, thus the down-most hole
has the coordinate $n-1-\beta_1$.

Notice that $\beta_1+1$ is the length of partition lambda,
$\ell(\lambda)$.

One may give an alternative description of coordinates of
particles on the Maya diagram related to the vector
$|\lambda,n\rangle$. Take any integer $N$ which is not less than
$\ell(\lambda)$. Then all site with coordinates
$n-N-1,n-N-1,\dots$ are occupied. On the sites $n-N,n-N+1,\dots$
there are $N$ particles which occupy sites with coordinates
 \be\label{lambda-h}
 h_i=\lambda_i-i+N,\quad i=1,\dots,N,\quad N\ge\ell(\lambda)
  \ee
All other sites including and above $n-N$ are empty.

 Dual vector is
 \begin{equation}\label{dual-basisFock-vector}
 \langle\lambda,n| := (-1)^{\sum_{i=1}^k\beta_i}
 \langle n| f_{-1-\beta_1}{\bar f}_{\alpha_1}\cdots
 f_{n-1-\beta_k}{\bar f}_{n+\alpha_k}\cdots
 \end{equation}
One can verify the  orthonormality condition
 \be\label{orthonormality}
\langle\lambda,l|\nu,n\rangle=\delta_{l,n}\delta_{\lambda,\nu}
 \ee

Along these notations we shall identify partitions, Maya diagrams
and basis Fock vectors. Being identical to basis Fock vectors,
Maya diagrams may be multiplied by numbers and may form (formal)
linear combinations.

{\bf Action of $\hat{gl}(\infty)$ on Maya diagrams}

It was old observation that quadratic operators of type $f_i{\bar
f}_k$, $i,k=1,\dots,N$, form $gl(N)$ Lie algebra. In the papers of
Kyoto school it was studied within symmetry analysis of integrable
equations like Toda lattice equation. One of the observations made
in the wonderful series devoted to hierarchies of integrable
equations (in particular, see \cite{DJKM},\cite{JM}) is that
$N=\infty$ case needs a special consideration. When there are an
infinite number of fermionic modes one needs normal ordering.
Then, natural combination are
 \be\label{E-ik}
E_{i,k}:=f_i{\bar f}_k-\langle 0|f_i{\bar f}_k |0\rangle\; \qquad
-\infty <i,k<+\infty
 \ee
which form an algebra called $\hat{gl}(\infty)$ (the algebra of
infinite matrices with central extension). We do not need
properties of this algebra. Let us only note that such algebras
play essential role in the construction of hierarchies of
integrable classical (not quantum) equations, where the algebra of
infinite matrices appears as an underlying symmetry algebra of
integrable equations.

This algebra naturally acts on the fermionic Fock space.

For our purposes we are interested in the action of $E_{i,k}$  on
the basis Fock vectors, which may be viewed as an action on Maya
diagrams. The point is that this action transforms one Maya
diagram to the other one multiplied by a sign factor. Then any
product of operators (\ref{E-ik}) do the same.

Keeping in mind definition of basis Fock vector and applying
(\ref{E-ik}) one can verify that the action of $E_{i,k}$ on Maya
diagram is as follows.

First, let us notice that the second term in the definition of
operator $E_{i,k}$ is vanishing for $i\neq k$. Let us consider
this case first.
 Each $E_{i,k}$ acts on a Maya diagram trivially (as
multiplication by zero of the related basis Fock vector) in two
cases. They occur either if $k$-th site is empty, or/and if $i$-th
site is already occupied by a particle  as on Figure 2.1. Otherwise
$E_{i,k}$ action places black ball from its $k$-th site to the $i$
site, multiplying the new Maya diagram by $(-1)^{c_{ik}}$, where
$c_{ik}$ is a number of black balls between $i$-th and $k$-th
sites, in other words, the number of particles which the particle
originally located at $k$-th site "jumps over" when moving to its
new, $i$-th, site, see Figure 2.2. An arrow on this figure shows
that the particle hope from the site $k$ to the site $i$.

\bfig
\begin{picture}(120,150)\label{}
 \put(45,119){.}
 \put(45,109){.}
 \put(45,99){.}
 \put(45,89){.}
 \put(45,79){.}
 \put(45,69){.}

   \put(7,75){E}\put(15,71){i,k}\put(29,78){.} \put(45.5,57){\circle{5}} \put(73,75){= 0,}

 \put(45,45){.}
 \put(45,35){.}

 \put(55,53){k}

   \put(30,10){1.  Elimination of Maya diagrams}
\end{picture}
\begin{picture}(120,150)\label{}
  \put(45,119){.}
   \put(45,109){.}

   \put(7,75){E}\put(15,71){i,k}\put(29,78){.} \put(45.5,97){\circle*{5}} \put(73,75){= 0,}

  \put(45,85){.}
  \put(45,75){.}
 \put(45,65){.}
 \put(45,55){.}
 \put(45,45){.}
 \put(45,35){.}

 \put(55,95){i}

\end{picture}
\begin{picture}(120,150)\label{}

   \put(45,109){.}

   \put(7,75){E}\put(15,71){i,k}\put(29,78){.} \put(45.5,97){\circle{5}}

   \put(45.5,55.5){\circle*{5}} \put(73,75){=}

  \put(45,85){.}
  \put(45,75){.}
  \put(45,65){.}

 \put(45,45){.}

 \put(55,95){i}
 \put(55,53){k}

   \put(136,109){.}

   \put(136.5,97){\circle*{5}}
   \put(136.5,55.5){\circle{5}}

  \put(136,85){.}
  \put(136,75){.}
 \put(136,65){.}

 \put(136,45){.}

 \put(146,95){i}
 \put(146,53){k}

 \put(89,75){(-1)}\put(103,83){c}\put(108,79){ik}
 \put(131,77.5){\oval(18,40)[l]}
 \put(122,75){\vector(0,1){5}}

   \put(43,10){2. Nontrivial action}
\end{picture}

\label{fig2}
\caption{ Elementary $gl(\infty)$ actions on Maya diagrams}
\efig

The action of operators $E_{i,i}$ on Maya diagrams may be called
testifying action. They acts trivially on the vacuum Maya diagram
of level $0$.  For $i<0$ the operator $E_{i,i}$  eliminates any
Maya diagram which (similar to the vacuum diagram) contains a
particle at the site $i<0$. If the site $i$ is empty then
$E_{i,i}$ acts as the multiplication by unity. For $i\ge 0$ the
operator $E_{i,i}$ eliminates any Maya diagram which (as the
vacuum diagram of the level $0$) contains the white ball at the
site $i$. If there is a particle at the site $i$ then $E_{i,i}$
acts as multiplication by unity.

Because Maya diagrams are in one-to-one correspondence with basis
Fock vectors we can act on them by sums and products of operators
(\ref{E-ik}).

The element
\begin{equation}
Q=\sum_{i\in\mathbb{Z}} E_{i,i}
\end{equation}
plays a special role. One can easily check that operator $Q$ acts
a Maya diagram of a level $n$ by the multiplication by $n$. It is
called the charge operator.

{\bf Graphs}

Let us take vertices of a Maya diagram as vertices of a graph,
where vertices are labelled by integers and each pair of vertices
may be connected by a pair of opposite directed arrows. Given
\begin{equation}
A=\sum_{i\neq k} a_{i,k}E_{i,k}\in \hat{gl}(\infty),
\end{equation}
 we assign the weight $a_{i,k}$ to the arrow
which starts at the vertex $k$ and ends at the vertex $i$. For the
sake of simplicity, we shall not consider diagonal terms,
$E_{i,i}$. Also let us not draw arrows of zero weight. In this way
we obtain some weighted graph, related to given $A=\in
\hat{gl}(\infty)$. For instance, for
\begin{equation}\label{Brownian}
A=\sum_{i=n-2}^{n} (E_{i+1,i}+E_{i,i+1})
\end{equation}
we obtain the graph on Figure 3.1.

\bfig
\begin{picture}(60,150)

 \put(48,50){\vector(0,1){15}}
 \put(48,70){\vector(0,1){15}}
 \put(48,90){\vector(0,1){15}}

 \put(52,65){\vector(0,-1){15}}
 \put(52,85){\vector(0,-1){15}}
 \put(52,105){\vector(0,-1){15}}

  \put(50,47.5){\circle{5}}
   \put(50,67.5){\circle{5}}
 \put(50,87.5){\circle{5}}
  \put(50,107.5){\circle{5}}

 \put(60,45){n-2}
 \put(60,65){n-1}
 \put(60,85){n}
 \put(60,105){n+1}

  \put(40,0){1. Graph}
\end{picture}
\begin{picture}(60,150)
 \put(50,27.5){\circle*{5}}
  \put(50,47.5){\circle*{5}}
   \put(50,67.5){\circle*{5}}
 \put(50,87.5){\circle{5}}
  \put(50,107.5){\circle{5}}
   \put(50,127.5){\circle{5}}
 \put(60,25){n-3}
 \put(60,45){n-2}
 \put(60,65){n-1}
 \put(60,85){n}
 \put(60,105){n+1}
 \put(60,125){n+2}
  \put(40,13){}
    \put(50,22.5){\circle{1}}
    \put(50,19){\circle{1}}
    \put(50,15.5){\circle{1}}
       \put(50,132.5){\circle{1}}
       \put(50,136){\circle{1}}
       \put(50,139.5){\circle{1}}
  \put(40,00){2.}
\end{picture}
\begin{picture}(60,150)\label{maya-2}
 \put(50,27.5){\circle*{5}}
  \put(50,47.5){\circle*{5}}
   \put(50,67.5){\circle{5}}
 \put(50,87.5){\circle{5}}
  \put(50,107.5){\circle*{5}}
   \put(50,127.5){\circle{5}}
 \put(60,25){n-3}
 \put(60,45){n-2}
 \put(60,65){n-1}
 \put(60,85){n}
 \put(60,105){n+1}
 \put(60,125){n+2}
    \put(50,22.5){\circle{1}}
    \put(50,19){\circle{1}}
    \put(50,15.5){\circle{1}}
       \put(50,132.5){\circle{1}}
       \put(50,136){\circle{1}}
       \put(50,139.5){\circle{1}}
   \put(40,00){3.}
\end{picture}
\begin{picture}(60,150)
 \put(48,50){\vector(0,1){15}}
 \put(48,70){\vector(0,1){15}}
 \put(48,90){\vector(0,1){15}}

  \put(52,65){\vector(0,-1){15}}
 \put(52,85){\vector(0,-1){15}}
 \put(52,105){\vector(0,-1){15}}

  \put(50,47.5){\circle*{5}}
   \put(50,67.5){\circle*{5}}
 \put(50,87.5){\circle{5}}
  \put(50,107.5){\circle{5}}

 \put(60,45){n-2}
 \put(60,65){n-1}
 \put(60,85){n}
 \put(60,105){n+1}

  \put(40,0){4.}
\end{picture}
\begin{picture}(60,150)
 \put(48,50){\vector(0,1){15}}
 \put(48,70){\vector(0,1){15}}
 \put(48,90){\vector(0,1){15}}

 \put(52,65){\vector(0,-1){15}}
 \put(52,85){\vector(0,-1){15}}
 \put(52,105){\vector(0,-1){15}}

  \put(50,47.5){\circle*{5}}
   \put(50,67.5){\circle{5}}
 \put(50,87.5){\circle{5}}
  \put(50,107.5){\circle*{5}}

 \put(60,45){n-2}
 \put(60,65){n-1}
 \put(60,85){n}
 \put(60,105){n+1}

  \put(40,0){5.}
\end{picture}

\label{fig3}
\caption{Maya diagrams projected on graphs}
\efig

In the present paper we are interested in motion of particles
along arrows of the graph whose vertices coincide with vertices of
Maya diagram. Having in mind this picture
 one may forget about all vertices of Maya diagrams which are
not linked by vertices. For instance, the Maya diagrams depicted
on Figures 3.2 and 3.3, projected on the graph depicted on Figure 3.1,
respectively, are as on Figures 3.4 and 3.5. If the number of arrows
is finite we shall ignore the rest part of a Maya diagram.

\subsection {Young diagram versus Maya diagram. Interface between
$2D$ area versus configuration of particles on the $1D$ lattice}.

For discrete systems the difference between $1D$ and $2D$ systems
of hard core particles is rather conventional in the following
sense: any $1D$ lattice gas configuration may be converted into a
Young diagram viewed as an example of a condensate of hard core
$2D$ particles.

Let us figure out an example:

\bfig
\begin{picture}(100,170)
 \put(28,70){\vector(0,1){15}}\put(32,85){\vector(0,-1){15}}
  \put(30,47.5){\circle{5}}
   \put(30,67.5){\circle*{5}}
 \put(30,87.5){\circle{5}}
  \put(30,107.5){\circle{5}}
   \put(30,127.5){\circle*{5}}
 \put(40,45){-2}
 \put(40,65){-1}
 \put(40,85){0}
 \put(40,105){1}
 \put(40,125){2}

  \put(29,130){.}
    \put(29,135){.}
   \put(29,140){.}

\put(65,83){\vector(1,0){25}} \put(90,76){\vector(-1,0){25}}

 \put(118,70){\vector(0,1){15}}\put(122,85){\vector(0,-1){15}}
  \put(120,47.5){\circle{5}}
   \put(120,67.5){\circle{5}}
 \put(120,87.5){\circle*{5}}
  \put(120,107.5){\circle{5}}
   \put(120,127.5){\circle*{5}}
 \put(130,45){-2}
 \put(130,65){-1}
 \put(130,85){0}
 \put(130,105){1}
 \put(130,125){2}

 \put(119,130){.}
    \put(119,135){.}
   \put(119,140){.}

  \put(20,00){1. Upward/downward hop}
\put(5,-10){of a particle on a Maya diagram }
\end{picture}
\begin{picture}(100,170)

 \put(149.9,70){\line(0,1){40}}\put(150.2,70){\line(0,1){40}}

 \put(169.9,70){\line(0,1){40}}\put(170.3,70){\line(0,1){40}}

 \put(189.9,70){\line(0,1){40}}\put(190.7,90){\line(0,1){20}}

 \put(209.9,90){\line(0,1){20}}\put(210.7,90){\line(0,1){20}}

 \put(150,69.9){\line(1,0){40}}\put(150,70.6){\line(1,0){20}}

 \put(150,109.8){\line(1,0){60}}\put(150,110.5){\line(1,0){60}}

 \put(150,89.9){\line(1,0){60}}\put(150,90.5){\line(1,0){60}}

\put(210,60){\line(5,1){18}}\put(215,40){\line(5,1){18}}
\put(210,60){\line(1,-5){4}}\put(228,63){\line(1,-5){4}}

 \put(185,85){\vector(4,-3){40}}

 \put(220,50){\vector(-4,3){37}}

\put(180,75){*}

  \put(100,00){2. Adding/removing a box to a Young diagram}
\end{picture}

\medskip
\caption{Equivalent representations of elementary steps}

\label{fig4}
\efig

These two figures show the relation between the motion of
particles on Maya diagrams and a motion of the interface.

One may say, that each hop of a $1D$ particle upward is related to
an increasing of the area of the corresponding Young diagram
 (a condensation of $2D$ particles to the Young diagram), while a
hop downward results in the inverse process of 'evaporation of
particles' and the decreasing of the area the Young diagram. (A
figure similar to Fig. 4.2 may be found in \cite{DJKM} without
relations to the topic of the present paper: to stochastic
motion.)

\subsection{Generating vector for Maya diagrams}

There are vectors,  coherent states,  $\l n|e^{H({\bf t})}$ and
$e^{{\bar H}({{\bf t}})}|n\r$, which depend on parameters ${\bf
t}=(t_1,t_2,\dots)$, see Appendix A.3, and which generate all Fock
space with fixed given sea level $n$,

\begin{equation}\label{coherent}
\langle n|e^{H(\bf{t})} =\sum_{\lambda\in P}\langle
\lambda,n|s_\lambda({\bf{t}}) ,\qquad e^{{\bar
H}({\bf{t}})}|n\rangle =\sum_{\lambda\in P}|\lambda,n\rangle
s_\lambda({\bf{t}})
\end{equation}
where $s_\lambda({\bf{t}})$ is the Schur polynomial \cite{Mac}.
The Schur polynomial is a polynomial in many variables
$t_1,t_2,\dots$, which are labelled by partitions. The appearance
of the Schur polynomial is typical for integrable systems.

\subsection{Random turn motion on graphs.  Permutation factor and weight of path}

Consider the graph related to
\be
A=\sum_{i\neq k} \; a_{i,k}E_{i,k},\qquad E_{i,k}=f_i\bar{f}_k \;
,\quad i\neq k
\ee
which is  a set of vertices, some of which may be linked by
arrows:
 an arrow
$(i,k)$ starts on a vertex $k$ and ends on a vertex $i$. We assign
a weight $a_{i,k}$ to each arrow $(i,k)$. In case $a_{i,k}=0$
there is no arrow. The matrix $a$ with entrances $a_{i,k}$  is
called connecting matrix for a graph. For simplicity of
consideration we omit diagonal terms.

We shall consider a discrete dynamical motion along a Maya diagram
with an initial diagram $\nu$ given by
\begin{equation}\label{A-motion}
|\nu\rangle \to A|\nu\rangle \to \cdots \to
A^{\textsc{t}}|\nu\rangle
\end{equation}
Here $\time$ is viewed as discrete time variable.

As we have explained in the Introduction we have different paths
connecting two different configurations, see (\ref{one-path});
each path is related to a sequence of intermediate partitions.

Consider $A^T$ which is a sum of monomials, each one is a product
of T terms $ a_{i_\alpha,k_\alpha}E_{i_\alpha,k_\alpha},\;
\alpha=1,\dots,\time$. Each { path} of process (\ref{A-motion}) is
generated by a unique monomial applied to a initial Maya diagram.
The number $\time$ may be referred as duration of path.  It may be
described as set of consequent events, single hops. Each time
instant, say, $\alpha$, a particle hops along an arrow
$(i_\alpha,k_\alpha)$: one hop at each time instant. Say, monomial
$a_{i_\time,k_\time}E_{i_\time,k_\time}\cdots
a_{i_1,k_1}E_{i_1,k_1}$, provided it acts nontrivially on $\nu$,
at first time instant describes the hop of a particle along arrow
$(i_1,k_1)$ and at last time instant, $\time$, describes the hop
of particle along an arrow $(i_\time,k_\time)$. The product
weights of $\time$ arrows which were crossed by particles,
$\prod_{\alpha=1}^\time a_{i_\alpha,k_\alpha}$, is called the {\em
weight} of path.

Each pair of configurations may be connected by a number of paths
each of which has its own weight.

 The fact that $A^\time$ is a sum of weighted monomials means randomness of the
motion described by (\ref{A-motion}).

Let us call it random turn motion along a graph, given by $A$.

The random motion where in each time instant only one chosen at
random particle hops is called random turn motion.

To each path we shall assign {\em permutation factor} as follows.

Given acting nontrivially monomial, we can follow each particle
involved into the motion along graph. Let us enumerate particles
located on the initial Maya diagram the from the top of diagram,
and keep these (personal) numbers of particles on the target Maya
diagram (where each particle 'remembers' it's given number). Let
us re-enumerate particles of the target Maya diagram from the top:
this may be also obtained by a certain permutation of personal
numbers which we assigned to the particles in the beginning of the
motion. If the sign of the permutation is even then the
permutational factor of the path is equal to $+1$. If the sign is
odd the permutation factor of the path is equal to $-1$.

Thus, to a each path  we assign a weight, a permutation factor,
and a duration.

\subsection{Graphs and tau functions. Tau function as counting function}

Consider the random turn motion described in the previous
subsection.

Consider all paths of duration $\textsc{t}$ starting with a
configuration described by a partition $\lambda'$ and ending on a
configuration  $\lambda$. The sum of weights of all such paths
with positive permutation factor we denote by
\be
W_{\lambda'\to\lambda}^{(+)}(\time)
\ee
The sum of weights of all such paths with negative permutation
factor we denote by
\be
W_{\lambda'\to\lambda}^{(-)}(\time)
\ee

Let us note that permutation sign appears due to the fermionic
approach to the problem. In problems of random motion of hard-core
particles along graphs which know nothing about fermions, one may
wonder about a total weight, which is
\be
W_{\lambda'\to\lambda}(\time)
 =W_{\lambda'\to\lambda}^{(+)}(\time)
+W_{\lambda'\to\lambda}^{(-)}(\time)
\ee

From the previous consideration we have
 \be
\langle \lambda,n|A^\time
|\lambda',n\rangle=W_{\lambda'\to\lambda}^{(+)}(\time)
-W_{\lambda'\to\lambda}^{(-)}(\time)
 \ee

By (\ref{coherent}) we relate these numbers to the so-called TL
tau function \cite{DJKM},\cite{UT} in the following way
\begin{equation}\label{}
\tau_A(n,\bt,{\bar \bt}):=\langle
n|e^{H({\bt})}e^{zA}e^{\bar{H}(\bar{\bt})}
|n\rangle=\sum_{\time=0}^\infty \sum_{\lambda,\lambda'}\;
\frac{z^\time}{\time!}\left(W_{\lambda'\to\lambda}^{(+)}(\time)
-W_{\lambda'\to\lambda}^{(-)}(\time) \right)
s_\lambda({\bt})s_{\lambda'}(\bar{\bt})
\end{equation}
where $A$ is
\begin{equation}\label{}
A=\sum_{i\neq j} a_{i,j}f_i\bar{f}_j \
\end{equation}
and $z,{\bt}=(t_1,t_2,\dots),\bar{\bf
t}=(\bar{t}_1,\bar{t}_2,\dots)$ are formal parameters, and
$f_i,\bar{f}_i$ are free fermions.

\begin{equation}\label{H}
H({\bt})=\sum_{m=1}^\infty H_mt_m,\quad \bar{H}(\bar{\bt})=
\sum_{m=1}^\infty H_{-m}\bar{ t}_m,\quad
H_m=\sum_{i=-\infty}^\infty f_i\bar{f}_{i+m}
\end{equation}

Thus, we say, that the power of $z$ counts the number of steps.
$\textsc{t}$-step process is
\be
|\nu\rangle \to A|\nu\rangle \to \cdots \to
A^{\textsc{t}}|\nu\rangle
\ee
The variable $\textsc{t}$ is treated as the (discrete) time of the
random process described by $A$.

The most simple case is random processes where
 \be\label{condition}
 W_{\lambda'\to\lambda}^{(-)}(\time) = 0
 \ee
for all pairs $\lambda'$ $\lambda$. Then tau function may be
interpreted as counting function for weights
$W_{\lambda'\to\lambda}(\time)$. Then it follows that
$W_{\lambda'\to\lambda}(\time)$ is a subject to certain discrete
bilinear equations (Hirota equations), which we shall write down
not here.

However we can weaken condition (\ref{condition}) keeping the
interpretation of tau function as counting function. Consider two
examples where particles may hop only to the nearest neighboring
sties. One example is presented on the figures 5.
 Figures 5.1 and 5.3 depict graphs where initial step
function, $\nu=0$, configuration will decay. In figure 5.1 at each
time instant a particle chosen at random will hop upward along an
arrow if the neighboring  place is not occupied. In figure 5.2,
motion is impossible: in this configuration particles are locked.
Each finite perturbation of $\nu=0$  will come to $\nu=0$ in a
final number of steps. In Figure 5.3 not only particles are chosen
at random but also arrows are to be chosen at random: if the
neighboring  vertices are free, chosen at random particle may hop
either upward or downward. It is called non-intersecting random
turn walkers, (on this topic see \cite{Baik},
\cite{Forr1},\cite{BF} and \cite{AvM}). These examples of random
motion keep the condition (\ref{condition}) because it is ordered
motion and the signs of each path is positive. For the choice of
$A$ as in Figures 5.1-5.3, namely, $A=H_1,H_{-1},H_1+H_{-1}$ the
weight of any path is either zero, or one. It means that tau
function is a generating function for number of ways  to get
configuration $\lambda$  starting with a configuration $\lambda'$
in $\time$ steps.

\bfig
\begin{picture}(100,150)
 \put(50,30){\vector(0,1){15}}
 \put(50,50){\vector(0,1){15}}
 \put(50,70){\vector(0,1){15}}
 \put(50,90){\vector(0,1){15}}
 \put(50,110){\vector(0,1){15}}
 \put(50,130){\vector(0,1){15}}
 \put(50,27.5){\circle*{5}}
  \put(50,47.5){\circle*{5}}
   \put(50,67.5){\circle*{5}}
 \put(50,87.5){\circle{5}}
  \put(50,107.5){\circle{5}}
   \put(50,127.5){\circle{5}}
 \put(60,25){n-3}
 \put(60,45){n-2}
 \put(60,65){n-1}
 \put(60,85){n}
 \put(60,105){n+1}
 \put(60,125){n+2}
  \put(40,00){1. $H_{-1}$}
\end{picture}
\begin{picture}(100,150)
 \put(50,45){\vector(0,-1){15}}
 \put(50,65){\vector(0,-1){15}}
 \put(50,85){\vector(0,-1){15}}
 \put(50,105){\vector(0,-1){15}}
 \put(50,125){\vector(0,-1){15}}
 \put(50,145){\vector(0,-1){15}}
 \put(50,27.5){\circle*{5}}
  \put(50,47.5){\circle*{5}}
   \put(50,67.5){\circle*{5}}
 \put(50,87.5){\circle{5}}
  \put(50,107.5){\circle{5}}
   \put(50,127.5){\circle{5}}
 \put(60,25){n-3}
 \put(60,45){n-2}
 \put(60,65){n-1}
 \put(60,85){n}
 \put(60,105){n+1}
 \put(60,125){n+2}
  \put(40,0){2. $H_{1}$}
\end{picture}
\begin{picture}(100,150)
 \put(48,30){\vector(0,1){13}}
 \put(48,50){\vector(0,1){13}}
 \put(48,70){\vector(0,1){13}}
 \put(48,90){\vector(0,1){13}}
 \put(48,110){\vector(0,1){13}}
 \put(48,130){\vector(0,1){13}}

 \put(52,45){\vector(0,-1){13}}
 \put(52,65){\vector(0,-1){13}}
 \put(52,85){\vector(0,-1){13}}
 \put(52,105){\vector(0,-1){13}}
 \put(52,125){\vector(0,-1){13}}
 \put(52,145){\vector(0,-1){13}}

 \put(50,27.5){\circle*{5}}
  \put(50,47.5){\circle*{5}}
   \put(50,67.5){\circle*{5}}
 \put(50,87.5){\circle{5}}
  \put(50,107.5){\circle{5}}
   \put(50,127.5){\circle{5}}
 \put(60,25){n-3}
 \put(60,45){n-2}
 \put(60,65){n-1}
 \put(60,85){n}
 \put(60,105){n+1}
 \put(60,125){n+2}
  \put(30,0){3. $H_{-1}+H_{1}$}
\end{picture}

\caption{Random turn motion along graphs}
\label{fig5}
\efig

Second example is a walk on a ring, where particles are allowed to
hop to the nearest neighboring site.
  How to obtain such graphs
is depicted on figure 6: figures 6.3 and 6.4 are obtained
respectively from figure 6.1 and 6.2. On figures 6.2 (or, the same
6.4) condition (\ref{condition}) is violated because when a
particle hops along the arrow linking vertices $n$ and $n+5$
permutation of particles is odd in case the total number of
particles on the ring is even. Thus each path containing odd
number of hops along arrows linking sites $n$ and $n+5$ has
negative permutation factor. Nevertheless if we assign positive
weight to all arrows except arrows linking $n$ and $n+5$ sites,
tau function will count the transition weights between any pair of
configurations. Say, if arrows linking $n$ and $n+5$ have weight
$-1$, while all other arrows have weight $1$ tau function counts
number of ways linking any pair of configuration on the ring.

\bfig
\begin{picture}(100,150)
 \put(49,30){\vector(0,1){15}}
 \put(49,50){\vector(0,1){15}}
 \put(49,70){\vector(0,1){15}}
 \put(49,90){\vector(0,1){15}}
 \put(49,110){\vector(0,1){15}}

 \put(52,45){\vector(0,-1){15}}
 \put(52,65){\vector(0,-1){15}}
 \put(52,85){\vector(0,-1){15}}
 \put(52,105){\vector(0,-1){15}}
 \put(52,125){\vector(0,-1){15}}

  \put(50,27.5){\circle*{5}}
  \put(50,47.5){\circle*{5}}
   \put(50,67.5){\circle*{5}}
 \put(50,87.5){\circle{5}}
  \put(50,107.5){\circle{5}}
   \put(50,127.5){\circle{5}}

 \put(50,77.5){\oval(18,96.5)[l]}
 \put(48,77.5){\oval(20,102.5)[l]}
 \put(41,80){\vector(0,1){15}}
 \put(38,90){\vector(0,-1){15}}

 \put(60,25){n}
 \put(60,45){n+1}
 \put(60,65){n+2}
 \put(60,85){n+3}
 \put(60,105){n+4}
 \put(60,125){n+5}
 \put(40,0){1.}
\end{picture}
\begin{picture}(100,150)
 \put(49,30){\vector(0,1){15}}
 \put(49,50){\vector(0,1){15}}
 \put(49,70){\vector(0,1){15}}
 \put(49,90){\vector(0,1){15}}
 \put(49,110){\vector(0,1){15}}

 \put(52,45){\vector(0,-1){15}}
 \put(52,65){\vector(0,-1){15}}
 \put(52,85){\vector(0,-1){15}}
 \put(52,105){\vector(0,-1){15}}
 \put(52,125){\vector(0,-1){15}}

  \put(50,27.5){\circle{5}}
  \put(50,47.5){\circle*{5}}
   \put(50,67.5){\circle*{5}}
 \put(50,87.5){\circle{5}}
  \put(50,107.5){\circle{5}}
   \put(50,127.5){\circle{5}}

 \put(50,77.5){\oval(18,96.5)[l]}
 \put(48,77.5){\oval(20,102.5)[l]}
 \put(41,80){\vector(0,1){15}}
 \put(38,90){\vector(0,-1){15}}

 \put(60,25){n}
 \put(60,45){n+1}
 \put(60,65){n+2}
 \put(60,85){n+3}
 \put(60,105){n+4}
 \put(60,125){n+5}
 \put(40,0){2.}
\end{picture}
\begin{picture}(100,100)
 \put(52,70){\vector(2,1){15}}
 \put(70,77.5){\circle{5}}
 \put(68,75){\vector(-2,-1){15}}
 \put(50,67.5){\circle{5}}
 \put(51,65){\vector(0,-1){15}}
 \put(49,50){\vector(0,1){15}}

 \put(91,50){\vector(0,1){15}}
 \put(88,68){\vector(-2,1){15}}
  \put(89,65){\vector(0,-1){15}}
  \put(90,67.5){\circle{5}}
 \put(73,78){\vector(2,-1){15}}

 \put(90,47.5){\circle*{5}}
 \put(88.5,44.5){\vector(-2,-1){15}}
 \put(72,39){\vector(2,1){15}}

 \put(50,47.5){\circle*{5}}
 \put(66.5,37){\vector(-2,1){15}}
 \put(53,47){\vector(2,-1){15}}

 \put(70,37.5){\circle*{5}}

 \put(98,45){n}
 \put(62.5,26){n+1}
 \put(27,45){n+2}
 \put(27,65){n+3}

 \put(62.5,84){n+4}
  \put(98,65){n+5}

  \put(60,0){3.}

\end{picture}
\begin{picture}(100,100)
 \put(52,65){\vector(2,1){15}}
 \put(70,72.5){\circle{5}}
 \put(68,70){\vector(-2,-1){15}}
 \put(50,62.5){\circle{5}}
 \put(51,60){\vector(0,-1){15}}
 \put(49,45){\vector(0,1){15}}

 \put(91,45){\vector(0,1){15}}
 \put(88,63){\vector(-2,1){15}}
  \put(89,60){\vector(0,-1){15}}
  \put(90,62.5){\circle{5}}
 \put(73,73){\vector(2,-1){15}}

 \put(90,42.5){\circle{5}}
 \put(88.5,39.5){\vector(-2,-1){15}}
 \put(72,34){\vector(2,1){15}}

 \put(50,42.5){\circle*{5}}
 \put(66.5,32){\vector(-2,1){15}}
 \put(53,42){\vector(2,-1){15}}

 \put(70,32.5){\circle*{5}}

 \put(98,40){n}
 \put(62.5,22){n+1}
 \put(27,40){n+2}
 \put(27,60){n+3}

 \put(62.5,78){n+4}
  \put(98,58){n+5}

  \put(60,0){4.}
\end{picture}

\caption{ For the case depicted in 2 and 4, in order to keep the
probabilistic interpretation, one has to assign negative weights
to the arrows that link $n+5$ and $n$} \label{fig6} \efig

 Processes like those depicted in Figures 5 and 6 are also known as discrete
 versions of simple exclusion processes (totally asymmetric (d-TASEP) for figures 5.1, 5.2,
 and symmetric for figure 5.3), which may be considered either on the
line as on figures 1-3, or on a ring as in the Figure 6. This type
of models was introduced by M.E.Fisher.  It may be called discrete
time asymmetric simple exclusion process (d-ASEP) in case the
probability rate for hops in the opposite directions are different
and does not depend upon site numbers.  The solvability of ASEP
was proven by H. Spohn in \cite{GS1, GS2} by method of Bethe
anzats and a huge literature is devoted to the study of different
aspects of this model which, in non-equilibrium physics,  plays a
role comparable to the role of Ising model in statistical physics.

The problem of evaluation of correlation functions, asymptotic
behavior we hope to consider separately using fermionic approach.

The dependence of transition probabilities on parameters (weights
of arrows) will be considered in subsection 2.6. Before this
consideration of non-equilibrium system of particles we find that
a consideration of fermions in thermal equilibrium will be
helpful.

\subsection{Fermions in thermal equilibrium}

Consider the following statistical ensemble: there is an infinite
set of levels labelled by integers. Each level $i$ has energy
${U}_i$ and may be occupied by a fermion. In thermodynamics the
probability to occupy a site $i$ is proportional to $\exp
(-{U}_i)$, where ${U}_i$ is energy level numbered by $i$ (we put
the Boltzmann constant times temperature to be one). In case
${U}_i$ is a monotonous function of the site number we can use
Maya diagrams of some level, say $n$. At zero temperature all
lower site are occupied up to this level (in solid state physics
called Fermi level). The bottom limit of Maya diagram is fully
packed by particles, while top limit is free of particles. A
fragment of
 the picture is sketched out in
the following figure

\begin{picture}(100,150)
  \put(48.5,27.5){.} \put(49,32.5){.} \put(49,37.5){.}
  \put(50,47.5){\circle*{5}}
   \put(50,67.5){\circle*{5}}
 \put(50,87.5){\circle{5}}
  \put(50,107.5){\circle*{5}}
   \put(49,127.5){.} \put(49,122.5){.} \put(49,117.5){.}

 \put(60,45){i-2} \put(93,45){exp (}\put(115,45){-{U}}\put(127,40){i-2}\put(140,45){)}
 \put(60,65){i-1} \put(93,65){exp (}\put(115,65){-{U}}\put(127,60){i-1}\put(140,65){)}
 \put(60,85){i}   \put(93,85){exp (}\put(115,85){-{U}}\put(127,80){i}\put(130,85){)}
 \put(60,105){i+1}\put(93,105){exp (}\put(115,105){-{U}}\put(127,100){i+1}\put(140,105){)}

 \put(20,0){Fig.1. Fermionic levels where some sites are occupied}
\end{picture}

\begin{picture}(100,20)
\end{picture}

 Each configuration of fermions distributed among levels
  has its weight given by the product of the Gibbs factors related to
the occupied sites. The contribution of the fragment depicted on
the figure is $e^{-{U}_{i-2}-{U}_{i-1}-{U}_{i+1}}$.

These are non-interacting fermions. The equilibrium ensemble is
described by its partition function.

First, let us introduce the notation which we shall use throughout
the paper
 \be\label{T-lambda}
{U}_{\lambda}(n)= \; \sum_{i=1}^\infty \;
\left(U_{\lambda_i-i+n}-U_{-i+n}\right)
 \ee

 Notice that for the
vacuum configuration of any level $n$ we have $U_0(n)=0$.

 The normalized partition function is then a sum over partitions
\begin{equation}\label{ff-energy-0}
e^{-\textmd{F}_0}=\sum_{\lambda} e^{-{U}_{\lambda}(n)}
\end{equation}
 where $\textmd{F}_0$ is the free energy of the system of
non-interacting fermions living on a Maya diagram of a level $n$.

Remark. Let notice, that we can cut Maya diagram from the bottom
via the following procedure. We send all energies of sites located
bellow, say, site $N$, to minus infinity. Without the loss of
generality we take $N=0$. Then, we have in particular,
 \be
 \label{cut}
e^{U_{-1}-U_0}=0
 \ee
In this way we restrict our consideration by finite number of
fermions, $n$. It is available in case our system is related to
one-dimensional solid state system where there are finite number
of particles. In such cases  $n$ will be the number of particles
equal to Fermi level. If only $n$ fermions are involved we have
  \be\label{T-lambda-Fermi}
{U}_{\lambda}(n)= \; \sum_{i=1}^n \;
\left(U_{\lambda_i-i+n}-U_{-i+n}\right),
 \ee
 where $\lambda_i-i$ is the coordinate of the particle numbered by
 $i$ ($i=1,\dots,n$) on a
 Maya diagram of a level $n$ which is cut at zero site.
 We count particles from the top and their coordinates from the
 bottom.

We shall consider both cases, $N=0$ and $N\to -\infty$, the first
will be described with the help of semi-infinite Maya diagram and
the second by usual Maya diagram.

{\bf Modified Coulomb potential ${\textsc{E}}_\lambda$}. Now,
suppose that fermions pair-wise interact, that is, now, the energy
of the system is $U_{\lambda}(n)$ plus the energy of pair-wise
interaction. We choose this interaction as Coulomb interaction (if
sites are interpreted as level under Fermi surface then, this is
Coulomb interaction in the momentum space)
\begin{equation}\label{pair-wise}
q^2\log (i-j),
\end{equation}
which describes the repulsion of fermions of a charge $q$ at sites
$i$ and $j$. The problem that Coulomb energy
 goes to $+\infty$ for semi-infinite lattice. In order to avoid
 the divergency of the energy for semi-infinite lattice we shall modify
 it taking the energy of particles in a configuration $\lambda$ in
 form
  \be\label{Coulomb-modified}
{\textsc{E}}_\lambda=-\log s_\lambda({\bt_\infty}):=\log
\frac{\prod_{i<j}^L(h_i-h_j)}{\prod_{i=1}^L h_i!}, \quad
h_i=\lambda_i-i+L,\; i=1,\dots,L,\; L\ge\ell(\lambda)
  \ee
  Notation $s_\lambda({\bt_\infty})$ is taken from the Appendix
  A.1, see (\ref{schurhook}).
  \br\label{on-Coulomb-modified}
 This expression contains pair-wise Coulomb interaction
(\ref{pair-wise}) in the enumerator and an external electric
potential in the denominator. As one may check this expression
vanishes on the vacuum configuration for any value of $L\ge 0$.
Let us mark that $L$ is not a parameter:
 the energy does not depend on the choice of $L$ provided it exceeds
 the length of a partition.
It is reasonable to recall the meaning of numbers $i,\lambda_i$
and $h_i$. The number $i$ numerate particles counted in the
downward direction. The number $\lambda_i$ measure the shift of
the particle $i$ in the upward direction from it's homesite in the
vacuum configuration. Given $L$ the number $h_i$ is the coordinate
of the particle numbered by $i$ which is counted from the origin
located on the site $n-L$ on the Maya diagram of level $n$. The
point that the interaction (\ref{Coulomb-modified}) does not
depend on the choice of this origin which one can send to
$-\infty$. It is perfectly adopted for the lattice gas with fully
packed "bottom" limit. Moreover, one can show that the vacuum
configuration is the configuration with minimal energy equal to is
zero.

\er

 The related free energy
 will be denoted by $\textmd{F}_{q}$.

In this case the partition function of the system is
\begin{equation}\label{ff-energy}
e^{-\textmd{F}_{q}(n)}=\sum_\lambda
e^{-U_\lambda-q^2\;{\textsc{E}}_\lambda+q^2|\lambda|\log(t_1{\bar
t}_1)}=\sum_\lambda e^{-U_\lambda}
\left(s_\lambda({\bt_\infty})\right)^{q^2}(t_1{\bar
t}_1)^{|\lambda|}
\end{equation}
where we introduce dependence on axillary parameters $t_1{\bar
t}_1$ for further convenience.

There are special cases, $q^2=1$ and $q^2=2$, where
(\ref{ff-energy}) may be identified with tau functions of quite
different integrable systems. (The case $q^2=4$ is also related to
integrable systems, however we will not need it at present paper).
 We shall consider these cases separately.

For $q^2=2$ the slightly modified series (\ref{ff-energy}) is
equal to
\[
e^{-\textmd{F}_{2}(n,t_1,{\bar t}_1)}:=\sum_\lambda
e^{-U_\lambda-q^2\;{\textsc{E}}_\lambda+|\lambda|\log(t_1{\bar
t}_1)}=\sum_\lambda e^{-U_\lambda}(t_1{\bar t}_1)^{|\lambda|}
\left(s_\lambda({\bt_\infty})\right)^{2}
\]
\begin{equation}\label{hyp-geom-tau}
=c_n^{-1}\langle n|e^{t_1H_1}e^{\sum_{i\ge 0 }{U}_i f_i{\bar
f}_i-\sum_{i<0 }{U}_i {\bar f}_i f_i}e^{{\bar t}_1H_{-1}}|n\rangle
\end{equation}
which is a Toda lattice (TL) tau function. (Here $c_n$ is a
normalization constant chosen in a way that the right hand side is
equal to unity for $t_1={\bar t}_1=0$, see (\ref{T-vac})). The
equality directly follows from (\ref{coherent}) and
(\ref{orthonormality}) if we take into account (\ref{lambda-T})
and finely use (\ref{choicetinfty'}) and (\ref{schurhook}).
 This formula is true for both cases $N=-\infty$ and $N=0$. To
obtain the last case we need keep in mind relation (\ref{cut}).

Let us refer to the paper \cite{LS} where the Coulomb gas on the
lattice was related to the tau functions of  different type.



In general, the following vacuum expectation value which is known
to be an example of the Toda lattice tau function, namely
\begin{equation}\label{hgtf}
\tau(n,{\bt},{U},\bar{\bt})=\langle n|e^{H({\bf t})}e^{\sum_{i\ge
0 }{U}_i f_i{\bar f}_i-\sum_{i<0 }{U}_i {\bar f}_i
f_i}e^{\bar{H}(\bar{\bt})}|n\rangle=c_n\sum_\lambda
e^{-U_\lambda}s_\lambda(\bt)s_\lambda(\bar{\bt})
\end{equation}
yields the partition function for the system of fermions in a
configuration $|\lambda,n\r$ whose pair-wise interaction
(\ref{pair-wise}) is replaced by interaction of all particles via
the potential given by
 \be\label{Schur-interaction}
\log s_{\lambda}({\bt}) + \log s_{\lambda}(\bar{\bt})
 \ee
which, in general, is not of a pair-wise type. The second equality
of (\ref{hgtf}) is derived from relations
(\ref{coherent}),(\ref{orthonormality}) and (\ref{lambda-T}),
exactly in the same way as  previous formula (\ref{hyp-geom-tau})
but without specification of Schur functions. The pair-wise type
interactions one may obtain via specification of parameters
${\bt}$ and $\bar{\bf t}$ which enter the tau function
(\ref{hgtf}), via formulas (\ref{schurhookq})-(\ref{choicet(a)q'})
of Appendix A.2. (Such specifications  were considered in
\cite{OSch1},\cite{OSch2}, \cite{OS} and \cite{hypsol}  in quite
different contexts).


\br

To complete the talk about links between the Coulomb lattice gas
and tau functions we note, that if we choose a special
parametrization of the set $\{U_i,i\in\mathbb{Z}\}$: via sets of
variables ${\tilde t}:=({\tilde t}_1,{\tilde t}_2,\dots)$ and
${\tilde t}^*:=({\tilde t}_{-1},{\tilde t}_{-2},\dots)$ as
follows:
 \be\label{Tt-tilde-t}
{U}_i =\sum_{m\neq 0}^\infty i^m {\tilde t}_m ,\quad
i\in\mathbb{Z}
 \ee
we obtain (see \cite{hypsol}) that $e^{-\textmd{F}_{2}}$ is a TL
tau function in variables of a certain dual hierarchy.

Similar parametrization of the parameters $U$ which enter tau
function (\ref{hyp-geom-tau}) was used in the context of
completely different problems in \cite{ON} and in \cite{ABW}. In
\cite{ABW} relation (\ref{Tt-tilde-t}) has a meaning of the
dispersion law for fermions in solid state physics, where $i$
plays the role of momentum and $U_i$ plays the role of energy of
fermions. Authors thanks P. Wiegmann for explaining \cite{ABW}
before it was published.

\er


The case $q^2=1$ is most important in view of its application in
the next section, and is less trivial from the point of view of
what it may be in integrable hierarchies.  Given level of vacuum
configuration $n$ consider
\begin{equation}\label{kappa1-vev}
\varrho_n({U}) :=\sum_{\lambda\atop \ell(\lambda)\le n}\langle
\lambda,n|e^{\sum_{i\ge 0 }{U}_i f_i{\bar f}_i-\sum_{i<0 }{U}_i
{\bar f}_i f_i}e^{{H}_{-1}}|0,n\rangle
\end{equation}
\[
=c_n\sum_{\lambda\atop \ell(\lambda)\le
n}e^{-U_\lambda(n)}s_\lambda({\bf
t}_\infty)=c_n\sum_{k=1}^n\sum_{h_1>\cdots>h_n\ge
0}\prod_{i=1}^{n} \frac{e^{-{U}_{h_i}+{U}_{-i+n}}}{h_i!}
{\prod_{i<j}^{n}(h_i-h_j)}
\]
where $h_i=\lambda_i-i+n$. The second equality is obtained from
formulas (\ref{lambda-T}) and (\ref{lemma-H}) and the last
equality from (\ref{schurhook}) of Appendix A.2, and where $c_n$
is given by (\ref{T-vac}) (We shall omit this constant $c_n$
below). Then replacing summation over the cones $h_1>\cdots
>h_n\ge 0$ by summation over non-ordered non-negative
$h_1,\dots,h_n$ which may be done due to the permutational
symmetry of the terms (and getting the factor $(n!)^{-1}$) we
arrive at the partition function for $n$ fermions with the
electric charge $1$:
\begin{equation}\label{pos-matr}
\varrho_n({U}) =\frac{1}{n!}\sum_{h_1,\dots,h_n\ge 0
}\prod_{i=1}^{n} \frac{e^{-{U}_{h_i}+{U}_{-i+n}}}{h_i!}
{\prod_{i<j}^{n}|h_i-h_j|}= e^{-{\textmd{F}_{1}(n)}}
\end{equation}
Notice that we do not need condition (\ref{cut}) because the
reduction to the $n$ particle partition function is achieved by
imposing the condition $\ell(\lambda)\le n$.

\br The interesting fact is the following. Let us write the grand
partition function
\begin{equation}\label{gen-for-pos-matr}
\varrho(\mu,{U}) =\sum_{n=0}^\infty e^{n\mu} \varrho_n({U})
\end{equation}
where $\mu$ is a chemical potential. Then one can show that this
is an example of the infinite-soliton tau function of a version of
the BKP hierarchy, suggested by V.Kac and J. van de Leur
\cite{KdL-2} (called them "charged" BKP hierarchy). This BKP
hierarchy is different from the BKP hierarchy presented in
\cite{DJKM},\cite{JM}.

The fermionic representation is a specification of the fermionic
representation found in \cite{L1} in the context of the study of
Pfaff lattice \cite{AMS} and its generalizations. For the sake of
simplicity we shall write down the fermionic representation only
for $\varrho_n({U})$, where $n$ is an even number. It is
\begin{equation}\label{ferm-Leur-sol}
\varrho_n({U}) =n!\langle n|e^{H({\tilde t})}e^{\frac12
\sum_{i,j\ge 0 } \frac{1}{i!j!}f(i){
f}(j)sign(j-i)}e^{-\bar{H}({\tilde t}^*)}|0\rangle
\end{equation}
where
\[
f(i)=\sum_{-\infty}^{+\infty} i^mf_m
\]
and where $H({\tilde t})$ and $\bar{H}({\tilde t}^*)$ are given by
(\ref{H}) and sets of variables ${\tilde t}:=({\tilde t}_1,{\tilde
t}_2,\dots)$ and ${\tilde t}^*:=({\tilde t}_{-1},{\tilde
t}_{-2},\dots)$ are related to the set
${U}:=({U}_{0},{U}_1,\dots)$ via
 \begin{equation}\label{ferm-Leur-sol-t}
{U}_i =2\sum_{m\neq 0}^\infty i^m {\tilde t}_m
 \end{equation}
where two semi-infinite sets of parameters ${{\tilde t}_m}$ are
called higher times of the coupled charged BKP hierarchy. Let us
figure out the equidistant energy levels case:
 \be\label{tilde-t-one}
{U}_{i-1}-{U}_{i}=-2{\tilde t}_1
 \ee
  which is described by all ${\tilde t}_m$ vanish except ${\tilde t}_1$.

 Notice that (\ref{ferm-Leur-sol-t})  is not one-to-one correspondence,
 given sets
${\tilde t}$ and ${\tilde t}$ uniquely define the variables $U$
but not vice versa.

One need to make some comments on (\ref{ferm-Leur-sol}). We obtain
it by developing the exponential term in the middle of the vacuum
expectation value into the Taylor series. Then, as one can see,
only $\frac12 n$-th term contributes. After one uses certain
tricks invented in \cite{L1} he gets (\ref{ferm-Leur-sol}).

In the same way, using results of \cite{L1} we obtain that
$e^{-\textmd{F}_{q}}$ is also a tau function. We shall not write
precise formulas for this case. Thus, for special values of
electric charges, $q=1,\sqrt{2}$ (and also for the case $q=2$
which we omit in the present paper) we relate Coulomb particles to
tau functions.

\er

\subsection{Random turn walk in an external potential. Decay of the step function}

Let us describe a certain model. Consider the random turn walk
related to the weighted graph given by
 \be A({U})=A_+({U})+
A_-({U}),
 \ee
 and where
 \be\label{random-turn-Brownian}
A_+({U})=\sum_{i\in\mathbb{Z}}e^{{U}_{i-1}-{U}_i}f_i{\bar
f}_{i-1},\quad A_{-}({U})=
\sum_{i\in\mathbb{Z}}e^{{U}_{i}-{U}_{i-1}}f_{i-1}{\bar f}_{i},
 \ee

 We use the same letters $U=\{ U_i \}$ as in the previous
 subsection  in order to pay
 attention to a certain similarity of
 the answers for probabilities to find particles in given
 configurations obtained in rather  different problems: particles in
the thermal equilibrium and  particles subjected to the chosen
random motion.

According to the previous consideration our model is the
following. There is the infinite graph, which is the
one-dimensional lattice $\mathbb{Z}$ (which we view as a  set of
integer points on a vertical line), with all neighboring sites
linked by pairs of oppositely directed arrows. Each site may be
either empty, or filled by at most one particle. The 'bottom'
limit of this graph is fully packed by particles, while the 'top'
limit is free of particles. At each time step, in the 'middle' of
the graph we have some configuration of  particles which are
subject to random motion.

 In each unit time interval a particle chosen at random hops to the
neighboring position provided it is free. It may hop either
upward, or downward with different probabilities. The rate of the
hop of a particle from the site $i-1$ to the site $i$ is given by
 \be\label{rate-up}
\quad up:\qquad \qquad r(i)= e^{-{U}_{i}+{U}_{i-1}}
 \ee
while the rate of the hops from the site $i$ to the site $i-1$ is
given by
 \be\label{rate-down}
down:\qquad \qquad  r^*(i)= e^{-{U}_{i-1}+{U}_{i}}
 \ee
which is the number inverse to (\ref{rate-up}).

 One may
say that we consider random turn walking particles in an external
field, which affects the rates of the hops. Notice that this model
should be considered as a generalization of the so-called random
turn model which was introduced in \cite {F}. (See e.g.,
\cite{BF},\cite{Forr2}, for other approaches.)

A simplifying auxiliary picture may be the following. Each
particle moves in the potential field given by ${U}$ where
${U}_{i}$ is referred as the potential of a site $i$. The rate of
a hop depends on the difference of the potentials between final
and target sites as given by (\ref{rate-up}) and
(\ref{rate-down}).

The alternative picture is obtained via the Young diagram of a
Maya diagram. The process under consideration is a random changing
of the shape of a Young diagram, viewed as random gluing and
evaporating of boxes to it in such a way that only one box is
added (or removed) at each time step. Each box has a 'gluing
energy' ${U}_{\textsc{r}}-{U}_{\textsc{r}-1}$, given by the
distance, $\textsc{r}$, of the box to the main diagonal of the
Young diagram. (This distance is called a content of a box of a
Young diagram \cite{Mac}). An example is given below, where a Maya
diagram of level $0$ and the related Young diagram are drawn (the
Young diagram is drawn by bold lines). The distances (the
"contents") of boxes marked by x on the figure are $\textsc{r}=-1$
for the lower box and $\textsc{r}=2$ for the upper box (these
numbers coincide with the height coordinates of black balls on the
first figure). For boxes marked by *, as going up and right we
respectively obtain $\textsc{r}=-2,0,3$.

\bfig
\begin{picture}(100,170)
 \put(37,30){\vector(0,1){13}}\put(42.5,45){\vector(0,-1){13}}
 \put(37,50){\vector(0,1){13}}\put(42.5,65){\vector(0,-1){13}}
 \put(37,70){\vector(0,1){13}}\put(42.5,85){\vector(0,-1){13}}
 \put(37,90){\vector(0,1){13}}\put(42.5,105){\vector(0,-1){13}}
 \put(37,110){\vector(0,1){13}}\put(42.5,125){\vector(0,-1){13}}
\put(37,130){\vector(0,1){13}}\put(42.5,145){\vector(0,-1){13}}

  \put(40,47.5){\circle{5}}
   \put(40,67.5){\circle*{5}}
 \put(40,87.5){\circle{5}}
  \put(40,107.5){\circle{5}}
   \put(40,127.5){\circle*{5}}
    \put(40,147.5){\circle{5}}
 \put(50,45){-2}
 \put(50,65){-1}
 \put(50,85){0}
 \put(50,105){1}
 \put(50,125){2}
 \put(50,145){3}

  \put(0,00){1.  Random turn walk}
 \put(0,-10){of particles}
 \put(0,-20){on a Maya diagram.}

\end{picture}
\begin{picture}(100,170)

 \put(149.7,50){\line(0,1){60}}\put(150.4,70){\line(0,1){40}}

 \put(169.7,70){\line(0,1){40}}\put(170.4,70){\line(0,1){40}}

 \put(189.7,70){\line(0,1){40}}\put(190.4,90){\line(0,1){20}}

 \put(209.7,90){\line(0,1){20}}\put(210.4,90){\line(0,1){20}}

 \put(230,90){\line(0,1){20}}

\put(150,50){\line(1,0){20}}

 \put(150,69.7){\line(1,0){20}}\put(150,70.4){\line(1,0){20}}

\put(150,109.7){\line(1,0){60}}\put(150,110.4){\line(1,0){80}}

\put(150,89.7){\line(1,0){60}}\put(150,90.4){\line(1,0){60}}

\put(210,60){\line(5,1){18}}\put(215,40){\line(5,1){18}}
\put(210,60){\line(1,-5){4}}\put(228,63){\line(1,-5){4}}

 \put(222,46){\vector(-4,3){39}}
\put(222,46){\vector(-4,1){56}} \put(222,46){\vector(0,3){50}}

\put(180,74){*} \put(160,55){*}\put(220,95){*}

\put(159,78){x}\put(199,98){x}

 \put(70,00){2.   Random adding/removing a box to a Young diagram}
 \put(70,-10){related to the up/downward hops of particles on Maya}
 \put(70,-20){diagram. At unit time instant either a box has to be added}
 \put(70,-30){at any  of vacant places marked by star, or a box marked}
 \put(70,-40){by x has to be removed.}

\end{picture}

{\vskip 40 pt}

\caption{Two realizations of random turn walk} \efig

Let us note that the one directed process of only adding of boxes
(obtained from the general case by the imposing of condition that
all $U_{i}\;\ll \;U_{i-1}$) is well studied, then the problem is
equivalent to the enumeration of the so-called standard Young
tableau \cite{Mac}, see below.

Let us evaluate transition probabilities.

For the sake of simplicity we take Dirac level $n$ to be zero and
denote $U_\lambda(n)$ by $U_\lambda$ where we use notation
(\ref{T-lambda}).

First of all let us note that the weight of the transition from an
initial configuration $|\lambda'\rangle$ to the final
configuration $|\lambda\rangle$ in $\textsc{t} $
 steps is given by the formula
 \be
W_{\lambda'\to\lambda}(\time) =e^{{U}_{\lambda'}-{U}_{\lambda}}
\textsf{N}_{\lambda,\lambda'}(\textsc{t})
 \ee
 where $\textsf{N}_{\lambda,\lambda'}(\textsc{t})$ is the number of ways to
 come from the position $\lambda'$ to the position $\lambda$ in $\textsc{t} $
 steps.  This
 follows from the choice of the weight of a single hop in form
 (\ref{rate-up})-(\ref{rate-down}) and from the definition of the
 weight.

 For the fermions in the state of the thermal equilibrium we have the partition
 function
 \be
Z=\sum_{\lambda} e^{-{U}_{\lambda}}
 \ee
 Each $Z^{-1}e^{-{U}_{\lambda}}$ yields the probability
 to find fermions in a state $|\lambda\rangle$.

 In our case of random turn walk, the hard core particles
  which start from a given initial configuration $|\lambda'\rangle $
  are described via
 \be
Z_{\lambda'}(\textsc{t})=\sum_{\lambda}
e^{{U}_{\lambda'}-{U}_{\lambda}}\textsf{N}_{\lambda,\lambda'}(\textsc{t})
 \ee
  (One may call $\textsf{N}_{\lambda,\lambda'}(\textsc{t})$ kinematic entropy
  factor.)
Then
 \be
P_{\lambda'\to\lambda}(\time)=Z_{\lambda'}(\textsc{t})^{-1}e^{{U}_{\lambda'}-{U}_{\lambda}}
\textsf{N}_{\lambda,\lambda'}(\textsc{t})
 \ee
 yields the probability to find  the hard core particles  in a state $|\lambda\rangle$ as
a result of  $\textsc{t}$-step random turn walk of an initial
configuration $|\lambda'\rangle$  in an external potential
$\{{U}_i ,i\in\mathbb{Z}\}$.

 In particular, in case the random turn walk  starts with the vacuum
 configuration, then
 \be\label{forced-Brownian-partition-func}
Z_{0}(\textsc{t})=\sum_{\lambda}
e^{-{U}_{\lambda}}\textsf{N}_{\lambda,0}(\textsc{t})
 \ee
 where $\textsf{N}_{\lambda,0}(\textsc{t})$ is the number of
 ways to gain the configuration $|\lambda\rangle$ in
 $\textsc{t}$ steps if  the initial configuration is the vacuum
 one.

 The probability to find the hard-core particles (subject to random turn walk
 in a potential
$\{{U}_i ,i\in\mathbb{Z}\}$)  in a
 configuration $|\lambda\rangle$ is
 \be
P_{0\to\lambda}(\time)=Z_{0}(\textsc{t})^{-1}e^{-{U}_{\lambda}}
\textsf{N}_{\lambda,0}(\textsc{t})
 \ee
 yields the probability
 to find the hard core particles in a state $|\lambda\rangle$ as a result of
 $\time$-step random turn walk of an initial configuration $|\lambda'\rangle$.

As we shall show below
\begin{equation}\label{textsfN}
\textsf{N}_{\lambda,0}(\time)
 =2^{ \frac{|\lambda|-\time}{2} }
\frac{\time!}{(\frac{\time-|\lambda|}{2})!}\cdot
 s_{\lambda}({\bt}_\infty)
\end{equation}
where $s_{\lambda}({\bt}_\infty)$ is the Schur function
 evaluated in a special point,
  ${\bt}_\infty$=(1,0,0,\dots), see Appendix A.3 (\ref{schurhook}).

Let us obtain
(\ref{forced-Brownian-partition-func})-(\ref{textsfN}) via the
fermionic representation. Notice that we take
\begin{equation}\label{heisA1}
[A_{-}({U}),A_+({U})]=1
\end{equation}
which simplifies calculations.

We are interested in the evaluation of
\begin{equation}\label{lambda'-lambda-transition}
W_{\lambda'\to\lambda}(\textsc{t})=\tau_{\lambda,\lambda'}(\textsc{t})=\langle
\lambda|{(A_+({U})+A_{-}({U}))^{\textsc{t}}}|\lambda'\rangle
\end{equation}
which gives the weight of the random process where an initial
configuration of hard core particles $|\lambda'\rangle $ comes to
the final configuration $\langle \lambda| $ in $\textsc{t} $
steps.

By (\ref{heisA1}) we have
\begin{equation}\label{decA}
e^{z(A_++ A_{-})}=e^{\frac{z^2}{2}}e^{zA_+}e^{z A_{-}}
\end{equation}

At the present paper we shall study the simplest case:
$\lambda'=0$. Then, our model describes a decay of the  step
configuration as a function of the discrete time $\textsc{t}$. Let
us take into account that $A_-(U)|0\r=0$.

By (\ref{decA}), by Appendix A.5 and also by (\ref{schurhook}) we
obtain \be \langle \lambda|e^{z(A_+ +
A_{-})}|0\rangle=e^{\frac{z^2}{2}} z^{|\lambda|}\cdot
e^{-{U}_{\lambda}}
 s_{\lambda }({\bt}_\infty)
\ee

Now let us decompose both sides in Taylor series in $z$ to find
the right hand side of (\ref{lambda'-lambda-transition}) where
$\lambda'=0$.

For  time duration $\textsc{t}=2m+|\lambda|$ we obtain
\begin{equation}\label{time-duration}
W_{0\to\lambda}(\time)=\textsc{t}! \left(\frac{1}{2^m m!}\right)
{e^{-{U}_{\lambda}} } s_{\lambda }({\bt}_\infty)=\textsc{t}!2^{
\frac{|\lambda|-\textsc{t}}{2} }
\frac{1}{(\frac{\textsc{t}-|\lambda|}{2})!}\cdot
{e^{-{U}_{\lambda}} } s_{\lambda}({\bt}_\infty)
\end{equation}
which, for $ {U}=0$, gives the number of ways to get a
configuration $|\lambda\rangle$ starting from vacuum configuration
$|0\rangle$ in $\textsc{t}$-steps (for $\textsc{t}-|\lambda|=2m$
is even and non-negative). The index $m$ counts how many times
particles, which are involved in the random process, move
downward.
 The left hand side vanishes if $\textsc{t}-|\lambda|$
is odd.

Also, notice that the length of partition $\lambda$ can not exceed
the duration of motion $\time$. The equality $\ell(\lambda)=\time$
is related to the motion (or, the same, to the "path" as was
explained in the subsection 2.3) where a single (the upmost on the
vacuum Maya diagram) particle moves, and each hop is upward one.
Thus in this very case the distance it has pass (which is equal to
the first part of the partition $\lambda$, or, the same, equal to
the length of the first row of Young diagram) is equal to $\time$.

\;

Let us consider different cases described by relation
(\ref{time-duration}).

(1) For $\lambda=0$ the Boltzmann factor $e^{U_0}=1$ and the
weight is equal to the number of ways to return back
 \be\label{r=0weight}
W_{0\to 0}(\time)=\textsf{N}_{0,0}(\time)
 =(\textsc{t}-1)!! = e^{\frac12\time \log\time +\cdots}
 \ee
where the last relation describes the large $\time$ limit.

(2) The case ${\textsc{t}}=|\lambda|$ corresponds to the non-stop
forward motion: all jumps of the particles are in upward
 direction, and described by
 $W_{0\to \lambda}(|\lambda|)=\langle \lambda|{A_{+}^{|\lambda|}}|0\rangle$.
 If one turns to
the description of the motion via Young diagrams, one may see that
 the number
 \be\label{N-tasep-case}
 \textsf{N}_{\lambda,0}(|\lambda|)=|\lambda|!s_{\lambda}({\bt}_\infty)=:d(\lambda)
 \ee
describes the number of ways to create a Young diagram of given
shape $\lambda$ by gluing at random box by box in such a way that
each time we have a Young diagram, see fig.7. (In other words it
is the number of the so-called standard tableau  of the shape
$\lambda$, see Appendix A.1 and for details see \cite{Mac}). Thus
we have
 \be\label{tasep-case}
 W_{0\to\lambda}(|\lambda|)= {e^{-{U}_{\lambda}} } |\lambda|!\  s_{\lambda}({\bf
t}_\infty)={e^{-{U}_{\lambda}} }d(\lambda),
 \ee

(3) The case $\ell(\lambda)=1$  means that the final state is a
single particle configuration (however, in the middle of the
process more particles may be involved, thus, the problem is
different from the random walk of a single particle restricted to
the half-line, which in $\time\to\infty$ limit yields Brownian
motion on the half-line). Thus, in this case $\lambda=(\lambda_1)$
where $\lambda_1$ is equal to the shift of the upmost particle
from its position in the initial vacuum configuration. Suppose the
hopping rate
 \be\label{constant-rate}
e^{-U_{i}+U_{i-1}}=r(i)=r
  \ee
does not depend on $i$ (which was figured out in
(\ref{tilde-t-one}), $r=\exp  (-2{\tilde t}_1)$). As a result of
simple evaluation one shows that in $\time\to\infty$ limit the
dominant term in the subset of $\ell(\lambda)=1$ configurations
$\lambda$ is related to the partition
${\tilde\lambda}=({\tilde\lambda}_ 1)$ given by
\[
{\tilde\lambda}_1=r\sqrt{\time}+\frac{1-r^2}{2}+
\frac{1}{r\sqrt{\time}}\left(r^2+\frac{(1-r^2)^2}{8}
\right)+O\left(\time^{-\frac32}\right)
\]
The weight for one-particles configurations in large $\time$ limit
near ${\tilde\lambda}$  is given by the formula which resembles
formula for the Brownian motion
\[
W_{0\to\lambda} \simeq \time! \exp \left(
-\frac{(\lambda-r\sqrt{\time})^2}{2\sqrt{\time}}\right)
\]
As we see the variance is given by $\time^{\frac14}$. At last we
note, that in case the rate (\ref{rate-up}) depends on site, then
in a wide class of rates in large $\time $ limit
${\tilde\lambda}_1$ may be evaluated as the solution of
${\tilde\lambda}_1=r({\tilde\lambda}_1)\sqrt{\time}$. For
instance, for Gauss potential, $U_i=\frac12 i^2$, one obtains
${\tilde\lambda}_1\sim \log\time$.

(4) The case where ${\textsc{t}}-|\lambda|$ is large enough. Via
Stirling's formula we obtain
 \be
W_{0\to\lambda}(\time) \simeq
\sqrt{2}(\textsc{t})^{\frac{\textsc{t}}{2}}
e^{-\frac{\textsc{t}}{2}}
 \left(\sqrt{\frac{2\textsc{t}}{\sqrt{e}}}\right)^{|\lambda|}
e^{O(\frac{1}{\textsc{t}})} {e^{-{U}_{\lambda}} }
s_{\lambda}({\bt}_\infty),\quad |\lambda| \ll \textsc{t}
 \ee

\;

\br

Let us notice that the sum in the right hand side  of
(\ref{forced-Brownian-partition-func}) may be considered as a
certain correlation function of a {\bf discrete} version of the
{\bf orthogonal ensemble of random matrices}. To see it we use the
same trick as in \cite{OS}. First, according to (\ref{schurhook})
we write
 \be\label{Schur-T-vandermond}
s_{\lambda}({\bt}_\infty)=\frac{\prod_{i<j}^\time(h_i-h_j)}{\prod_{i=1}^\time
h_i!}\; ,\qquad h_i=\lambda_i-i+\time\ge 0,\quad i=1,\dots,\time
 \ee
Then we use the fact that we summarize a symmetric in the
variables $h_i$ function multiplied by the right hand side of
(\ref{Schur-T-vandermond}). In this case we can use permutations
of $h_1,\dots,h_\time$ to replace the sum over the cones
$h_1>\cdots
>h_\time \ge 0$ by sum over all $h_i\ge 0$ getting a factor $1$ over
$\time!$. At last we obtain
\begin{equation}\label{orthog-enssemble}
Z_0(\time)=c\sum_{h_1,\cdots,h_\textsc{t}\ge 0
}\Gamma\left(1+\frac{\textsc{t}^2+\textsc{t}}{4}-\frac12\sum_{i=1}^\textsc{t}h_i\right)^{-1}
e^{\sum_{i=1}^\textsc{t}V(h_i)} \prod_{i,j=1}^\textsc{t}|h_i-h_j|
\end{equation}
\begin{equation}\label{induced Coulomb}
=c\sum_{h_1,\cdots,h_\textsc{t}\ge 0 }
\Gamma\left(1+\frac{\textsc{t}^2+\textsc{t}}{4}-\frac12\sum_{i=1}^\textsc{t}h_i\right)^{-1}
e^{-E(h,1)}
\end{equation}
\be =\sum_\lambda \tau_{\lambda,0}(\textsc{t}) \ee
 where $c=2^{-\frac{\time^2+\time}{2}}$ and  where
\begin{equation}\label{}
V(h_i)= -{U}_{h_i}+{U}_{\time -i}-\log \Gamma(h_i+1)+\frac{\log
2}{2}  h_i
\end{equation}
and $E(h,q),h=(h_1,\dots,h_\time)$ is the electrostatic energy of
$\textsc{t}$ charges $q$ located on the one-dimensional lattice
and having coordinates $h_1,\dots,h_\textsc{t}$:
\begin{equation}\label{U(h,q)}
E(h,q)= \sum_{i=1}^\textsc{t}({U}_{h_i}-{U}_{\textsc{t}-i})-
\frac{\log 2}{2}  h_i-q^2\log
\prod_{i<j}^\textsc{t}|h_i-h_j|+q^2\log \Gamma(h_i+1)
\end{equation}
The number of terms in sums (\ref{orthog-enssemble}) and
(\ref{induced Coulomb}) is finite thanks to the Gamma function
insertion. However in ${\textsc{t}\to \infty}$ limit, the sum
ranges over all positive integers $h_i$.

Notice that  ${E(h,q)}$ vanishes on the vacuum configuration
$\lambda=0$.

Terms in the right hand side of (\ref{U(h,q)}) are to be
interpreted as follows:  the probability rates given by ${U}$ are
related to the external electric potential (which may depend on a
site coordinate), gives rise to the first term. The third term is
the Coulomb interaction of particles with the unit charge. The
last term has a meaning of an external electric field which
provides the vanishing of the Coulomb energy of the vacuum
configuration of the particles. Two last terms appeared due to the
Schur function in the right hand side of (\ref{time-duration}) and
descended from the hard core interaction of the particles. The
(discrete) time of the random turn walk is equated with the total
number of Coulomb particles, or, the same, with the 'size' of
orthogonal matrix whose eigenvalues are presented by the positive
integers $h_1,\dots,h_\textsc{t}$.

The $\tau_{\lambda,0}(\textsc{t})$ gives the weight of the random
process where the vacuum initial configuration  of hard core
particles decays to a final configuration given by coordinates
$h_1,\dots,h_\textsc{t}$ in $\textsc{t} $ steps.

\er

\;

\;

Now let us turn to the problem of finding large $\time$ limit. In
a usual way, we variate positions of particles in order to obtain
the dominant term in sum (\ref{orthog-enssemble}).

First let us remind that the length, say $R$, of the partition we
are looking for does not exceed $\time$. Introducing the density
of particles $\sigma\le 1$ (which the number of particles for one
site) as a function of the variable $h_i=\lambda_i-i+R$  (the
origin of Maya diagram is related to $h=R$)  in a way
 \be
\int_0^\infty \sigma(h)dh=R
 \ee
we come to the equation for $\sigma$ which defines the dominant
configuration of particles at time $\time >>1$ in the continues
limit
 \be\label{equilibrium-rate}
\log \frac{r(h-\time)}{h} + P \int_0^\infty
\frac{\sigma(x)dx}{h-x} +\frac12 \log\left(
\time+\frac{R^2}{2}-\frac{R}{2} - \int_0^\infty x\sigma(x)dx
\right)=0
 \ee
where $P\int$ stands for the principal value.

Let us note that this equation may have no solutions.

\begin{figure}
\begin{center}
\includegraphics[scale=0.66]{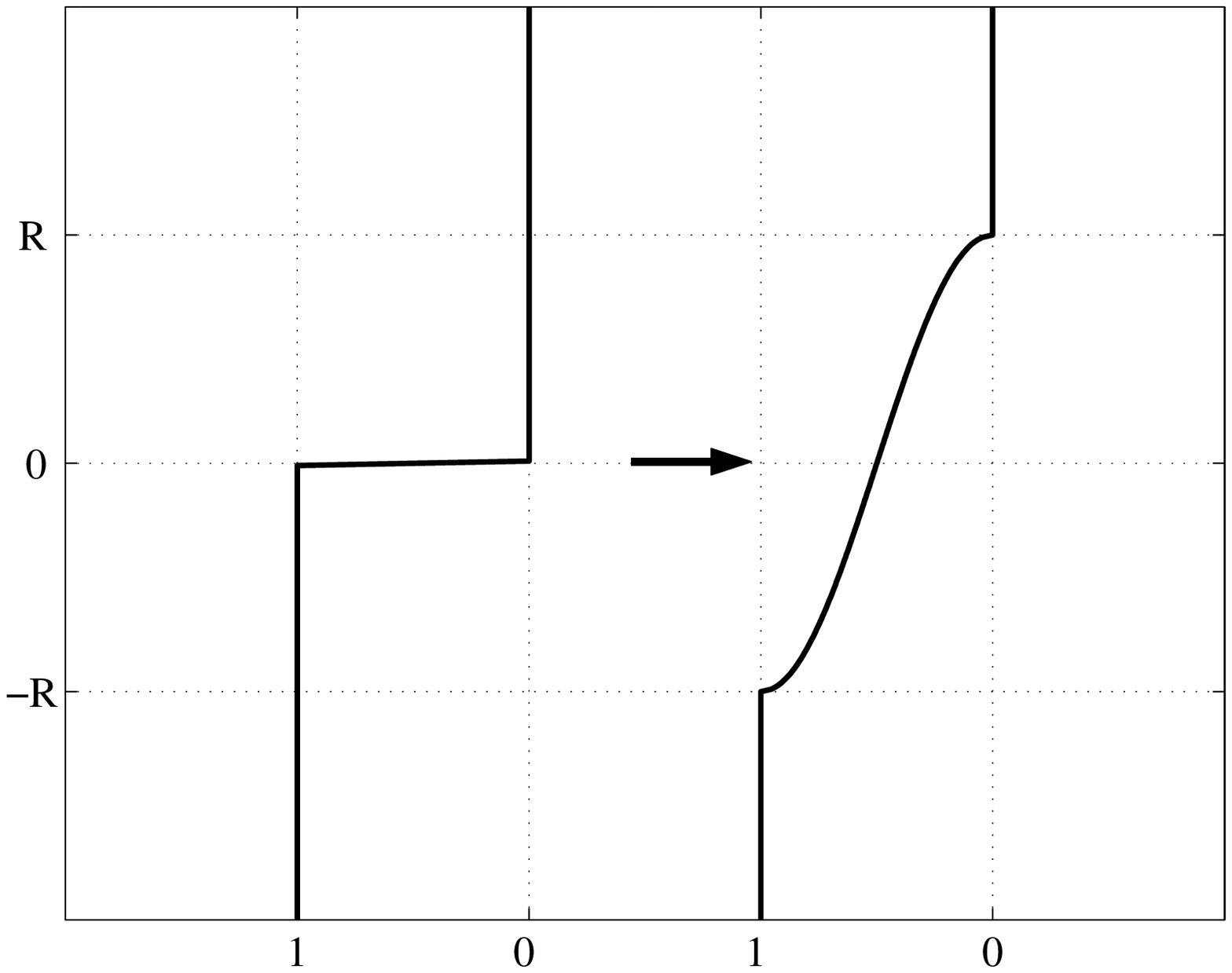}
\caption{ Decay of step function for the case of a constant
hopping rate in $\time\to\infty$ limit} \label{sigma-const-rate}
\end{center}
\end{figure}

This equation one can solve in case the hopping rate
(\ref{rate-up}) does not depend on site: $r(i)=r$, see
(\ref{constant-rate}). With the help of the formula
 \be\label{arcsinus-density}
\int_0^{2R} \; \frac{\frac12-\frac 1\pi\arcsin\left(\frac hR
-1\right)}{x-h}dh=\log\frac{2x}{R}
 \ee
(which may be extracted from \cite{Kerov-Vershik-2}, our proof the
reader will find in the Appendix)
 we obtain the following solution to (\ref{equilibrium-rate})
 \be\label{constant-rate-density}
\sigma(h)= \frac12-\frac {1}{\pi} \arcsin \left(\frac hR
-1\right),\quad h\in [0,2R]
 \ee
 and
 \be\label{constant-rate-density-2}
 \sigma(h)= 1,\quad h<0; \qquad
\sigma(h)=0,\quad h>2R ,
 \ee
  where the length $R$ of the partition is found by substitution of
(\ref{constant-rate-density})-(\ref{constant-rate-density-2}) into
(\ref{equilibrium-rate}) and is equal to
 \be\label{R}
R=2\sqrt{\frac{\time}{1+r^{-2}}},
 \ee
see fig.\ref{sigma-const-rate} which describes the decay of
initial vacuum Maya diagram of level $0$ in large $\time$ limit.

Let us consider the partition $\lambda=\lambda(\time)$ and it's
Young diagram related to the asymptotic density function
$\sigma(h)$. Then solution (\ref{constant-rate-density})-(\ref{R})
correspond to a Young diagram symmetrical under the reflection
with respect to the main diagonal (this follows from the symmetry
between particles and holes which exists in case of constant
rate). The length of the partition
$\ell\left(\lambda(\time)\right)$, it's area $|\lambda(\time)|$
and the number of boxes on the main diagonal
$k\left(\lambda(\time)\right)$ are given by the following formulae
 \be\label{length}
\ell\left(\lambda(\time)\right)=R = 2\sqrt{\frac{\time}{1+r^{-2}}}
 \ee
 \be\label{area}
|\lambda(\time)|=\int_{0}^{2R}h\sigma(h)dh-\frac{R^2}{2}+\frac{R}{2}=
\frac{R^2}{4}+\frac{R}{2}=\frac{\time}{1+r^{-2}}+\sqrt{\frac{\time}{1+r^{-2}}}
 \ee
 \be\label{Frobenius-k}
k\left(\lambda(\time)\right)=\frac{2}{\pi}\sqrt{\frac{\time}{1+r^{-2}}}
 \ee
We obtain (\ref{Frobenius-k}) using the fact that the last number
is equal to the number of particles which passed the origin (the
origin is related to $h=R$) after duration $\time$. This number is
obtained by evaluating the following integral:
 \be\label{Frobenius-k-2}
k\left(\lambda(\time)\right) =\int_R^{2R}\sigma(h)dh=\frac{R}{\pi}
 \ee
 The derivative
 $\frac{dk\left(\lambda(\time)\right)}{d\time}=
 \frac{1}{\pi}({\time}(1+r^{-2}))^{-\frac12}$
 yields current of particles through the origin related to
 asymptotic configuration.

The number of downward steps $m(\time)$ of particles (see
(\ref{time-duration})) related to the asymptotic configuration is
 \be\label{m-time}
m(\time)=\frac{\time-|\lambda|}{2}=\frac{\time}{2(1+r^2)}-\frac12
\sqrt{\frac{\time}{1+r^{-2}}}
 \ee

These formulae give simple answers for particular cases: (a)
completely symmetric simple exclusion process, $r=1$, (b) locking
potential $r\to 0$ resulting to $\lambda(\time)=0$ (c) totally
asymmetric exclusion process (d-TASEP) where particles hop only
upward, $r\to\infty$. Two last cases were considered in the list
of examples which follows (\ref{time-duration}).

Let us write down a formula for the Schur function evaluated for
the asymptotic partition $\lambda=\lambda(\time)$
 \be
s_\lambda({\bf
t}_\infty)=e^{R^2\left(\int_0^2{\tilde\sigma}(u)(u-\log
uR)du+\frac12\int\int
{\tilde\sigma}(u){\tilde\sigma}(u')\log(uR-u'R)dudu'\right)+\cdots}=e^{-\frac14
R^2\log R+R^2O(1)}
 \ee
$$
=e^{-\frac {\time}{2(1+r^{-2})}\log \time+\cdots}
$$
 where by dots we denote minor terms in $\time\to\infty$ limit and
where ${\tilde\sigma}(u):=\sigma( uR)$. Together with
(\ref{textsfN}) and (\ref{m-time}) it yields the number of ways to
get the asymptotic configuration in $\time$ steps via random turn
walk in large $\time$ limit as
 \be
 \textsf{N}_{\lambda(\time),0}(\time)=\frac{T!}{2^{m(\time)}m(\time)!}
 s_{\lambda(\time)}({\bf
t}_\infty)=T!e^{-\frac{\time}{2(1+r^2)}\log\time
-\frac{\time}{2(1+r^{-2})}\log\time
+\cdots}=e^{\frac{\time}{2}\log\time +\cdots}
 \ee
 Notice that in the large $\time$ limit the leading term of
 $\log\textsf{N}_{\lambda(\time),0}(\time)$ does not
 depend on the rate $r$,
 though the asymptotic configuration $\lambda(\time)$ depends. In the large $\time$ limit
 the leading term of $\log\textsf{N}_{\lambda(\time),0}(\time)$  is the same for the locking
 case $r\to 0$  where answer is given by (\ref{r=0weight}) and in the d-TASEP
 case $r\to \infty$ where answer is given by (\ref{N-tasep-case}).

It will be interesting to compare considered numbers (length,
weight, current,...) evaluated on the asymptotic configuration
with their average values.

Let us mark that the shape of the asymptotic Young diagram given
by (\ref{constant-rate-density})-(\ref{constant-rate-density-2})
coincides with the shape found by Kerov and Vershik
\cite{Kerov-Vershik} in the problem of study of limiting shape of
random partitions with fixed weight $|\lambda|=N$ distributed
according to the so-called Plancherel measure. We obtain the
identification if we put $N=R^2$. As soon as we arrive at
(\ref{time-duration}) this fact is not so striking because the
Plancherel measure is equal to $N!s_\lambda({\bf t}_\infty)^2$.
Thus, it is similar to the variational problem for d-TASEP where
(\ref{time-duration}) is replaced by (\ref{tasep-case}). Then, the
variational problem for d-ASEP may be split into two parts: given
weight of partition and taking it into account via Lagrangian
multiplier to variate the shape of the partition and obtain
(\ref{constant-rate-density})-(\ref{constant-rate-density-2})
 (this is
quite similar to the d-TASEP problem where the weight is fixed as
$|\lambda|=\time$), then to variate with respect to the weight.
This might be the other way to get formulae
(\ref{constant-rate-density})-(\ref{R}).


The problem of evaluation of correlation functions for various
versions of random turn walk is studied in the forthcoming paper
\cite{HOasymp}.

\subsection{Weights which depend on mutual configuration: coupling
with Coulomb-type weight}

As we have seen the case ${U}=0$ may be interpreted as a free
random turn walk of the non-intersecting particles on the lattice.
Here "free" means that each particle hops either upward or
downward with the same probability (provided, the target site is
empty). Then, as we have seen, in the answer for the weight of
transition between initial vacuum configuration to a configuration
given by (integer) coordinates $h$,  hard-core interaction gives
rise to the factor $\Delta(h)$. Then (\ref{induced Coulomb}) can
be interpreted as a partition function for the Coulomb gas of
particles (each one has the unit charge) living on the lattice.

For non-vanishing ${U}$ the weights of hops also depend on the
location of the hop. In Coulomb gas picture ${U}_i$ is interpreted
as an electric potential at a point $i$.

Now let us consider a model of random motion where the weight of
each step is additionally dependent on the particle configuration
as follows. Consider the 'coupling' of the model studied in the
previous subsection 2.6 to the modified Coulomb interaction
(\ref{Coulomb-modified}) where particles posses an {\em arbitrary}
charge $q$. Namely, suppose that the weight of the hop is
additionally multiplied by a factor which depends on the
configuration of the particles according to formula
\begin{equation}
\left(\frac{h_{i}}{h_{i}+1}\prod_{j=1,2,\dots \atop j\neq
i}^{}\frac{h_{i}+1-h_j}{h_{i}-h_j}
 \right)^{q^2}
\end{equation}
for the hop of $i$-th particle from the site $h_i$ to the site
$h_i+1$ (provided this site is empty), and according to
\begin{equation}
\left(\frac{h_{i}}{h_{i}-1}\prod_{j=1,2,\dots \atop j\neq
i}^{}\frac{h_{i}-1-h_j}{h_{i}-h_j} \right)^{q^2}
\end{equation}
for the hop of $i$-th particle from the site $h_i$ to the site
$h_i-1$ (provided this site is empty).

Then instead of (\ref{time-duration}) we obtain
\begin{equation}\label{time-duration-Coulomb}
W_{0\to\lambda}(\time;q)=\textsc{t}! \left(\frac{1}{2^m m!}\right)
{e^{-{U}_{\lambda}} } s_{\lambda
}({\bt}_\infty)^{1+q^2}=\textsc{t}!2^{
\frac{|\lambda|-\textsc{t}}{2} }
\frac{1}{(\frac{\textsc{t}-|\lambda|}{2})!}\cdot
{e^{-{U}_{\lambda}} } s_{\lambda}({\bt}_\infty)^{1+q^2}
\end{equation}

 Normalization function
 is a sum of the weights over all transitions of the vacuum
configuration of particles which are described by the
non-intersecting random turn walk in the Coulomb potential of time
duration $\textsc{t}$. After the corresponding changing of the
weights of steps according to the appearance of the Coulomb
interaction with a charge $q$, we get the resulting normalization
function as
 \begin{equation}\label{normalization-q}
Z(\time,q):=\sum_{\lambda\in
P}\left(s_\lambda({\bt_\infty})\right)^{q^2}\langle \lambda| (
A_{-}+ A_{+})^{\textsc{t}}|0\rangle
\end{equation}
\begin{equation}\label{beta-series}
 =
\sum_{h_1,\dots,h_\textsc{t}=1}^\infty  \frac{1}{\Gamma(\frac
{\textsc{t}-|\lambda(h)|}{2}+1)}\left(\frac{1 }{2}\right)^{\frac
{\textsc{t}-|\lambda(h)|}{2}}
 \prod_{i=1}^\textsc{t}
e^{-V(h_i)} \prod^\textsc{t}_{i,j}\left| h_i-h_j \right|^{\beta}
\end{equation}
In (\ref{beta-series}) $h$ is (non-ordered) set of positive
numbers, $h_1,\dots,h_\textsc{t}$, $|\lambda(h)|:=h_1+\cdots
+h_\textsc{t}+\frac12 \textsc{t}(\textsc{t}-1) $, and
 \be\label{beta-q}
\beta=1+q^2
 \ee
 \be
V(h)={U}_h+\beta\log h!
 \ee

\;

Thus, by choosing the rates of hops of the random turn walk we
arrive to effective electric charges model in thermal equilibrium
with particles forced by

(1) external field given by ${U}$

(2) hard-core interaction between particles

(3) Coulomb interaction between particles with coupling constant
$q$

 Note that if $q\neq 0$ we have no anymore need to mention that
our particles are hard-core ones, since the Coulomb repulsion
force does not allow to occupy the same place by more than one
particle.

One can see that the special cases, $q=0,1,\sqrt{3}$, result in
links with discrete analogues of orthogonal ($\beta=1$), unitary
($\beta=2$) and symplectic ($\beta=4$) matrix ensembles
\cite{Mehta} which as it is known may be related to the integrable
hierarchies.

Then, for constant rate the asymptotic configuration will be the
same
(\ref{constant-rate-density})-(\ref{constant-rate-density-2}), see
fig.(\ref{sigma-const-rate}),  where now
 \be\label{R-Coulomb}
R^{1+q^2}=2\sqrt{\frac{\time}{1+r^{-2}}}
 \ee
As we see the size of the asymptotic domain shrinks because of the
adding of the modified Coulomb interaction which tries to confine
particles near the vacuum configuration, see Remark
\ref{on-Coulomb-modified}. For asymptotic configuration
$\lambda(q,\time)$ we straightforwardly obtain
 \[
\ell(\lambda(q,\time))=\left(\frac {4\time}{1+r^{-2}}
\right)^{\frac{1}{2(1+q^2)}},\quad|\lambda(q,\time)|=\frac14\left(\frac
{4\time}{1+r^{-2}} \right)^{\frac{1}{1+q^2}}+\frac12\left(\frac
{4\time}{1+r^{-2}} \right)^{\frac{1}{2(1+q^2)}}
 \]
 \[
k\left(\lambda(q,\time)\right)=\frac{1}{\pi}\left(\frac
{4\time}{1+r^{-2}} \right)^{\frac{1}{2(1+q^2)}},\quad
m(q,\time)=\frac{\time}{2}- \frac18\left(\frac {4\time}{1+r^{-2}}
\right)^{\frac{1}{1+q^2}}-\frac14\left(\frac {4\time}{1+r^{-2}}
\right)^{\frac{1}{2(1+q^2)}}
 \]
which gives answers for the asymptotic partition length, weight,
for the number of particles which passed the origin and for the
number of backward steps. For $q^2> 0$ in the large $\time$ limit
and  we also obtain
\[
\log s_{\lambda(q,\time)}({\bf t}_\infty)=-\frac{1}{2(1+q^2)}
\left(\frac {4\time}{1+r^{-2}} \right)^{\frac{1}{1+q^2}}\log
\time+\cdots
\]
\[
\log
\textsf{N}_{\lambda(q,\time),0}(\time)=\frac{\time}{2}\log\time
+\cdots
\]

At last let us mark that the introduction of the free parameter
$q$ is similar to the introduction of a parameter $s$ in
\cite{LAW}, where a notion of an entropy of stochastic dynamical
systems was studied. We hope to consider this problem in future.

\subsection{Determinantal formulae: Wick theorem and Gessel-Viennot formulae}

Consider $A_+({U})$ of (\ref{random-turn-Brownian}). Then
 \begin{equation}\label{equality}
\frac{1}{N!}\langle \lambda' | A_+({U})^N |\lambda \rangle=
\langle \lambda' | e^{A_+({U})} |\lambda \rangle ,\quad
N=|\lambda'|-|\lambda|
 \end{equation}

The form of right hand side allows to apply the Wick theorem:
 \begin{equation}\label{WGV}
\langle \lambda' | e^{A_+({U})} |\lambda \rangle=\det\left(
\langle m | {\bar f}_{m+h'_j} e^{A_+({U})} f_{m+h_i}|m \rangle
\right)_{i,j=1,\dots,m}=\det\left(\frac{e^{{U}_{m+h_i}-{U}_{m+h'_j}}}{(h'_j-h_i)!}
\right)_{i,j=1,\dots,m}
 \end{equation}
 where
 \be\label{hh'll'}
h_i=\lambda_i-i+m,\quad h'_i=\lambda'_i-i+m,\quad m\ge
\ell(\lambda')\ge \ell(\lambda)
 \ee
where we use
\begin{equation}\label{}
\langle 0|{\bar
f}_{i+N}e^{A_+(U)}f_i|0\rangle=\frac{1}{N!}e^{{U}_{i}-{U}_{i+N}}
 \end{equation}

The right hand side also
 allows to consider tau functions
 \begin{equation}\label{}
\tau({\bt},{U},{\bar{\bt}})=\langle 0|e^{H({\bf
t})}e^{A_+({U})}e^{{\bar H}({\bar{\bt}})}|0\rangle
 \end{equation}
as a generating function for the transition weights $\langle
\lambda' | A_+({U})^N |\lambda \rangle$.

\br
 Let us mark the formula
\begin{equation}\label{via-skew-Schur}
\langle
\lambda'|e^{A_+(U)}|\lambda\rangle=s_{\lambda'/\lambda}({\bf
t}_\infty)\prod_{i=1}^{\ell(\lambda')} e^{-{U}_{h_i}+{U}_{h'_i}}
\end{equation}
where ${\bt}_\infty=(1,0,0,\dots)$ and \be \lambda_i=h_i-i+m,\quad
\lambda'_i=h'_i-i+m,\quad
m=\max(\ell(\lambda),\ell(\lambda'))=\ell(\lambda')
\ee.
 \er

 The left hand side of (\ref{equality}) is interpreted as a
 $N$-step random turn process, described above.

 Let us consider an example.

{\bf Gessel-Viennot formula for binomial determinants }

Consider two sets of positive integers $a_k>\cdots >a_1\ge 0$ and
$b_k>\cdots >b_1 \ge 0$, where $a_i\ge b_i,\; i=1,\dots,k$.
Following \cite{GV} let us study the following determinant
 \begin{equation}\label{}
\det {{a_1\ \dots \ a_k}\choose {b_1\ \dots \ b_k}}:=\det
\left({a_i\choose b_j}\right)_{i,j=1,\dots,k}
 \end{equation}
 called binomial determinant. Let us identify sets of $b_{k-i}$ and $a_{k-i}$
 with sets of $h_i$ and $h'_i$ related by (\ref{hh'll'}) to partitions $\lambda$ and $\lambda'$
   that is
\begin{equation}\label{A-B-lambda-mu}
\lambda_i=b_{k-i}+i-k,\quad  \lambda'_i=a_{k-i}+i-k,\quad
i=1,\dots,k
\end{equation}
Take
\be [x]_k:=\left(\frac 11 \sum_{i=1}^k x_i , \frac 12
\sum_{i=1}^k x_i^2 ,\dots \right) \quad [y]_k:=\left( \frac 11
\sum_{i=1}^k y_i, \frac 12 \sum_{i=1}^k y_2 ,\dots \right)
\ee

It is known \cite{Mac} that Schur functions $s_{\lambda}(\bt)$,
where $\bt=[x]_k$, vanish on partitions whose length exceed $k$:
$\ell(\lambda)>k$.

Then, as a specification of (\ref{WGV}) and by using formulas
(\ref{coherent}), we get that the following tau function
\begin{equation}\label{tau-an-2MM}
\tau(k,[x]_k,[y]_k)=\langle k|e^{H([x]_k)}e^{A}e^{{\bar
H}([y]_k)}|k\rangle
\end{equation}
\[ =\sum_{\lambda,\lambda'\in P\atop
\ell(\lambda),\ell(\lambda')\le k} \langle
k|e^{H([x]_k)}|\lambda',k\rangle \langle\lambda,k|e^{{\bar
H}([y]_k)}|k\rangle \langle\lambda',k|e^{A}|\lambda,k\rangle
 \]
 \be =\sum_{\lambda,\lambda'\in P\atop
\ell(\lambda),\ell(\lambda')\le k}
s_{\lambda'}([x]_k)s_\lambda([y]_k)\det {{a_1\ \dots \ a_k}\choose
{b_1\ \dots \ b_k}}, \ee where
\begin{equation}\label{A-Gesel-Vientot}
A=\sum_{i\in\mathbb{Z}} if_i{\bar f}_{i-1},
\end{equation}
is a generating function for the binomial determinants. This $A$
is related to the motion along the following graph

\bfig
\center{
\begin{picture}(150,170)
 \put(50,30){\vector(0,1){15}} \put(50,50){\vector(0,1){15}} \put(50,70){\vector(0,1){15}}
 \put(50,90){\vector(0,1){15}} \put(50,110){\vector(0,1){15}}\put(50,130){\vector(0,1){15}}
 \put(50,27.5){\circle{5}}  \put(50,47.5){\circle*{5}}\put(50,67.5){\circle{5}}
 \put(50,87.5){\circle{5}}  \put(50,107.5){\circle{5}}\put(48.3,105.4){x}
 \put(50,127.5){\circle{5}}\put(60,25){i-1} \put(60,45){i}
 \put(60,65){i+1} \put(100,85){There are (i+1)(i+2)(i+3) ways to}
\put(60,85){i+2} \put(100,65){come from a point i to the point
i+3}\put(60,105){i+3}\put(60,125){i+4}
\put(40,34){i}\put(32,54){i+1}\put(32,74){i+2}\put(32,94){i+3}\put(32,114){i+4}\put(32,134){i+5}
\end{picture}
}
\caption{The numbers to the left are the weights of the
arrows}
\label{fig9}
\efig

 \br
 A tau function similar to (\ref{tau-an-2MM}) was used in \cite{HO2} for the study
 of the so-called two matrix model.
 \er

The graph related to (\ref{A-Gesel-Vientot})  is sketched out
above. The weight of an arrow may be considered as the number of
identical arrows of unit weight. Thus, the product of weights of
successive arrows connecting two vertices yields the number of
ways connecting these vertices.

The number of ways connecting points $k$ and $i$ on the Maya
diagram is counted by $\frac{i!}{j!}=\langle 0|{\bar f}_i A^{i-j}
f_j |0\rangle$. Therefore,
 \be \langle 0|{\bar f}_ie^A f_j
|0\rangle =\left( = \frac{1}{(i-j)!}\langle 0|{\bar f}_i A^{i-j}
f_j |0\rangle = \right) ={i\choose j}
 \ee

From (\ref{via-skew-Schur}) we have the relation obtained in
\cite{GV}:
\begin{equation}\label{}
\det \left({h'_i \choose h_j }\right)_{i,j=1,\dots,k}=
\frac{(n)_{\lambda}}{(n)_{\lambda'}}s_{\lambda'/\lambda}({\bt}_\infty)
\end{equation}
where $ $ is the so-called skew Schur function (\ref{skewSchur}).

 Gessel and Viennot suggested to interpret the binomial
determinant as the number of nonintersecting ways connecting
points $a_i$ with points $b_i$, where $i=1,\dots,k$ (in our
approach connecting a pair of Maya diagrams) as it is depicted at
the right hand side figure \cite{GV}. Each step is either down or
left directed:

\bfig

\begin{picture}(100,170)
 \put(30,30){\vector(0,1){15}}
 \put(30,50){\vector(0,1){15}}
 \put(30,70){\vector(0,1){15}}
 \put(30,90){\vector(0,1){15}}
 \put(30,110){\vector(0,1){15}}
 \put(30,27.5){\circle{5}}
  \put(30,47.5){\circle*{5}}
   \put(30,67.5){\circle*{5}}
 \put(30,87.5){\circle{5}}
  \put(30,107.5){\circle*{5}}
   \put(30,127.5){\circle{5}}
 \put(40,25){0}
 \put(40,45){1}
 \put(40,65){2}
 \put(40,85){3}
 \put(40,105){4}
 \put(40,125){5}

  \put(29,130){.}
    \put(29,135){.}
   \put(29,140){.}

\put(65,80){\vector(1,0){25}}

\put(120,30){\vector(0,1){15}}
 \put(120,50){\vector(0,1){15}}
 \put(120,70){\vector(0,1){15}}
 \put(120,90){\vector(0,1){15}}
 \put(120,110){\vector(0,1){15}}
 \put(120,27.5){\circle{5}}
  \put(120,47.5){\circle{5}}
   \put(120,67.5){\circle*{5}}
 \put(120,87.5){\circle{5}}
  \put(120,107.5){\circle*{5}}
   \put(120,127.5){\circle*{5}}
 \put(130,25){0}
 \put(130,45){1}
 \put(130,65){2}
 \put(130,85){3}
 \put(130,105){4}
 \put(130,125){5}

 \put(119,130){.}
    \put(119,135){.}
   \put(119,140){.}

  \put(0,00){1. The transition weight is given by}
\put(-10,-10){the number of paths to the next figure}
\end{picture}
\begin{picture}(100,170)
 \put(250,30){\line(0,1){15}}
 \put(250,50){\line(0,1){15}}
 \put(250,70){\line(0,1){15}}
 \put(249.7,90){\line(0,1){15}}\put(250.4,90){\line(0,1){15}}
 \put(250,110){\line(0,1){15}}

\put(250,27.5){\circle{5}} \put(248,29){\line(-1,1){16}}
 \put(248,48){\line(-1,0){15}}
  \put(250,47.5){\circle{5}}\put(228,48){\line(-1,1){16}}
\put(230,47.5){\circle*{5}} \put(248,88){\line(-1,0){55}}
\put(248,88){\line(-1,0){55}} \put(230,68){\line(0,1){40}}
\put(229.7,50.5){\line(0,1){17.5}}\put(230.4,50.5){\line(0,1){17.5}}

   \put(250,67.5){\circle*{5}}
\put(247,67.7){\line(-1,0){17.5}}\put(247,68.4){\line(-1,0){17.5}}
\put(229.5,68){\line(-1,0){17}} \put(210,67.5){\circle*{5}}
\put(209.7,70.5){\line(0,1){17.5}}\put(210.4,70.5){\line(0,1){17.5}}
\put(210,88){\line(0,1){20}}

\put(250,87.5){\circle{5}}
   \put(190,87.5){\circle{5}}\put(208,69){\line(-1,1){16}}
\put(247,87.7){\line(-1,0){37}}\put(247,88.4){\line(-1,0){37}}\put(210,88){\line(-1,0){17.5}}
\put(190,90.5){\line(0,1){17.5}}

  \put(250,107.5){\circle*{5}}
  \put(170,107.5){\circle*{5}}\put(188,89){\line(-1,1){16}}
\put(247,108.5){\line(-1,0){17.5}}
\put(230,108.1){\line(-1,0){57.5}}\put(230,108.8){\line(-1,0){57.5}}
\put(229.7,108.5){\line(0,1){18.5}}\put(230.4,108.5){\line(0,1){18.5}}
\put(181.5,94){*}

\put(230,127.7){\line(1,0){18.5}}\put(230,128.4){\line(1,0){18.5}}
\put(250,127.5){\circle*{5}}\put(168,109){\line(-1,1){16}}
\put(239,113){*}\put(239,93){*}\put(219,93){*}\put(199,93){*}\put(199,75){*}

 \put(260,25){0}
 \put(260,45){1}
 \put(260,65){2}
 \put(260,85){3}
 \put(260,105){4}
 \put(260,125){5}

 \put(215,45){1}
 \put(195,65){2}
 \put(175,85){3}
 \put(155,105){4}

 \put(135,45){\vector(1,1){15}}\put(145,60){t}
 \put(135,45){\vector(1,-1){15}}\put(150,30){x}

  \put(130,00){2. Paths connecting a pair of Maya diagrams}
  \put(130,-10){may be viewed as the interface of growing area}
\end{picture}

{\vskip 10 pt}
\caption{Counting connecting paths}
\label{fig10}
\efig

Notice that if we introduce the space-time axis $(x,t)$ as it is
figured on the picture Fig.3 we observe non-intersecting
one-dimensional random walk of $k$ vicious walkers, moving with
speed $\pm1$, changing  at random the direction at each time
instant $\Delta t=\frac12$. The particles start at $t=0$ their
motion from positions $a_1,\dots,a_k$. Later a particle by
particle meets a 'light' signal spreading with the speed $-1$
which starts motion in the origin at $t=0$. A particle which
started at $a_i$ meets the signal at point $\frac12 b_i$.

Considering random path to be an interface one can relate the
right hand figure to the randomly growing area below the path. By
stars we mark the last portion of area (ice covering of a step
scala). Then the parameter $k$ in the tau function
(\ref{tau-an-2MM}) may be viewed as discrete time.

These problems will be considered in the forthcoming paper.


If we replace $e^A$ by $e^A\cdots e^A=e^{\time A}$ in the
expression for tau function (\ref{tau-an-2MM}) we obtain
$\time$-ple mapping of partitions described in the next section.
In certain sense the model considered in this section interpolates
between random turn and vicious walkers models, which we consider
in the next section.

\section{Action of $\widehat{GL}(\infty)$
group elements on Fock vectors and random processes}

Below we shall construct certain versions of stochastic motion on
the space of partitions, which may be also called random slices
("rocks") and random layering. Namely, one may view  stochastic
dynamics on partitions as models of random slices, which may be
visualized as a set of partitions placed one above another
according to its time evolution. Each three-dimensional figure of
height $\time$ obtained in this way present a path
(\ref{one-path}) and may be compared with a rock inscribed into a
corner between two walls. In this section we present only
descriptions of the models in terms of fermions without study of
these models.

\subsection{Determinantal processes}

Now consider the case where each $o^{j,j-1}$ in
(\ref{fock-sequence}) has a special form
 \be\label{choose-group}
o^{j,j-1}=g^{j,j-1}:=e^{\sum_{i,k} a_{ik}^{j,j-1}f_i{\bar
f}_k},\quad o(\time)= g(\time):=g^{\time,\time-1}\cdots g^{1,0}
 \ee
These exponentials are known to be elements of
${\widehat{GL}}(\infty)$ group, see \cite{DJKM},\cite{JM}, where
the importance of $\widehat{GL}(\infty)$  for constructing of
integrable equations together with their solutions was
established.

Then
 \[
  \tau_\time(x,y)=\langle 0| {\bar
f}(x_N^{-1})\cdots {\bar f}(x_1^{-1})\ g(\time)\ { f}(y_1)\cdots {
f}(y_N)|0\rangle\ \Delta(x)^{-1}\Delta(y)^{-1}
 \]
 is known \cite{DJKM},\cite{JM} to be a TL tau function. In our case we want
 tau functions  to be dependent on the axillary discrete variable $\time
 $ via (\ref{choose-group}).

In what follows we shall assume the following simplifying
condition
 \be\label{finite-matrices}
a_{ik}^{(j,j-1)}=0,\quad i,k<0,\quad j=1,2,\dots
 \ee

 Let us remind definitions of the Schur function
\[ s_\nu(x)=\det \left(x_a^{h_b}\right)\Delta(x)^{-1},\quad
h_i=\nu_i-i+N \] and basis Fock vectors
\[
|\nu,N \rangle=f_{h_N}\cdots f_{h_1}| 0 \rangle,\quad \langle
\nu,N |=\langle 0 | {\bar f}_{h_1}\cdots {\bar f}_{h_N}
\]
Denoting the matrix elements of $g(\time)$:
  \be\label{lambda-g-nu}
\tau_{\lambda,\nu}(\time)=\langle \lambda,N| g(\time)
|\nu,N\rangle ,
 \ee
we may  rewrite tau function in form
 \[
 \tau_\time(x,y)=\sum_{\lambda,\nu}\tau_{\lambda,\nu}(\time)s_\lambda(x)s_\nu(y)
 \]

 If we have (\ref{finite-matrices}), then Wick's theorem has especially
simple form:
  \be\label{matrix-elements}
 \langle \mu^{(j)}|g^{j,j-1}|\mu^{(j-1)}\rangle
=\det \ g^{j,j-1} (h^{(j)}_a,  h^{(j-1)}_b), \quad g^{j,j-1}
(a,b)= \langle 0|f_a g^{j,j-1} {\bar f}_b|0\rangle
 \ee

 \br\label{det-ens}
Thanks to the Wick theorem, we write (\ref{intermediate}) as
 \be\label{tau-group}
 \tau_\time(x,y)=\sum p_M^\time\left(\{h_a^{(j)} \}\right) \Delta(x)\Delta(y)
 \ee
where the sum ranges over $\time$ sets of all admissible
configurations $h_a^{(j)},\ j=0,\dots,\time$
 \be
\label{multi-level-black-box}
 p_M^\time\left(\{h_a^{(j)} \}\right)=\det \left(x_a^{h_b^{(0)}}\right)
\det  \left(y_a^{h_b^{(\time)}}\right) \prod_{j=0}^{\time-1}
\det\left( g^{j+1,j} (h^{(j+1)}_a,h^{(j)}_b)\right)
 \ee

Now,
 \be\label{discrete-rpd}
   P_M^\time\left(\{h_a^{(j)} \}\right)=
\frac{ p_M^\time\left(\{h_a^{(j)} \}\right)}{\tau_\time(x,y)}
=\frac{p_M^\time\left(\{h_a^{(j)} \}\right)}
{\sum_{\lambda,\nu}\tau_{\lambda,\nu}(\time)s_\lambda(x)s_\nu(y)}
 \ee
almost coincides with the formula (1.10) of \cite{H-IMRN} (under
some specification of functions $\psi$ and $\phi$ in this paper,
and under certain notational replacements: say, $h_a^{(j)}$ of
(\ref{discrete-rpd}) are  $x_a^{(j)}$ of \cite{H-IMRN} and
$\tau_\time(x,y)$  of (\ref{discrete-rpd}) is  $Z_{M,m}$ of
\cite{H-IMRN}), where it describes the so-called reduced
probability density . The only difference between formula
(\ref{discrete-rpd}) and formula (1.10) of \cite{H-IMRN} is that,
in our case, variables $h_a^{(j)}$ take discrete values as they
are related to the parts of the intermediate partitions, while the
main example of \cite{H-IMRN} is the multimatrix model where these
variables coincide with eigenvalues of random matrices (then, $a$
serves for the number of an eigenvalue, and $j$ serves for a
number of a matrix in the chain of coupled matrices).

More generally (\ref{discrete-rpd}) describes a discrete analogue
of the so-called multi-level  determinantal ensembles where the
"levels" are numbered by  $j$ in (\ref{multi-level-black-box}).

Any $o^{j,j-1}$ in form (\ref{choose-group})
 gives rise to a multi-level   determinantal ensemble \cite{soshnikov}.

\er

In our approach to random processes the tau function
$\tau_\time(x,y)$ plays the role of the generating function for
$\tau_{\lambda,\nu}(\time)$. The normalization function is
\[
 Z_{\nu}(\time)=\sum_\lambda \tau_{\lambda,\nu}(\time)
\]

In the present paper we are interested not in the ratio
$P_M^\time\left(\{h_a^{(j)} \}\right)$ mentioned in the remark
(see (\ref{discrete-rpd}))  but in:
 \be\label{transition-probability}
P_{\nu\to\lambda}(\time)=
\frac{\tau_{\lambda,\nu}(\time)}{Z_{\nu}(\time)}
 \ee
yielding the normalized weight (which may be interpreted as the
probability in case all weights are positive) to arrive to a
configuration $\lambda$ in $\time$ steps if an initial
configuration is $\nu$.

Let us note that the most natural way to consider
(\ref{multi-level-black-box}) is to use the so-called
multi-component fermions and multi-component integrable
hierarchies. This will be done below.

\subsection{Multi-component fermions}

For certain problems it is suitable to re-write
(\ref{multi-level-black-box})  with the help of multi-component
fermions,
\[
[f_n^{(j)},f_m^{(i)}]_+=[{\bar f}_n^{(j)},{\bar
f}_m^{(i)}]_+=0,\quad [f_n^{(j)},{\bar
f}_m^{(i)}]_+=\delta_{j,i}\delta_{n,m}
\]
(see  Appendix A.2 for details). The multicomponent fermions were
used in \cite{DJKM},\cite{JM} to construct hierarchies of
multicomponent integrable equations.

We shall use the following notations. Given set of partitions,
$\nu^{(j)},\ j=0,\dots,\time $, we introduce a basis Fock vector
\[
|\nu^{(0)},n^{(0)};\dots;\nu^{(\time)},n^{(\time)}\rangle :=
\Pi^{(\time)}\cdots\Pi^{(1)}|0,\dots,0\rangle
\]
where
\[
\Pi^{(j)}=f_{h_N^{(j)}}^{(j)}\cdots f_{h_1^{(j)}}^{(j)},\quad
h_i^{(j)}=\nu_i^{(j)}-i+n^{(j)},\quad n^{(j)}=\ell(\nu_i^{(j)})
\]

The dual Fock vector we denote by $\langle
\nu^{(0)},n^{(0)};\cdots;\nu^{(\time)},n^{(\time)}|$:
\[
\langle \lambda^{(0)},l^{(0)};\dots;\lambda^{(\time)},l^{(\time)}|
\nu^{(0)},n^{(0)};\dots;\nu^{(\time)},n^{(\time)}\rangle
=\delta_{\lambda^{(0)},\nu^{(0)}}\cdots
\delta_{\lambda^{(\time)},\nu^{(\time)}}
\]

\;

Now, we replace (\ref{fock-sequence}) by
 \be\label{multi-fock-sequence}
 |\nu,N;0,0;\dots;0,0\rangle \to
o^{1,0}|\nu,N;0,0;\dots;0,0\rangle \to \cdots \to
o^{\time,\time-1}\cdots o^{1,0}|\nu,N;0,0;\dots;0,0\rangle
 \ee
 where
 \be\label{N=ell}
 N=\ell(\nu^{(0)})
 \ee
 and
where each operator $o^{j+1,j}$ is an intertwining operator
between different Fock spaces $F^{(j+1)}$ and $F^{(j)}$ :
 \be\label{G-j+1-j}
 {o}^{j+1,j}={G}^{j+1,j}:=e^{ \sum_{i,k\ge 0}
g_{k,i}^{j+1,j}f_k^{(j+1)}{\bar f}_i^{(j)}}
 \ee

Then (\ref{multi-fock-sequence}) describes the same process on
partitions as (\ref{fock-sequence}) conditioned by
(\ref{choose-group}) and by (\ref{finite-matrices}) provided that
$g_{k,i}^{j+1,j}$ coincides with $g^{j+1,j}(k,i)$ of
(\ref{matrix-elements}).

 Denoting
 \be
  G(\time) = G^{\time,\time-1}\cdots
G^{1,0}
 \ee
we re-write tau function (\ref{tau-group}) as
 \be\label{tau=tau-multi}
\tau_\time(x,y)=(N!)^\time \langle
0,\dots,0|f^{(\time)}(x_1)\cdots f^{(\time)}(x_N) G(\time) {\bar
f}^{(1)}(y_N)\cdots {\bar f}^{(1)}(y_1) | 0,\dots,0 \rangle
\Delta(x)^{-1}\Delta(y)^{-1}
 \ee
 \[
 =\sum_{\lambda,\nu}\tau_{\lambda,\nu}(\time)s_\lambda(x)s_\nu(y)
 \]
where $\tau_{\lambda,\nu}(\time)$  of (\ref{lambda-g-nu}) is
reexpressed as
  \be\label{lambda-g-nu-multicomponent}
\tau_{\lambda,\nu}(\time)=\langle 0,0;\dots;0,0;\lambda,N|
G(\time) |\nu,N; 0,0;\dots;0,0 \rangle
 \ee
Tau function (\ref{tau=tau-multi}) is a specification of the
multi-component tau function introduced in \cite{DJKM} (see also
\cite{KdL}), and almost coincides with the fermionic
representation of the milti-matrix model presented in
\cite{paper5}.

Regardless to the choice of representation for
$\tau_{\lambda,\nu}(\time)$ which is obtained either from the
one-component fermionic expectation value (\ref{lambda-g-nu}), or
from the multi-component one, (\ref{lambda-g-nu-multicomponent}),
the formulae (\ref{transition-probability}) yields the transition
probability to come to a configuration $\lambda$ if the initial
configuration is $\nu$.

{\em Example: $N$ vicious walkers (discrete Brownian motion of $N$
hard core particles)}


Let us apply it to the model of $N$ vicious walkers introduced by
M.Fisher in \cite{F} (actually he mainly considered the so-called
lock step model where all walkers start on odd (or even) numbered
coordinates on the lattice. At each tick of the clock each walker
moves either up or downward subject to no two walkers occupying
the same site at the same time. In various contexts this model was
studied in \cite{Baik},\cite{Forr1},\cite{Forr2}, \cite{TW}.

 In this case for
(\ref{G-j+1-j}) we take
 \be\label{half-line-Brownian-motion}
G^{j,j-1}=\exp \sum_{i=0}^\infty \left(
e^{-U^{(j)}_{i+1}+U^{(j-1)}_{i}}f_{i+1}^{(j)}{\bar f}_i^{(j-1)} +
e^{-U^{(j)}_{i-1}+U^{(j-1)}_{i}} f_{i-1}^{(j)}{\bar
f}_i^{(j-1)}\right)
 \ee

The number $j$ plays the role of physical time, the parameter
$\time$ coincides with the duration of the process. Application of
operator $G^{j,j-1}$ to a Fock vector (=Maya diagram)
 describes the
unit-time transition of $N$ particles, each of which hops to the
nearest site  conditioned that in both initial and final
configurations there are no sites occupied by more than one
particle. In (\ref{half-line-Brownian-motion}) $U$ plays the role
of an external field: a particle located at a site $i$ hops upward
with the rate $e^{-U^{(j)}_{i+1}+U^{(j-1)}_{i}}$ and downward with
the rate $e^{-U^{(j)}_{i-1}+U^{(j-1)}_{i}}$.

Then (\ref{multi-fock-sequence}) describes the motion of hard-core
particles which start with a configuration $\nu$. As it follows
from (\ref{vak}) the particles are constrained to move only along
the positive half-line. Particles are not influenced by this
restriction in case their initial configuration (given by $\nu$)
is far enough from the origin.

The exponent in (\ref{half-line-Brownian-motion}) should be
compared with expression (\ref{random-turn-Brownian}) which was
used earlier to get the random turn motion.


{\em Vicious walkers on the circle}

We need to replace (\ref{half-line-Brownian-motion}) by
 \be\label{circle-Brownian-motion}
 G^{j,j-1}=\exp g^{j,j-1},\quad g^{j,j-1}= e^{-U^{(j)}_{0}+U^{(j-1)}_{n}}f_{0}^{(j)}{\bar
f}_n^{(j-1)} + e^{-U^{(j)}_{n}+U^{(j-1)}_{0}} f_{n}^{(j)}{\bar
f}_0^{(j-1)}
 \ee
\[ +\sum_{i=0}^{n-1} \left(
e^{-U^{(j)}_{i+1}+U^{(j-1)}_{i}}f_{i+1}^{(j)}{\bar f}_i^{(j-1)} +
e^{-U^{(j)}_{i-1}+U^{(j-1)}_{i}} f_{i-1}^{(j)}{\bar
f}_i^{(j-1)}\right)
 \]

\;

{\bf On correlation functions}. An advantage of the multicomponent
approach is the simple representation for the correlation
functions in form of {\bf determinants}.

Given set of partitions $\rho^{(j)},\ j=1,\dots,\time-1$, whose
lengths, $\ell(\rho^{(j)})$, do not exceed $N$, one may address
the following questions:

(a) what is the weight of the process where all paths
(\ref{one-path}) are constrained by the condition that each number
$\rho^{(j)}_i,\; i=1,\dots,\ell(\rho^{(j)})\le N $ is a part of
$j$-th intermediate partition $\mu ^{(j)}$ belonging to the path.

Let us consider 'characteristic operators'
 \be \label{characteristic-1}
\chi^{(j)}:=\chi(\rho^{(j)})=f^{(j)}_{h^{(j)}_1} \cdots {
f}^{(j)}_{h^{(j)}_N} {\bar f}^{(j)}_{h_N^{(j)}} \cdots {\bar
f}^{(j)}_{h_1^{(j)}}
 \ee
where $h^{(j)}_i=\rho^{(j)}_i-i+N$. Introduce
 \be
o_{\chi}(\time)=
o^{\time,\time-1}\chi^{(\time-1)}o^{\time-1,\time-2}\ \cdots \
o^{2,1}\chi^{(1)}o^{1,0}
  \ee

Then, the weight we are looking for is  the following correlation
function
 \be
\texttt{K}_{\nu\to\lambda}(\rho^{(1)},\dots,\rho^{(\time-1)})=
\frac{\langle 0,0;\dots;0,0;\lambda,N| o_\chi(\time)
|\nu,N;0,0;\dots;0,0\rangle}{Z_\nu(\time)}=
\frac{\det\left(\langle 0|{\bar f}^{(\time)}_{h_i} o_{
\chi}(\time) f^{(0)}_{h_k'}|0\rangle\right)}{Z_\nu(\time)}
 \ee
where $h_i=\lambda_i-i+N$ and $h_i'=\nu_i-i+N$ ($i=1,\dots,N$),
and where we use the notation $|0\rangle:=|0,0;\dots;0,0\rangle$.
 The last equality is due to the Wick theorem.

(b) what is the weight of the process where  all paths
(\ref{one-path}) constrained by the condition that no one part of
an  $j$-th intermediate partition $\mu ^{(j)}$ belonging to the
path coincides
 with any of parts of the $\rho^{(j)}$, where $j=1,\dots,\time-1$.

Let us consider 'characteristic operators'
 \be \label{characteristic-2}
{\bar \chi}^{(j)}:={\bar \chi}(\rho^{(j)})= {\bar
f}^{(j)}_{h_N^{(j)}} \cdots {\bar f}^{(j)}_{h_1^{(j)}}
f^{(j)}_{h^{(j)}_1} \cdots { f}^{(j)}_{h^{(j)}_N} \ee where
$h^{(j)}_i=\rho^{(j)}_i-i+N$. Introduce
 \be
o_{\bar \chi}(\time)= o^{\time,\time-1}{\bar
\chi}^{(\time-1)}o^{\time-1,\time-2}\ \cdots \ o^{2,1}{\bar
\chi}^{(1)}o^{1,0}
  \ee

The weight of this process  is the following correlation function
 \be
\texttt{K}_{\nu\to\lambda}({\bar\rho}^{(1)},\dots,{\bar\rho}^{(\time-1)})=
\frac{\langle 0,\dots,0,\lambda|  o_{\bar \chi}(\time)
|\nu,0,\dots,0\rangle}{Z_\nu(\time)}= \frac{\det\left(\langle
0|{\bar f}^{(\time)}_{h_i}  o_{\bar \chi}(\time)
f^{(0)}_{h_k'}|0\rangle\right)}{Z_\nu(\time)}
 \ee
 where $h_i=\lambda_i-i+N$ and $h_i'=\nu_i-i+N$ ($i=1,\dots,N$).
 The last equality is due to the Wick theorem.

Equivalent representation is given by
 \be
\texttt{K}_{\nu\to\lambda}({\bar\rho}^{(1)},\dots,{\bar\rho}^{(\time-1)})=\frac{\langle
0,0;\rho^{(\time-1)},n^{(\time-1)};\dots;\rho^{(1)},n^{(1)};\lambda,N
|
o(\time)|\nu,N;\rho^{(\time-1)},n^{(\time-1)};\dots;\rho^{(1)},n^{(1)};0,0
\rangle}{Z_\nu(\time)}
 \ee
 where
$n^{(\time-1)}=\ell(\rho^{(\time-1)})$

This form allow to write it via the Wick theorem as
 \[
\texttt{K}_{\nu\to\lambda}({\bar\rho}^{(1)},\dots,{\bar\rho}^{(\time-1)})=\frac{
\det \left( \langle 0|{\bar f}_{h_i}\;o(\time)\; f_{h_k} |0\rangle
\right)}{Z_\nu(\time)}
 \]

By an analogy with the so-called spectral correlation functions in
random matrices, we call the weights
$\texttt{K}_{\nu\to\lambda}(\rho^{(1)},\dots,\rho^{(\time-1)})$
and
$\texttt{K}_{\nu\to\lambda}({\bar\rho}^{(1)},\dots,{\bar\rho}^{(\time-1)})$
 correlation functions.

\subsection{Random layering. Chains of Darboux transformations}

Below we consider simple application of certain $GL_\infty$
transformations, sometimes called Darboux transformations. For
$U=0$ below and for the case of growing partitions these examples
are mainly reformulation of different known random models
considered in literature.

(I) {\bf Random lay}

Let us consider four basic examples of $\widehat{GL}(\infty)$
operators. They are basic in the sense that their non-vanishing
matrix elements $o_{\lambda,\lambda'}$ are related to pairs of
Young diagrams different by adding/removing layers (strips) to
Young diagrams: vertical or horizontal ones. Thus, we pick up four
different types of $o$, say, $o^{(i)}=\exp A^{(i)},\; i=1,2,3,4$.
Let us consider these four cases separately.

(1) The exponential of
 \be\label{sigma=1}
A^{(1)}(x,U):=-A^{(1)}(-x,U)=-\sum_{k=1,2,\dots}\sum_{i\in
\mathbb{Z}} e^{-U_{i}+U_{i-k}}(-x)^{k}f_{i}{\bar f}_{i-k}
 \ee
yields laying of a given Young diagram by a vertical strip, see
fig. 11,  as it is given by
\[
o^{(1)}(x,U)|\lambda'\rangle=\sum_\lambda |\lambda\rangle \;
o^{(1)}_{\lambda,\lambda'}(x,U)
\]
where
\[
o^{(1)}_{\lambda,\lambda'}(x,U):=\langle \lambda
|o^{(1)}|\lambda'\rangle=e^{U_{\lambda'}-U_{\lambda}}x^{|\lambda|-|\lambda'|}
\]
is non-vanishing only if the difference between $\lambda$ and
$\lambda'$ is the so-called vertical strip \cite{Mac}. Example:

\bfig

\begin{picture}(300,120)
 \put(9.7,50){\line(0,1){40}}\put(10.4,50){\line(0,1){40}}

 \put(29.7,50){\line(0,1){40}}\put(30.4,50){\line(0,1){40}}

 \put(49.7,70){\line(0,1){20}}\put(50.4,70){\line(0,1){20}}

 \put(69.7,70){\line(0,1){20}}\put(70.4,70){\line(0,1){20}}

 \put(10,49.7){\line(1,0){20}}\put(10,50.4){\line(1,0){20}}

 \put(10,89.7){\line(1,0){60}}\put(10,90.4){\line(1,0){60}}

 \put(10,69.7){\line(1,0){60}}\put(10,70.4){\line(1,0){60}}

 \put(100,6){\line(0,1){84}}

 \put(120,6){\line(0,1){84}}

 \put(100,10){\line(1,0){20}}

 \put(100,90){\line(1,0){20}}

\multiput(100,30)(0,20){4}{\line(1,0){20}}

\multiput(100,20)(0,20){4}{\vector(-1,0){10}}

\multiput(110,-10)(0,5){3}{\circle{0.1}}

\multiput(108,18)(0,20){4}{x}

 \put(170,80){Fig.11a. Condensation of a lay onto a Young diagram }
 \put(170,70){$\lambda'$. Random 'rain' of boxes from rightward. These boxes}
 \put(170,60){may be fitted only to admissible places.}

\end{picture}

\begin{picture}(300,170)

\put(0.7,90){\line(0,1){40}}\put(1.4,90){\line(0,1){40}}

 \put(20.7,90){\line(0,1){40}}\put(21.4,90){\line(0,1){40}}

 \put(40.7,110){\line(0,1){20}}\put(41.4,110){\line(0,1){20}}

 \put(60.7,110){\line(0,1){20}}\put(61.4,110){\line(0,1){20}}

 \put(1,89.7){\line(1,0){20}}\put(1,90.4){\line(1,0){20}}

\put(1,129.7){\line(1,0){60}}\put(1,130.4){\line(1,0){60}}

 \put(1,109.7){\line(1,0){60}}\put(1,110.4){\line(1,0){60}}

 \put(89.7,70){\line(0,1){60}}\put(90.4,90){\line(0,1){40}}

 \put(109.7,90){\line(0,1){40}}\put(110.4,70){\line(0,1){60}}

 \put(129.7,110){\line(0,1){20}}\put(130.4,110){\line(0,1){20}}

 \put(149.7,110){\line(0,1){20}}\put(150.4,110){\line(0,1){20}}

 \put(90,70){\line(1,0){20}}

 \put(90,89.7){\line(1,0){20}}\put(90,90.4){\line(1,0){20}}

 \put(90,129.7){\line(1,0){60}}\put(90,130.4){\line(1,0){60}}

 \put(90,109.7){\line(1,0){60}}\put(90,110.4){\line(1,0){60}}

 \put(99,78){x}

 \put(179.7,90){\line(0,1){40}}\put(180.4,90){\line(0,1){40}}

 \put(199.7,90){\line(0,1){40}}\put(200.4,90){\line(0,1){40}}

 \put(219.7,110){\line(0,1){20}}\put(220.4,110){\line(0,1){20}}

 \put(239.7,110){\line(0,1){20}}\put(240.4,110){\line(0,1){20}}

 \put(180,89.7){\line(1,0){20}}\put(180,90.4){\line(1,0){20}}

 \put(180,129.7){\line(1,0){60}}\put(180,130.4){\line(1,0){80}}

 \put(180,109.7){\line(1,0){80}}\put(180,110.4){\line(1,0){60}}

 \put(180,110){\line(0,1){20}}

 \put(260,110){\line(0,1){20}}

 \put(249,118){x}

 \put(289.7,70){\line(0,1){60}}\put(290.4,90){\line(0,1){40}}

 \put(309.7,90){\line(0,1){40}}\put(310.4,70){\line(0,1){60}}

 \put(329.7,90){\line(0,1){40}}\put(330.4,110){\line(0,1){20}}

 \put(349.7,110){\line(0,1){20}}\put(350.4,110){\line(0,1){20}}

 \put(290,110){\line(0,1){20}}

 \put(290,70){\line(1,0){20}}

 \put(290,89.7){\line(1,0){40}}\put(290,90.4){\line(1,0){20}}

 \put(290,129.7){\line(1,0){60}}\put(290,130.4){\line(1,0){80}}

 \put(290,109.7){\line(1,0){80}}\put(290,110.4){\line(1,0){60}}

 \put(290,110){\line(0,1){20}}

 \put(370,110){\line(0,1){20}}

 \put(290,50){\line(1,0){20}}

 \put(290,30){\line(1,0){20}}

 \put(290,30){\line(0,1){20}}\put(310,30){\line(0,1){20}}

\put(290,65){\circle{0.4}}\put(290,60){\circle{0.4}}\put(290,55){\circle{0.4}}
\put(310,65){\circle{0.4}}\put(310,60){\circle{0.4}}\put(310,55){\circle{0.4}}

 \put(299,78){x}\put(359,118){x}
 \put(319,98){x}\put(299,38){x}

 \put(30,0){Fig.11b. Examples of diagrams of $\lambda$ where $o^{(1)}_{\lambda,\lambda'}$ is
 nontrivial. One layer (one strip)}
 \put(30,-10){is added at random to the Young diagram $\lambda'$. Boxes marked by x depict the}
 \put(30,-20){difference between $\lambda'$ and $\lambda$, called
 the vertical strip}

\end{picture}

\begin{picture}(430,60)
\end{picture}

{\vskip 10 pt} \caption{A layering by a vertical strip}
\label{fig11} \efig

Now turn to the related Maya diagrams which describes $1D$
configuration of the related lattice gas. Let us compare it with
the random turn walk, see Fig.7 in the subsection "Decay of the
step function". In the random turn walk model in each time instant
only one particle hops either one step upward or one step downward
to the nearest neighboring site. The Fig.11 above describes the
one step of a {\em group} of randomly chosen particles upward  to
the nearest neighboring sites which occurs at one time instant,
this hop being conditioned that after the hop each site is
occupied by no more than one particle. One may say that Fig.11b
describes 'trains' of particles which hop one step upward. The
potential $U$ gives rise to the gluing rate in $2D$ picture and
the  hopping rate in $1D$ picture which site-depending. The weight
to absorb $n$ boxes is proportional to $x^n$.

(2) The exponential of
 \be\label{sigma=2}
A^{(2)}(x,U):=-\sum_{k=1,2,\dots}\sum_{i\in \mathbb{Z}}
e^{-U_{i}+U_{i+k}}(-x)^{k}f_{i}{\bar f}_{i+k}
 \ee
 yields getting
away of a given Young diagram a vertical strip, that is
\[
o^{(2)}(x,U)|\lambda'\rangle=\sum_\lambda |\lambda\rangle \;
o^{(2)}_{\lambda,\lambda'}(x,U)
\]
where
\[
o^{(2)}_{\lambda,\lambda'}(x,U):=\langle \lambda
|o^{(2)}|\lambda'\rangle=e^{U_{\lambda'}-U_{\lambda}}x^{|\lambda'|-|\lambda|}
\]
is non-vanishing only if the difference between $\lambda'$ and
$\lambda$ is the vertical strip \cite{Mac}.

Presenting this process as a random motion of $1D$ particles on
Maya diagram we see that it describes a hop downward to the
nearest site of randomly chosen group of particles (they may form
'trains') at each time instant.

(3) The exponential of
 \be\label{sigma=3}
A^{(3)}(x,U)=\sum_{k=1,2,\dots}\sum_{i\in \mathbb{Z}}
e^{-U_{i}+U_{i-k}}x^{k}f_{i}{\bar f}_{i-k}
 \ee
yields laying of a given Young diagram by a horizontal strip, as
it is given by
\[
o^{(3)}(x,U)|\lambda'\rangle=\sum_\lambda  \; |\lambda\rangle \;
o^{(3)}_{\lambda,\lambda'}(x,U)
\]
where the transition weight
\[
o^{(3)}_{\lambda,\lambda'}(x,U):=\langle \lambda
|o^{(3)}|\lambda'\rangle=e^{U_{\lambda'}-U_{\lambda}}x^{|\lambda|-|\lambda'|}
\]
is non-vanishing only if the diagram $\lambda$ includes the
diagram $\lambda'$, and the difference between these diagrams is
the so-called horizontal strip \cite{Mac}.

Example is given by fig. \ref{fig12}:

\bfig

\begin{picture}(300,120)
 \put(9.7,50){\line(0,1){40}}\put(10.4,50){\line(0,1){40}}

 \put(29.7,50){\line(0,1){40}}\put(30.4,50){\line(0,1){40}}

 \put(49.7,70){\line(0,1){20}}\put(50.4,70){\line(0,1){20}}

 \put(69.7,70){\line(0,1){20}}\put(70.4,70){\line(0,1){20}}

 \put(10,49.7){\line(1,0){20}}\put(10,50.4){\line(1,0){20}}

 \put(10,89.7){\line(1,0){60}}\put(10,90.4){\line(1,0){60}}

 \put(10,69.7){\line(1,0){60}}\put(10,70.4){\line(1,0){60}}

 \put(10,0){\line(0,1){20}}


 \put(10,20){\line(1,0){105}}\put(10,0){\line(1,0){105}}

\multiput(30,0)(20,0){5}{\line(0,1){20}}

\multiput(20,20)(20,0){5}{\vector(0,1){10}}

\multiput(125,10)(5,0){3}{\circle{0.1}}

\multiput(17,8)(20,0){5}{x}

 \put(170,80){Fig.12a. Condensation of a lay onto a Young diagram }
 \put(170,70){$\lambda'$. Random 'rain' of boxes from downward. These boxes}
 \put(170,60){may be fitted only to admissible places, which means}
 \put(170,50){that new figure should be a Young diagram again}

\end{picture}

\begin{picture}(300,170)

\put(0.7,70){\line(0,1){40}}\put(1.4,70){\line(0,1){40}}

 \put(20.7,70){\line(0,1){40}}\put(21.4,70){\line(0,1){40}}

 \put(40.7,90){\line(0,1){20}}\put(41.4,90){\line(0,1){20}}

 \put(60.7,90){\line(0,1){20}}\put(61.4,90){\line(0,1){20}}

 \put(1,69.7){\line(1,0){20}}\put(1,70.4){\line(1,0){20}}

\put(1,109.7){\line(1,0){60}}\put(1,110.4){\line(1,0){60}}

 \put(1,89.7){\line(1,0){60}}\put(1,90.4){\line(1,0){60}}

 \put(89.7,50){\line(0,1){60}}\put(90.4,70){\line(0,1){40}}

 \put(109.7,70){\line(0,1){40}}\put(110.4,50){\line(0,1){60}}

 \put(129.7,90){\line(0,1){20}}\put(130.4,90){\line(0,1){20}}

 \put(149.7,90){\line(0,1){20}}\put(150.4,90){\line(0,1){20}}

 \put(90,50){\line(1,0){20}}

 \put(90,69.7){\line(1,0){20}}\put(90,70.4){\line(1,0){20}}

 \put(90,109.7){\line(1,0){60}}\put(90,110.4){\line(1,0){60}}

 \put(90,89.7){\line(1,0){60}}\put(90,90.4){\line(1,0){60}}

 \put(99,58){x}

 \put(179.7,70){\line(0,1){40}}\put(180.4,70){\line(0,1){40}}

 \put(199.7,70){\line(0,1){40}}\put(200.4,70){\line(0,1){40}}

 \put(219.7,90){\line(0,1){20}}\put(220.4,90){\line(0,1){20}}

 \put(239.7,90){\line(0,1){20}}\put(240.4,90){\line(0,1){20}}

 \put(180,69.7){\line(1,0){20}}\put(180,70.4){\line(1,0){20}}

 \put(180,109.7){\line(1,0){60}}\put(180,110.4){\line(1,0){80}}

 \put(180,89.7){\line(1,0){80}}\put(180,90.4){\line(1,0){60}}

 \put(180,90){\line(0,1){20}}

 \put(260,90){\line(0,1){20}}

 \put(249,98){x}

 \put(289.7,50){\line(0,1){60}}\put(290.4,70){\line(0,1){40}}

 \put(309.7,70){\line(0,1){40}}\put(310.4,50){\line(0,1){60}}

 \put(329.7,70){\line(0,1){40}}\put(330.4,90){\line(0,1){20}}

 \put(349.7,90){\line(0,1){20}}\put(350.4,70){\line(0,1){40}}

 \put(290,90){\line(0,1){20}}

 \put(290,50){\line(1,0){20}}

 \put(290,69.7){\line(1,0){61}}\put(290,70.4){\line(1,0){20}}

 \put(290,109.7){\line(1,0){60}}\put(290,110.4){\line(1,0){80}}

 \put(290,89.7){\line(1,0){80}}\put(290,90.4){\line(1,0){60}}

 \put(290,90){\line(0,1){20}}

 \put(370,90){\line(0,1){20}}

 \put(390,110){\line(1,0){20}}

 \put(390,90){\line(1,0){20}}

 \put(390,90){\line(0,1){20}}\put(410,90){\line(0,1){20}}

\put(375,109.07){\circle{0.4}}\put(380,109.07){\circle{0.4}}\put(385,109.07){\circle{0.4}}
\put(375,90.29){\circle{0.4}}\put(380,90.29){\circle{0.4}}\put(385,90.29){\circle{0.4}}

 \put(299,58){x}\put(359,98){x}
 \put(319,78){x}\put(339,78){x}
 \put(399,98){x}

 \put(30,10){Fig.12b. Examples of diagrams of $\lambda$ where $o^{(1)}_{\lambda,\lambda'}$ is
 nontrivial. One layer (one strip)}
 \put(30,00){is added at random to the Young diagram $\lambda'$. Boxes marked by x depict the}
 \put(30,-10){difference between $\lambda'$ and $\lambda$, called
 the horizontal strip}

\end{picture}

\begin{picture}(100,60)
\end{picture}

{\vskip 10 pt} \caption{A layering by a horizontal strip}
\label{fig12} \efig

Turning to a $1D$ configuration of the related lattice gas we see
that it is suitable to interchange roles of particle (pictured as
black balls) and free sites (white balls). In this dual picture
trains of holes hop one step downward at each time instant. The
rate of this process depends on the difference of potential in
initial and final positions and is proportional to $x^n$, where
$n$ is the number of absorbed boxes.

(4) The exponential of
 \be\label{sigma=4}
A^{(4)}(x,U):=\sum_{k=1,2,\dots}\sum_{i\in \mathbb{Z}}
e^{-U_{i}+U_{i+k}}x^{k}f_{i}{\bar f}_{i+k}
 \ee
yields the random process of getting away of a given Young diagram
a horizontal strip, that is
\[
o^{(4)}(x,U)|\lambda'\rangle=\sum_\lambda |\lambda\rangle \;
o^{(4)}_{\lambda,\lambda'}(x,U)
\]
where
\[
o^{(4)}_{\lambda,\lambda'}(x,U):=\langle \lambda
|o^{(4)}|\lambda'\rangle=e^{U_{\lambda'}-U_{\lambda}}x^{|\lambda'|-|\lambda|}
\]
is non-vanishing only if the difference between $\lambda'$ and
$\lambda$ is the horizontal strip \cite{Mac}.

\;

These are four cases. Now one can consider a chain of
transformations, each transformation is given by a set of
corresponding $U$ and $x$ which define gluing rates.

Examples of these chains were considered in the reviewing paper
\cite{Zchains} from the point of view of discrete Hirota
equations.

\;

(II) {\bf Chains of Darboux transformations.  Random layering and
"rocks"}

The chain of transformations of  initial Fock vector is as
follows:
 \be\label{chain}
  | \lambda'\rangle \to  e^{A^{\sigma_{1}}(x_{1})}|  \lambda'\rangle
\to  e^{A_\time^{\sigma_{\time}}(x_{\time})}\cdots
e^{A_1^{\sigma_{1}}(x_{1})}| \lambda' \rangle
 \ee
where $\sigma_j=1,2,3,4$, and where $\time$ is a discrete time.
Given $j=1,2,3,\dots,\time$, each $A_j^{\sigma_{j}}(x_{j})$ is
characterized by a given set of $\{ U_i^{(j)} ,i\in \mathbb{Z} \}$
which enters definitions
(\ref{sigma=1}),(\ref{sigma=2}),(\ref{sigma=3}) and
(\ref{sigma=4}).

Given set $\{ \sigma_j,j=1,2,3,\dots,\time \}$ and $\{
U_i^{(j)},j=1,2,3,\dots,\time ,i\in \mathbb{Z} \}$ gives rise to a
chain of Darboux transformations. Such chains may be viewed as
words of lengths $\time$ formed by four types of characters (a
'DNA' coding a set of random processes of adding or eliminating of
layers). Tau function is constructed as
\[
\tau(\time,\bt,\bar{\bt})=\langle 0|e^{H(\bt)}o(\time)e^{{\bar
H}({\bar \bt})}|0\rangle= \sum_{\lambda,\lambda'} \tau_{\lambda,
\lambda'}({\time})s_\lambda({\bt}) s_{ \lambda'}(\bar{\bt})
\]
\[
o(\time)= e^{A_\time^{\sigma_{\time}}(x_{\time})}\cdots
e^{A_1^{\sigma_{1}}(x_{1})}
\]
then, each word is related to a certain generalized Darboux
transformation of the tau function. This tau function may be
evaluated because we know all matrix elements of each 'letter'
$e^{A_j^{\sigma_{j}}(x_{j})}$. Let us note that tau functions of
this type were considered in \cite{OSch1}, \cite{Or1} as series in
skew Schur functions which in special cases provided examples of
Gelfand-Graev hypergeometric series.

This is a certain random process on partitions (and their Young
diagrams) describing random adding and elimination of lays to
their Young diagrams, each time step gives rise to an act of
either creation, or elimination of lays, as it is shown in figures
11-14.

For the process of time duration $\time$, the relative weight of a
configuration $\lambda$ is given by the ratio
 \[
P_{\lambda'\to\lambda}(\time)=
\frac{\tau_{\lambda,\lambda'}(\time)}{Z_{\lambda'}(\time)}
 \]
where
 \[ Z_{\lambda'}(\time)=\sum_\lambda
\tau_{\lambda,\lambda'}(\time)
 \]
is the sum of weights of all processes of duration $\time$ started
with an initial configuration $\lambda'$.

One may call the random process of creating and eliminating of
lays as model of random slices, or, a "rock" inscribed into a
corner between two walls, which may be viewed as a set of Young
diagrams placed one above another as sections of $3D$ figure
according to its time evolution. Each three-dimensional figure of
height $\time$ obtained in this way presents a path
(\ref{one-path}).

{\em Examples}. Choose  $\lambda'=0$, and let $\sigma_j=1$ (which
means that we shall consider growing Young diagrams) and
$U_i^{(j)}=U_i$ , for all $ j=1,\dots,\time $.

Let us note that for $U=0$ models considered below may be
connected to various models related to the so-called semi-standard
tableau (see Appendix A.1 and for details \cite{Mac}), like a
model of polynuclear growth (for instance, see \cite{F} for a
review).

 \be\label{chain-sigma=1}
A_j^{(1)}(x_j,U)=A^{(1)}(x_j,U)=-\sum_{k=1,2,\dots}\sum_{i\in
\mathbb{Z}} e^{-U_{i}+U_{i-k}}(-x_j)^{k}f_{i}{\bar f}_{i-k}
 \ee

In this model, $x_j,\ j=1,\dots,\time$ and $U_i,\ i\ge 0$ are sets
of parameters which define transition probabilities. Then the
process may be called a version of {\em random process of growing
Young diagram} which is as follows. The initial configuration is
the zero diagram. Each time step one lay (a set of strips) is
added at random to the previous Young diagram (in a way that new
figure is again a Young diagram). This is a Markov process. The
transition weight for each step, say, a step number $j$, where the
initial configuration (related to the moment $j-1$) is, say,
$\mu^{(j-1)}$, and the next configuration (related to the moment
$j$) is $\mu^{(j)}$, is equal to
 \be
W_{\mu^{(j-1)}\to\mu^{(j)}}(1)=e^{-U_{\mu^{(j)}}+U_{\mu^{(j-1)}}}
x_j^{|\mu^{(j)}-\mu^{(j-1)}|}
 \ee
This yields a un-normalized probability to add a lay (a strip) of
a weight (i.e. number of boxes) equal to $|\mu^{(j)}-\mu^{(j-1)}|$
to a Young diagram.

One can see that for this random process
 \be\label{Usemi-tablo}
W_{0\to\lambda}(\time)=
{e^{-U_\lambda}s_\lambda(x_1,\dots,x_\time)},\quad
Z_0(\time)=\sum_\lambda e^{-U_\lambda}s_\lambda(x_1,\dots,x_\time)
 \ee
Since $U_0=0$ and $s_0=1$ one can notice notice that
$1/Z_0(\time)$ yields the probability to create zero Young
diagram, $\lambda=0$, i.e. the probability to create no any
nonvanishing diagram).

Let us consider these formulae and find links to a few known
random topics.

(1) First, consider the Gauss potential $U_i=e^{\frac{ci^2}{2}}$
and chose $x_i=e^{ch_i'}$, where $c$ is a constant and
$h_1',\dots,h_\time'$ are ordered positive integers
$h_1'>\cdots>h_\time'$. Then the weight
 \be
 W_{0\to\lambda}(\time)=
 \frac{e^{c(h_1')^2+\cdots+c(h_\time')^2}}{\prod_{i<j}^\time(e^{ch_i'}-e^{ch_j'})}
 \det \left(e^{-c(h_i-h_j')^2}\right)_{i,j=1,\dots,\time},\quad
 h_i=\lambda_i-i+\time,
 \ee
unexpectedly coincides with the (un-normalized) transition
probability for $\time$ non-intersecting ("ordered")
 Brownian particles on the line whose initial and final
 coordinates are positive integers $h_1',\dots,h_\time'$ and
 $h_1,\dots,h_\time$. In large $\time$ limit it may be also related to
 the Itsykson-Zuber integral (relation between this integral and the
 Brownian motion is quite known topic, for instance see \cite{KT}).

Then there are few limiting cases for (\ref{Usemi-tablo}), which
are simple.

(2) The first is $x_1=\cdots =x_\time=x<1$. Then using
(\ref{choicet(a)'}) we obtain
\[
Z_0(\time)=e^{-U_{\time}}
x^{\frac12\time^2-\frac12\time}\prod_{i=0}^\time\frac{e^{U_{\time-i}}}{(\time
-i)!}\sum_{h_1,\dots,h_\time\ge 0} \prod_{i=1}^{\time}e^{-V_{h_i}}
\prod_{i,j=1}^{\time}|h_i- h_j|^\beta,\quad \beta=1
\]
where
\[
V_{h_i}= U_{h_i}-h_i\log x
\]
{\em Thus $Z_{0}(\time)$ is the partition function of a discrete
version of ensemble of random orthogonal matrices, which is also a
certain tau function: the tau function of the Pfaff lattice
introduced in \cite{AvM},\cite{KdL}.} Let us note that the
ensemble of random orthogonal matrices is well studied object, see
\cite{TW},\cite{Baik},\cite{Johansson}. For $U_i=0$ it is a
version of the well-known Laguerre ensemble.

(3) A similar result we obtain in case we take
$x_j=e^{\sqrt{-1}\left(\vartheta+(j-1)\phi \right)}$, where
$\phi,\vartheta$ are parameters. Then with the help of
(\ref{choicet(a)q'}) we obtain unitary orthogonal ensemble
 \be\label{}
 Z_{0}(\time)=\frac{e^{ \vartheta(\frac12\time^2-\frac12\time}){\sqrt{-1}}}{\time !}
 \prod_{i=1}^\time\frac{e^{U_{\time-i}}}{(q;q)_{\time
-i}}  \sum_{h_1,\dots,h_\time\ge 0}
\prod_{i=1}^{\time}e^{-V_{h_i}} \prod_{i,j=1}^{\time}|e^{\phi
h_i}- e^{\phi h_j}|^\beta,\quad \beta=1
 \ee
where
\[
V_{h_i}= U_{h_i}- h_i\vartheta{\sqrt{-1}}
\]
where q-factorials $(q;q)_{n}$ are defined in (\ref{Pochq}), and
$U_i$ may be chosen as series analogues to (\ref{Tt-tilde-t})
 \be\label{Tt-tilde-t-q}
{U}_i =\sum_{m\neq 0}^\infty e^{im\sqrt{-1}} {\tilde t}_m ,\quad
i\in\mathbb{Z}
 \ee
(Such parametrization was used in \cite{hypsol} for different
purposes. These three cases can be also compared with \cite{OS}).

At last let us write down a group of examples (found in
\cite{Mac}) where ${Z_{0}(\time)}$ (and, therefore, $P_{0\to
0}(\time)={Z_{0}(\time)}^{-1}$) may be explicitly evaluated :

(a) Take all $U_i=0$. Then, there is a simple formula
\[
 Z_{0}(\time)=\sum_\lambda s_\lambda(x_1,\dots,x_\time)=
\prod_{j=1}^\time (1-x_j)^{-1}\prod_{i<j}^\time(1-x_ix_j)^{-1}
\]

(b) Take all $e^{U_i}=1,\ i\le m$ and $e^U_i=0,\ i> m$. Then
\[
 Z_{0}(\time)=\sum_\lambda s_\lambda(x_1,\dots,x_\time)=
D_m/D_0,\quad D_m=\det\left(x_j^{2\time +m-i}-x_j^{i-1}\right)
\]
where sum is going over all partitions whose Young diagrams are
contained in the rectangle $\time\times m$.

(c) Take all $e^{U_i}=1,\ i\le m$ and $e^U_i=0,\ i> m$, as before,
and put
 $x_j=e^{(2\time -2j+1)\phi }$. Then
\[
 Z_{0}(\time)=\sum_\lambda s_\lambda(e^{(2\time-1)\phi },\dots,e^{\phi})=
\prod_{j=1}^\time \frac{e^{\phi(m+2j-1)}-1}{e^{\phi(2j-1)}-1}
\prod_{i<j}^\time
\frac{e^{2\phi(m+i+j-1)}-1}{e^{2\phi(m+i+j-1)}-1}= \sum_\Theta
N(\Theta)e^{|\Theta|\phi }
\]
where first sum is going over all partitions whose Young diagrams
are contained in the rectangle $\time\times m$.

The second sum ranges over all symmetric plane partitions $\Theta$
whose diagram are contained in the box $\time \times \time \times
m$. (Diagram of a
 plane partition is a 3D figure constructed from identical cubes,
whose sections by horizontal planes are Young diagrams, see
\cite{Mac}). Here $N(\Theta)$ is the number of symmetric plane
partitions of weight $|\Theta|$. Detailed explanations of the
right hand sides of formulae in the last three examples (together
with some other different examples of particular cases of
summation formulae for $\sum_\lambda s_\lambda$)
 may be found in \cite{Mac} from where they were borrowed by the authors.

\

Let us also note that the model of growing partition via layering
(\ref{chain-sigma=1}) is directly related to the problem of
enumerating of the so-called semi-standard tableau.

\section{Discussion}

In the present paper we want to figure out links between classical
integrable systems and random system. These links in a natural way
suggest to introduce site-dependent weights for hops of the
particles and provide certain analogies between hard-core
particles in equilibrium and non-equilibrium states. Along this
line we introduced a version of discrete ASEP with  site depending
hopping rates. We show that the normalization function for
probabilities of this model is related to classical integrable
hierarchy of type $B$ rather than to the hierarchy of the type $A$
(the Toda lattice hierarchy) we started with. The normalization
function plays the role similar to the role of partition function
for statistical ensembles. We present a compact formula which
describes the decay of the step function which also converts this
problem to the evaluation of a certain correlation function for a
discrete version of orthogonal matrix ensemble, where potential is
defined by the site-dependent hopping rates of our model.


The further project aims the following problems:

(1) to study a model which is a modification of the model
considered in the subsection 2.6. Namely, we replace the operator
$A_++A_-$  by a linear combination of the so-called Virasoro
generators $L_1,L_0$ and $L_{-1}$ which form $SL(2)$ algebra. This
model is also quite solvable, however we will show that it
presents different behavior

(2) to present a description of models of stochastic motion which
may be obtained from various integrable hierarchies (various
realizations of hierarchies of type A,B,C and D and their
multi-component versions) \cite{HLO}

(3) to study the asymptotic behavior  of transition probabilities
between different configurations and of certain correlation
functions \cite{HOasymp}

(4) In particular for random turn decay of step function to
evaluate the time dependence of : (a) the mean number of involved
particles, $< \ell(\lambda)>$ (b) the mean height $<|\lambda|>$
(c) the mean flow through the origin $<\textrm{k}(\lambda)>$
($\textrm{k}(\lambda)$ is the number of hooks, or, the same, the
number of Frobenius coordinates

(5) to consider d-ASEP with various boundary conditions. To study
the phenomenon of shock waves and phase transition of the first
order known for the problem with open boundaries
\cite{KPSchutz},\cite{iran}

(6) to study the phenomenon of thermalization which should occur
for finite graphs

(7) to understand links with other approaches to the $1D$
non-equilibrium phenomena, in particular with these of papers
\cite{GS1},\cite{GS2},\cite{Schutz},\cite{R}.

(8) to understand relations with the Bethe anzats method for the
ASEP model

\section*{Acknowledgements}
The starting point for the ideas developed in this work were
discussions between A.O. and  T. Shiota during a stay at Kyoto
university in 1999-2001. He wishes to thank A. Odzijevicz for
discussions and kind hospitality during his stay in Bialystok in
June 2005, where a large part of this work was done. The work was
completed and written in Montreal  (July 2005 and October 2006).
Results contained in this paper were presented by A.O.  at the
workshop ``Mathematics and Physics  of Growing Interfaces`` in
Santa Fe, January 2006. He also wishes to thank P. Zinn-Justin, J.
van de Leur, M. Mineev and V. B. Priezzhev for helpful discussions
and giving \cite{GV}, and to  O. Zaboronski and P. Wiegmann  who
called attention to papers on ASEP and chemical reactions \cite{R}
and S.I.Badulin for technical support.

  J. H. would like to thank  Herbert Spohn for helpful discussions  in
  Nov. - Dec. 2006, at the   Kavli Institute for Theoretical Physics,
  U.C. Santa Barbara program  on "Spectral geometry and field theory:
   from growth phenomena to disordered systems'' and also to thank
   the organizers of this program for the invitation to take part and the
   kind hospitality extended throughout his stay by the Kavli Institute.

\appendix

\section{Appendices}

\subsection{Partitions and Schur functions}

{\bf Partitions}. Polynomial functions in many variables are
parameterized by partitions. A {\em partition} is any (finite or
infinite) sequence of non-negative integers in decreasing
order:
\begin{equation}\label{partition}
\lambda = (\lambda_1, \lambda_2, \dots,\lambda_r,\dots )\ ,\quad
\lambda_1 \ge \lambda_2 \ge \dots \ge\lambda_r \ge \dots > 0
\end{equation}
The numbers $\lambda_i$  are called the {\em parts} of the $\lambda$.
The number of the parts is the {\em length} of
the $\lambda$, denoted $\ell(\lambda)$. The
sum of the parts denoted $|\lambda|$,  is called the {\em weight} of  $\lambda$.
 If $|\lambda| = n$, we say that  $\lambda$ is a {\em partition of} $n$.
 It is often convenient to extend the $\lambda_i$'s to an infinite sequence
 $\{\lambda_i\}_{i\in \Nb}$ where
\be
\lambda_i :=0 \quad {\rm if \ } i> \ell(\lambda)
\ee
The zero partition, with $\ell(\lambda)=0$ is denoted by $\bf 0$.
The set of all partitions, including $\bf0$, is denoted by $P$.

The {\em Young diagram} of a partition is defined as the set of
points (or {\it nodes}) $(i,j) \in \Zb^2$ such that $1\le j \le
\lambda_i$. The Young diagram is viewed as a subset of entries in
a matrix with $l(\lambda)$ rows and the $\lambda_1$ columns, with
the nodes denoted by squares aligned adjacently to form rows of
length $\lambda_1, \lambda_2, \dots$, stacked so that successive
rows downwards have  equal or diminishing lengths, as in the
example below \be \YD {4} {3} {1}  {0} \ee which is the diagram of
the partition $\lambda =(4,3,1)$, with weight $|\lambda|=8$ and
length $\ell(\llambda) = 3$. The partition whose diagram is
obtained by transposition of the diagram $\lambda$ with respect to
the main diagonal is called the conjugate partition and denoted by
$\nu$.

Another notation is due to Frobenius. Suppose that the main
diagonal of the diagram of $\lambda$  consists of $r$ nodes
$(i,i)\quad (1\le i\le r)$.  For $1\le i\le r$, let
$\alpha_i=\lambda_i-i$ be the number of nodes in the $i$th row to
the right of $(i,i)$,, and $\beta_i=\nu_i-i$ the number of nodes
in the $i$th column  below $(i,i)$.
 We then have
 \bea
 &\& \alpha_1>\alpha_2>\cdots >\alpha_r\ge 0 \cr
&\& \beta_1>\beta_2>\cdots >\beta_r\ge 0 \eea
 The Frobenius
notation for the partition $\lambda$ is then
\begin{equation}\label{Frob}
\lambda = \left( \alpha_1,\dots ,\alpha_r|\beta_1,\dots
,\beta_r\right)=\alpha |\beta ).
\end{equation}
The Frobenius notation may be viewed as a decomposition of a
diagram $\lambda$ into hooks with
corners situated along the main diagonal,
the largest hook being $\left( \alpha_1|\beta_1\right)$, the next one
$\left( \alpha_2|\beta_2\right)$, and so on down to the smallest
 $\left( \alpha_r|\beta_r\right)$.

 For example, the partition $(4,3,1)$ consists of two hooks $(3,2)$ and $(1,0)$:\\
\be
 \YD 4110  \quad {\rm and} \quad \YD20
 \ee
 and in Frobenius notation is written $(3,1|2,0)$.
 If  $\lambda=(\alpha|\beta)$, then $\lambda^t=(\beta|\alpha)$

 If we insert the increasing sequence of integers $1, 2, \dots |\lambda|$
 into a Young diagram such that $1$ is in the first box, and the numbers
 are  increasing to the right within each row and downward within
 each column, the result is called a (standard) Young Tableau. If we
 insert a sequence $1, 2, \dots n$ with $n \le |\lambda$ such that
 the numbers are nondecreasing to the right in each row,
 and decreasing downward in each column, this is called a
 {\it semi-standard} Young Tableau.

\noindent
 {\bf Schur functions}.

    The Schur functions $S_\lambda([x])$ associated
 with the partition $\lambda$ may be viewed either as a symmetric
 homogeneous polynomial in $N$ variables $(x_1, \dots , x_N)$,
 where $N$ may be any integer $\ge |\ell(\lambda)|$ or, equivalently
 a weighted homogeneous polynomial in the infinite sequence
 of variables $(t_1, t_2, \dots )$ defined by:

  \be
    s_\lambda([x]) := (\sum_{a=1}^N x_a, {1\over 2} \sum_{a=1}^N x^2_a, \dots
    , {1\over i}\sum_{a=1}^N x_a^i, \dots) := (t_1, t_2, \dots , t_i, \dots)
    \ee
     where each $t_i$  has weight $i$, having total weight $|\lambda|$.
     Because of the homogeneity condition, $S_\lambda$ can only depend on
     the finite set of variables $(t_1, \dots , t_{|\lambda|})$.
     Viewed as functions of the $x_a$'s, they may be defined
    as the following ratio of determinants (Jacobi-Trudi formula)
  \be
    S_\lambda ([x]) = {\det (x_i^{\lambda_j -j +N})\over \Delta(x_1, \dots, x_N)}
    \quad  1\le i,j \le N
    \label{Jacobi_Trudi_form}
    \ee
         where
           \be
    \Delta(x_1, \dots, x_N) = \prod_{i < j}^N(x_i-x_j) = \det (x_i^{\lambda_{N-j} })
    \ee
     is the Vandermonde determinant.
    For $N\le |\lambda|$, this is equivalent to the following combinatorial
    definition:
     \be
    S_\lambda([x]) = \sum x_1^{\mu_1} \dots x_N ^{\mu_N}
    \ee
    where the sum is over all semi-standard Young Tableau of shape
    $\lambda$ and $\mu_i$ is the number of times $i$ appears in
    the Tableau.

            Representation theoretically, the significance of this is that $S_\lambda([x])$
       is the character of the irreducible, rank $\lambda|$ tensor representation
       of $U(N)$ or $GL(N)$ whose symmetry properties are given by the
       Young diagram of $\lambda$. (i.e. the irreducible representation consisting
       of tensors  obtained by first   symmetrizing all components labelled  by the
       rows of the Young diagram,
       and then antisymmetrizing those  labelled by the columns. The $x_a$'a
       are viewed as eigenvalues of  the $U(N)$ or $GL(N)$ group element or
        one simply chooses diagonal elements $X=\diag(x_1, \dots, x_N)$ and
        $S_\lambda([x])$ is the trace of the representation evaluated at $X$.)

 Polynomial functions in many variables, like the Schur
functions, are parameterized by partitions.

Consider a semi-infinite set of variables ${\bf
t}=(t_1,t_2,t_3,\dots)$. Given partition $\lambda$, the Schur
function $s_\lambda({\bf t})$ is defined by
\begin{equation}\label{Schurt}
s_\lambda({\bf t})=\det\bigl(h_{\lambda_i-i+j}({\bf
t})\bigr)_{1\le i,j\le \ell(\lambda)}\ ,\quad\hbox{where}\quad
\sum_{k=0}^\infty z^kh_k({\bf t}) = \exp\sum_{m=1}^\infty z^mt_m \
,
\end{equation}
and, for $k<0$, we put $h_k=0$ . The $h_k({\bf t})$ is called the
elementary Schur function.

If
 \be {\bf t}={\bf
t}(x^{(n)})=(t_1(x^{(n)}),t_2(x^{(n)}),\dots),\quad
t_m(x^{(n)})=\frac 1m \sum_{i=1}^n x_i^m,
 \ee
 then definitions
(\ref{Schurt}) and (\ref{Jacobi_Trudi_form}) are equivalent
\cite{Mac}: \be s_\lambda({\bf t}(x^{(n)}))=\underline
s_\lambda(x^{(n)}). \ee
 From definition (\ref{Schurt}) it follows that $s_\lambda({\bf
t}(x^{(n)}))=0$ if $\ell(\lambda)>n$.

We use the underline in $\underline s_\lambda$ only to distinguish
the two definitions. If an $n\times n$ matrix $X$ has eigenvalues
$x_1$, \dots, $x_n$, we may denote $\underline
s_\lambda(x_1,\dots,x_n)$ by $s_\lambda(X)$, without underline,
since in this paper the Schur function with uppercase argument is
used only in this sense.

 {\bf Skew Schur function} is defined as follows
 \be\label{skewSchur}
s_{\lambda/\mu}=\det \left(h_{\lambda_i-\mu_j-i+j}\right)_{1\le
i,j\le n},\quad
 \ee
Properties of these polynomials are described in details in
\cite{Mac}.

 {\bf Schur functions evaluated at special points}.

We need notations:
\begin{equation}\label{Poch}
(a)_\lambda :=
(a)_{\lambda_1}(a-1)_{\lambda_2}\cdots(a-k+1)_{\lambda_k} \ ,
\quad (a)_m :=\frac{\Gamma(a+m)}{\Gamma(a)} \ ,
\end{equation}
\begin{equation}\label{Pochq}
(q^a;q)_\lambda :=
(q^a;q)_{\lambda_1}(q^{a-1};q)_{\lambda_2}\cdots(q^{a-k+1};q)_{\lambda_k}
\ , \quad (q^a;q)_m :=(1-q^a)\cdots(1-q^{a+m-1}) \ ,
\end{equation}
where $k=\ell(\lambda)$. We set $(a)_0=(q^a;q)_0=1$ and
$(a)_{-k}=(q^a;q)_{-k}=0$ for $k>0$.

We introduce the following notations \cite{OS}:
\begin{equation}\label{choicetinfty'}
{\bf t}_\infty=(1,0,0,0,\dots) \ ,
\end{equation}
\begin{equation}\label{choicet(a)'}
{\bf t}(a,1)=\Bigl(\frac{a}{1},\frac{a}{2},\frac{a}{3},\dots\Bigr)
\ ,
\end{equation}
\begin{equation}\label{choicetinftyq'}
{\bf t}(\infty,q)=(t_1(\infty,q),t_2(\infty,q),\dots),\quad
t_m(\infty,q)=\frac{1}{m(1-q^m)}\ ,\quad m=1,2,\dots\ ,
\end{equation}
\begin{equation}\label{choicet(a)q'}
{\bf t}(a,q)=(t_1(a,q),t_2(a,q),\dots)\ ,\quad
t_m(a,q)=\frac{1-(q^a)^{m}}{m(1-q^m)}\ ,\quad m=1,2,\dots
\end{equation}

Note that ${\bf t}(a,q)$ tends to ${\bf t}(\infty,q)$ (resp.\
${\bf t}(a,1)$) as $a\to\infty$ (resp.\ $q\to1$). As for ${\bf
t}_\infty$, if $f$ satisfies
$f(ct_1,c^2t_2,c^3t_3,\dots)=c^df(t_1,t_2,t_3,\dots)$ for some
$d\in {\mathbb Z}$, we have $\hbar^df({\bf t}(\infty,q))\to f({\bf
t}_\infty) $ as $\hbar:=\ln q \to0$. Below
$\Delta(h):=\prod_{i<j}(h_i-h_j)$.
\begin{Lemma} For a partition $\lambda=(\lambda_1,\lambda_2,\dots)$, let
$
h_i := n+\lambda_i-i$ $(1\le i \le n)$, where
$n\ge\ell(\lambda))$. Then
\begin{equation}\label{schurhook}
s_\lambda({\bf t}_\infty)=\frac{ \Delta(h)}{\prod^n_{i=1}h_i!} \ ,
\end{equation}
\begin{equation}\label{schurhookt(a)}
s_\lambda({\bf t}(a,1))=\frac{ \Delta(h)}{\prod^n_{i= 1}h_i!}
\prod_{i=1}^n \frac{\Gamma(a-n+h_i+1)}{\Gamma(a-i+1)}\ ,
\end{equation}
\begin{equation}\label{schurhookq}
s_\lambda({\bf t}(\infty,q)) =\frac{\Delta(q^h)}
{\prod_{i=1}^n(q;q)_{h_i}}\ ,
\end{equation}
\begin{equation}\label{schurhookqa}
s_\lambda({\bf t}(a,q))=\frac{\Delta(q^h)}
{\prod_{i=1}^n(q;q)_{h_i}}\prod_{i=1}^n (q^{a-i+1};q)_{h_i-n+i} \
,
\end{equation}
 Note that those quantities
(\ref{schurhook})--(\ref{schurhookqa}) are independent of the
choice of $n \ge \ell(\lambda)$. We also mark that for integer
positive $a$ the Schur functions (\ref{schurhookt(a)}) and
(\ref{schurhookqa}) vanish if $\ell(\lambda)>a$.
\end{Lemma}

\subsection{Fermionic Fock space, $gl(\infty)$ and $GL(\infty)$}

  The following is a summary regarding the one and  two-component free
  fermion algebra based on the introductory section of \cite{HO3}. The
reader may refer to \cite{JM},  \cite{DJKM} for further details.

In the following, $\AA$ denotes the complex Clifford algebra over
$ \mathbb{C}$ generated by
 \emph{charged free fermions} $\{f_i$, ${\bar f}_i\}_{i\in {\bf
Z} }$, satisfying the anticommutation relations
 \be\label{fermions}
[f_i,f_j]_+=[{\bar f}_i,{\bar f}_j]_+=0,\quad [f_i,{\bar
f}_j]_+=\delta_{ij}.
 \ee
where $[,]_+$ denotes the anticommutator.

Elements of the linear part \be W:=\left(\oplus_{m \in
\Zb}\mathbb{ C}f_m\right)\oplus \left(\oplus_{m\in \Zb}\mathbb{
C}{\bar f}_m\right)
 \ee will be referred to as a {\em free fermions}.  The
 fermionic free fields
\be \label{fermions-fourier}
    f(x):=\sum_{k\in\Zb}f_kx^k,\quad
    {\bar f}(y):=\sum_{k\in\Zb}{\bar f}_ky^{-k-1},
\ee may be viewed as generating functions for the $f_j,
\bar{f}_j$'s.

This Clifford algebra has a standard Fock space representation $F$
and dual space  ${\bar F}$ (see e.g. \cite{HO1, HO3}) which contain unique
vacuum states $|0\rangle$ and $ \langle 0|$ respectively
satisfying the properties
 \bea \label{vak}
f_m |0\rangle=0 \qquad (m<0),\qquad {\bar f}_m|0\rangle =0 \qquad
(m \ge 0) , \cr
 \langle 0|f_m=0 \qquad (m\ge 0),\qquad \langle
0|{\bar f}_m=0 \qquad (m<0) . \eea The {\em Fock spaces} $F$ and
${\bar F}$ are mutually dual, with the hermitian pairing defined
via the linear form $\langle 0| |0 \rangle$ on $\AA$ called the
{\em vacuum expectation value}.  This satisfies
 \bea
\label{psipsi*vac} \langle 0|1|0 \rangle&\&=1;\quad \langle
0|f_m{\bar f}_m |0\rangle=1,\quad m<0; \quad  \langle 0|{\bar
f}_mf_m
|0\rangle=1,\quad m\ge 0 ,\\
\label{end}
 \langle 0| f_n
 |0\rangle&\&=\langle 0|{\bar f}_n
 |0\rangle=\langle 0|f_mf_n |0\rangle=\langle 0|{\bar f}_m{\bar f}_n
 |0\rangle=0;
 \quad \langle 0|f_m{\bar f}_n|0\rangle=0, \quad m\ne n,.
\eea

Wick's theorem implies that for any finite set of elements $\{w_k
\in W\}$, we have \bea \label{Wick} \langle 0|w_1 \cdots
w_{2n+1}|0 \rangle &\&=0,\cr
 \langle 0|w_1
\cdots w_{2n} |0\rangle &\&=\sum_{\sigma \in S_{2n}} sgn\sigma
\langle 0|w_{\sigma(1)}w_{\sigma(2)}|0\rangle \cdots \langle 0|
w_{\sigma(2n-1)}w_{\sigma(2n)} |0\rangle . \eea Here  $\sigma$
runs over permutations for which $\sigma(1)<\sigma(2),\dots ,
\sigma(2n-1)<\sigma(2n)$ and $\sigma(1)<\sigma(3)<\cdots
<\sigma(2n-1)$.

If  $\{w_i\}_{ i=1,\dots,N}$, are linear combinations of the
$f_j$'s only, $j\in\mathbb{Z}$, and  $\{{\bar w}_i\}_{
i=1,\dots,N}$ linear combinations of the ${\bar f}_j$'s, $j
\in\mathbb{Z}$, then(\ref{Wick}) implies
 \be
 \label{Wick-det}
\langle 0|w_1\cdots w_{N}{\bar w}_N \cdots {\bar w}_1 |0\rangle
=\det\; (\langle 0| w_i{\bar w}_j|0\rangle)\ |_{i,j=1,\dots,N}
 \ee

Following \cite{DJKM},\cite{JM},  for all $ N\in \mathbb{Z}$, we
also introduce the states
 \be \label{1-vacuum}
  \langle  N|:=\langle 0|C_{N}
 \ee
where
 \bea
\label{1-vacuum'} C_{N}&\&:={\bar f}_0\cdots {\bar f}_{N-1}
 \quad {\rm if }\ N>0 \\
C_{N}&\&:={ f}_{-1}\cdots { f}_{N}
\quad {\rm if}\ N<0  \\
 C_{N}&\&:=1 \quad {\rm if}\ N=0
 \eea
and
 \be \label{1-vacuum-r}
    |N \rangle:={\bar C}_{N}|0\rangle
 \ee
where
 \bea
\label{1-vacuum'-r} {\bar C}_{N}&\&:=f_{N-1}\cdots
f_0 \quad {\rm if }\ N>0 \\
{\bar C}_{N}&\&:={\bar f}_{N}\cdots {\bar f}_{-1}
\quad {\rm if}\ N<0  \\
 {\bar C}_{N}&\&:=1 \quad {\rm if}\ N=0
 \eea
The states   (\ref{1-vacuum}) and (\ref{1-vacuum-r}) are referred
to as the left and right charged vacuum vectors, respectively,
with charge $N$.
From the relations
 \be
 \langle 0|  {\bar f}_{N-k} f(x_{i})|0\rangle
 =x_i^{N-k},\quad  \langle 0|
 { f}_{-N+k-1}  {\bar f}(y_{i})|0\rangle =y_i^{N-k},\quad k=1,2,\dots N,
 \ee
 and  (\ref{Wick-det}), it follows that
\bea\label{Delta-N-left} \langle N|f(x_1)\cdots
f(x_n)|0\rangle  &\&=\delta_{n,N}\Delta_N(x),\quad N\in \mathbb{Z},\\
\label{Delta-N-right} \langle -N|\bar{f}(y_1)\cdots
\bar{f}(y_n)|0\rangle&\&=\delta_{n,N}\Delta_N(y),\quad N\in
\mathbb{Z}. \eea

For free fermion generators with $|x|\ne|y|$,
 \be\label{f-barf}
    \langle 0|f(x){\bar f}(y)|0 \rangle
    =    \frac{1}{x-y}
 \ee
Note that the expression on the right hand side is
 actually defined, by (\ref{psipsi*vac}), as the infinite series
 $\sum_{n=0}^\infty y^nx^{-n-1}$
 which converges only inside $|x|<|y|$.
 However one can consider expression (\ref{f-barf}) for the whole
 region of $x$ and $y$ (when $|x|\ne|y|$) in the sense of
 analytical continuation.

From Wick's theorem it follows that
 \be\label{multi-f-bar-f-n-m}
    \langle  n-m|f(x_1)\cdots f(x_n){\bar f}(y_1)\cdots {\bar f}(y_m)|0 \rangle
=   \frac{\Delta_n(x)\Delta_m(y)}
    {\prod_{i=1,\dots,n\atop j=1,\dots,m}(x_i-y_j)}
 \ee

\subsection{Commuting flows, $\tau$ functions and Schur functions}

In the theory of integrable systems the following ${\hat
{gl}}(\infty)$ operators are important:
 \be\label{H-m}
H_m=\sum_{i=-\infty}^{+\infty} f_i{\bar f}_{i+m},\quad m=\pm 1,\pm
2,\dots
 \ee
 These operators form Heisenberg algebra relations
 \be\label{Heis}
 [H_m,H_n]=m\delta_{m+n,0}
 \ee
These oscillator algebra properties together with
 \be
H_m|n\r=0; \quad \l n|H_{-m}=0,\quad m>0, n\in\mathbb{Z}
 \ee
allow to refer vectors $\l 0|e^{H({\bf t})}$ and ${\bar
H}({\bar{\bf t}})$ as generalized coherent states which depend on
parameters
 ${\bf t}=(t_1,t_2,\dots)$ and ${\bar{\bf t}}=({\bar t}_1,{\bar
t}_2,\dots)$. Here
 \be
H({\bf t})=\sum_{m=1}^\infty H_mt_m,\quad {\bar H}({\bar{\bf
t}})=\sum_{m=1}^\infty H_{-m}{\bar t}_m
 \ee

We have the following fermionic representation of Schur functions
(which follows, as usual from Wick's theorem)

\noindent
{\bf Lemma 1} \cite{DJKM}
For  $\alpha_1>\cdots
>\alpha_k\ge 0$,   $\beta_1>\cdots
>\beta_k\ge 0$ the next formula is valid:
\begin{equation}\label{lemma-H}
\l 0|e^{H({\bf t})}{\bar f}_{-\beta_1-1}\cdots {\bar
f}_{-\beta_k-1}f_{\alpha_s}\cdots f_{\alpha_1}|0\r=
(-1)^{\beta_1+\cdots +\beta_k+k}s_{{\lambda}}({\bf t}) \ ,
\end{equation}
where in the Frobenius notation ${\lambda}=(\alpha_1,\dots ,
\alpha_{k}| \beta_{1},\dots , \beta_{k})$.

\noindent

Defining, as in eq. (\ref{random-turn-Brownian})

 \be
A_+({U}):=\sum_{i\in\mathbb{Z}}e^{{U}_{i-1}-{U}_i}f_i{\bar
f}_{i-1},
 \ee
 we have the following expression  for the matrix elements of
 its exponential  in terms of the Schur function evaluate at the special value
 ${\bf t}_\infty:= (1, 0 \dots)$

\noindent
{\bf Lemma 2} \cite{OSch1} \\
\begin{equation}\label{lemma-A}
\l\lambda,n|e^{A_+(U)}|n\r=e^{-U_\lambda(n)}s_{{\lambda}}({\bf
t}_\infty)
 \ee
 This relation follows from
 \be\label{}
A_+(U)=e^{H_0(U)}H_{-1}e^{-H_0(U)},\quad H_0(U):=\sum_{i<
0}U_i{\bar f}_if_i-\sum_{i\ge 0}U_if_i{\bar f}_i
 \ee
Indeed, by definitions of $H_0(U)$ and $\l \lambda,n|$ one
evaluates:
 \be\label{lambda-T}
\l \lambda,n|e^{H_0(U)}=c_ne^{-U_\lambda(n)}\l \lambda,n|
 \ee
 where $U_\lambda(n)$ is given by (\ref{T-lambda}), where $c_n$
 is defined by
  \be\label{T-vac}
 \l n|e^{H_0(U)}|n\r=  c_n,\quad
c_n=e^{-U_{0}\cdots-U_{n-1}},n>0;\quad c_n=e^{U_{-1}+\cdots
+U_{n}},n<0
 \ee
 Then, $(\ref{lemma-A})$ follows $(\ref{lambda-T})$ and
 $(\ref{lemma-H})$. Different detailed proof was written down in
 \cite{HO1}, where the notation $r$ of (\ref{rate-up}) was used.
 In papers \cite{hypsol},\cite{HO1} instead of
 $e^{-U_\lambda(n)}$ of the present paper the notation $r_\lambda(n)$ was
 used.

 {\bf Skew Schur function} is defined as follows
 \be
s_{\lambda/\mu}=\det \left(h_{\lambda_i-\mu_j-i+j}\right)_{1\le
i,j\le n},\quad
 \ee

 \noindent
 The following is a  generalization of (\ref{lemma-A}) in terms of skew Schur functions

\noindent
{\bf Lemma 3} \cite{OSch1} \\
\begin{equation}\label{scew-ferm}
\l \lambda,n|e^{A_+(U)}|\mu,
n\r=e^{U_\mu(n)-U_\lambda(n)}s_{{\lambda}/\mu}({\bf t}_\infty)
 \ee

\subsection{Derivation of (\ref{constant-rate-density}) and some other
formulae of subsection 2.6}

Formula (\ref{constant-rate-density}) results from
 \be\label{arcsinus-density-u}
\int_0^{2} \; \frac{\frac12-\frac 1\pi\arcsin\left( y
-1\right)}{u-y}dy=\log 2u,\quad u\in [0,2]
 \ee
where we put $u=\frac{h}{R}$. We have obtained formula
(\ref{arcsinus-density-u}) as follows. First, it is widely used in
random matrix theory (the proof may be found, say, in \cite{Mehta}
or in \cite{Forr1}) the relation called the Wigner semi-circle
law, which is the first relation of
 \be\label{Wigner}
\frac1\pi P\int_{-1}^{1}\frac{\sqrt{1-v^2}}{x-v}dv-x=0,\quad
x\in[-1,1],\quad
P\int_{-1}^{1}\frac{1}{\sqrt{1-v^2}}\frac{dv}{x-v}=0
 \ee
while the last relation is obtained from  the first one via taking
the derivative with respect to $x$ and via integrating by parts.
Now, keeping in mind $(\arcsin v)'=(1-v^2)^{-\frac12}$ and taking
the derivative of the l. h. s. of (\ref{arcsinus-density-u}) with
respect to $u$, using $\left((u-y)^{-1} \right)_u=
-\left((u-y)^{-1} \right)_y$ and integrating by parts after a
shift of the integration variable ($y-1=v$) we obtain
\[
\frac 1u+\frac1\pi P\int_{-1}^1
\frac{1}{\sqrt{1-v^2}}\frac{dv}{u-1-v}
\]
where as $u-1\in [-1,1]$ the second term vanishes due to
(\ref{Wigner}), while the first term (which originates from the
lower boundary term; upper boundary term vanishes) coincides with
the derivative of the r.h.s. of (\ref{arcsinus-density-u}) with
respect to $u$. Thus we have proven (\ref{arcsinus-density-u}) up
to a term, say $C$, independent of $u$. Let us find it. Consider
\[
P\int_0^{2} \; \frac{\frac12-\frac 1\pi\arcsin\left(y
-1\right)}{u-y}dy=P\int_0^{2} \; \left({\frac12-\frac
1\pi\arcsin\left(y -1\right)}\right)\left(-\frac{d}{dy}\right)\log
|{u-y}|dy
\]
\[
=\log u -\frac 1\pi
P\int_0^{2}\frac{\log|u-y|dy}{\sqrt{1-(y-1)^2}}=:\log u - C,\quad
u\in [0,2],
\]
where we integrated by parts. We should find the last term, $C$,
which (we know) is the constant we are looking for. We have
\[
C=\frac 1\pi
P\int_0^{2}\frac{\log|u-y|dy}{\sqrt{1-(y-1)^2}}=\frac1\pi
\int_{-\frac{\pi}{2}}^{\frac{\pi}{2}}\log (x-\sin \phi)d\phi
,\quad x\in[-1,1],
\]
which we evaluate at point $x=1$ (since it does not depends on $x$
we are free to chose any point of the interval $[-1,1]$):
\[
C=\frac1\pi \int_{-\frac{\pi}{2}}^{\frac{\pi}{2}}\log (1-\sin
\phi)d\phi=\frac 1\pi
\int_{-\frac{\pi}{2}}^{\frac{\pi}{2}}\sum_{n=1}^\infty \frac{d\phi
\sin^{2n}\phi}{2n}=-\sum _{n=1}^\infty \frac{2^{-2n}}{2n}{2n
\choose n}=
\]
\[
=-\int_{0}^{1}\frac{dz}{z}\left(({1-z^2})^{-\frac12}-1\right)
=-\int_0^{\frac{\pi}{2}} \frac{1-\cos\phi}{\sin
\phi}d\phi=2\log\cos \frac{\phi}{2}\; |_{0}^{\frac{\pi}{2}}=-\log
2
\]
which completes the proof of (\ref{arcsinus-density-u}).

Now let us obtain (\ref{area}).
\[
|\lambda|=\sum_{i=1}^R (h_i+i-R)=\sum_{i=1}^R h_i
-\frac{R^2}{2}+\frac{R}{2}\to
\int_{0}^{2R}h\sigma(h)dh-\frac{R^2}{2}+\frac{R}{2}
\]
Then
\[
\int_{0}^{2R}h\sigma(h)dh =\frac12\int_{0}^{2R}\left( \frac
12-\frac{1}{\pi}\arcsin\left( \frac{h}{R}-1\right)
\right)\left(\frac{dh^2}{dh}\right)dh
\]
\[
=-\frac{1}{2\pi R}\int_{0}^{2R}
\frac{h^2dh}{\sqrt{1-\left(\frac{h}{R}-1\right)^2}}=\frac{R^2}{2\pi}
\int_{-\frac\pi 2}^{\frac\pi 2}(1+\sin^2 \phi)d\phi =\frac34 R^2
\]

  Now let us derive (\ref{R}). We have
\[
\frac12 \log(\time-|\lambda|)=\time-\frac{R^2}{4}-\frac{R}{2}
\]
which we equate to $\log\frac{2}{rR}$ and obtain (\ref{R}).

At last let us obtain (\ref{Frobenius-k-2}):
 \[
\int_R^{2R}\sigma(h)dh=\int_R^{2R}\sigma(h)
\left(\frac{dh}{dh}\right) dh=-\frac{R}{2}-\frac{1}{\pi
R}\int_R^{2R}\frac{hdh}{\sqrt{ 1-\left(\frac
hR-1\right)^2}}=\frac{R}{\pi}
 \]

\end{document}